\documentclass[12pt,a4paper]{article}
\usepackage{graphicx}
\usepackage{amssymb}
\usepackage{amsmath}
\usepackage{bm}
\usepackage[dvipsnames]{xcolor}
\usepackage{theorem}
\usepackage{subfigure}
\usepackage{cite}
\usepackage{amsfonts}
\usepackage{bm}

\usepackage[sort&compress,numbers, merge]{natbib}

\definecolor{Orange}{cmyk}{0,0.61,0.87,0}
\definecolor{JungleGreen}{cmyk}{0.99,0,0.52,0}
\definecolor{OliveGreen}{cmyk}{0.64,0,0.95,0.40}
\definecolor{Brown}{cmyk}{0,0.81,1,0.60}
\definecolor{RoyalBlue}{cmyk}{0.71,0.53,0,0.12}

\newcommand{\be}{\begin{equation}}
\newcommand{\ee}{\end{equation}}
\newcommand{\bea}{\begin{eqnarray}}
\newcommand{\eea}{\end{eqnarray}}
\newcommand{\eq}[1]{Eq.~(\ref{#1})}
\newcommand{\la}{\langle}
\newcommand{\ra}{\rangle}
\newcommand{\amp}{\mathcal{A}}
\newcommand{\Mp}{M_{\rm Pl}}

\usepackage{tikz}
\usetikzlibrary{arrows,shapes}
\usetikzlibrary{trees}
\usetikzlibrary{matrix,arrows} 				
\usetikzlibrary{calc} 
\usetikzlibrary{positioning}				
\usetikzlibrary{calc,through}				
\usetikzlibrary{decorations.pathreplacing}  
\usepackage[tikz]{bclogo} 					
\usepackage{pgffor}							

\usetikzlibrary{decorations.pathmorphing}	
\usetikzlibrary{decorations.markings}
\usetikzlibrary{intersections}				
\tikzset{
    vector/.style={decorate, decoration={snake}, draw},
    graviton/.style={decorate, decoration={snake,amplitude=1.5pt}, draw},
    fermion/.style={postaction={decorate},
        decoration={markings,mark=at position .55 with {\arrow{>}}}},
    fermionbar/.style={draw, postaction={decorate},
        decoration={markings,mark=at position .55 with {\arrow{<}}}},
    fermionnoarrow/.style={},
    gluon/.style={decorate,
        decoration={coil,amplitude=4pt, segment length=5pt}},
    scalar/.style={dashed, postaction={decorate},
        decoration={markings,mark=at position .55 with {\arrow{>}}}},
    scalarbar/.style={dashed, postaction={decorate},
        decoration={markings,mark=at position .55 with {\arrow{<}}}},
    scalarnoarrow/.style={dashed,draw},
	provector/.style={decorate, decoration={snake,amplitude=2.5pt}, draw},
	antivector/.style={decorate, decoration={snake,amplitude=-2.5pt}, draw},
	    electron/.style={draw=black, postaction={decorate},
        decoration={markings,mark=at position .55 with {\arrow[draw=black]{>}}}},
	bigvector/.style={decorate, decoration={snake,amplitude=4pt}, draw},
	vectorscalar/.style={loosely dotted,draw=black, postaction={decorate}},
}

\usepackage[colorlinks=True, citecolor=blue, linkcolor=blue, urlcolor=blue,linktocpage]{hyperref}

\allowdisplaybreaks[1]

\makeatletter
\renewcommand\@dotsep{200}   
\makeatother

\usepackage{soul}

\begin{document}

\begin{titlepage}
\begin{center}
\hfill TUM-HEP-1363/21

\vspace{2.0cm}
{\Large\bf  
RG of GR from On-shell Amplitudes}

\vspace{1.0cm}
\renewcommand{\thefootnote}{\fnsymbol{footnote}}
{\small \bf 
Pietro Baratella$^{a}$,
Dominik Haslehner$^{a}$,
Maximilian Ruhdorfer$^{a,b}$,\\
Javi Serra$^{a}$ and
Andreas Weiler$^{a}$\,\footnote{E-mail: \href{mailto:javi.serra@tum.de}{javi.serra@tum.de}, \href{mailto:dominik.haslehner@tum.de}{dominik.haslehner@tum.de}, \href{mailto:m.ruhdorfer@cornell.edu}{m.ruhdorfer@cornell.edu}, \href{mailto:p.baratella@tum.de}{p.baratella@tum.de}, \\\href{mailto:andreas.weiler@tum.de}{andreas.weiler@tum.de}}
}

\vspace{0.7cm}
{\it\footnotesize
${}^a$Technische Universit\"{a}t M\"{u}nchen, Physik-Department, 85748 Garching, Germany\\
${}^b$Department of Physics, LEPP, Cornell University, Ithaca, NY 14853, USA
}

\vspace{0.9cm}
\abstract{\noindent We study the renormalization group of generic effective field theories that include gravity. We follow the on-shell amplitude approach, which provides a simple and efficient method to extract anomalous dimensions avoiding complications from gauge redundancies. As an invaluable tool we introduce a modified helicity $\tilde{h}$ under which gravitons carry one unit instead of two. With this modified helicity we easily explain old and uncover new non-renormalization theorems for theories including gravitons. We provide complete results for the one-loop gravitational renormalization of a generic minimally coupled gauge theory with scalars and fermions and all orders in $\Mp$, as well as for the renormalization of dimension-six operators including at least one graviton, all up to four external particles.}

\end{center}
\end{titlepage}
\renewcommand{\thefootnote}{\arabic{footnote}}
\setcounter{footnote}{0}

\tableofcontents
\newpage
%
%
\section{Introduction}
%
On-shell amplitude methods have proven extremely successful in studying the properties of four-dimensional elementary particle theories. This is especially true for those theories, like gauge and gravitational theories, that require the introduction of field redundancies when described in terms of an ordinary Lagrangian. In the on-shell formalism we are going to use --~where an amplitude is defined through its expression in terms of spinor-helicity variables~-- dealing with gauge bosons or gravitons is conceptually and practically the same as with scalars or Weyl fermions (see e.g.~\cite{Elvang:2013cua,Dixon:2013uaa,Cheung:2017pzi}).

Recently, a systematic treatment of low-energy effective (field) theories (EFTs) with amplitude methods has been initiated. This includes the identification of amplitude bases for higher-dimensional operators~\cite{Shadmi:2018xan,Ma:2019gtx,Aoude:2019tzn,Durieux:2019eor,Durieux:2019siw,Falkowski:2019zdo,Henning:2019enq}, as well as the use of unitarity methods to study the renormalization group (RG) structure of EFTs~\cite{Caron-Huot:2016cwu,Craig:2019wmo,Bern:2019wie,EliasMiro:2020tdv,Baratella:2020lzz,Jiang:2020mhe,Bern:2020ikv,Baratella:2020dvw}. Concerning in particular the study of quantum effects in EFTs, various interesting structures have emerged, which at the same time deepen our understanding of the RG and organize and simplify enormously computations: these include `helicity' selection rules~\cite{Cheung:2015aba} and `angular momentum' selection rules~\cite{Jiang:2021tqo}.

The purpose of this work is to extend these methods to cover generic EFTs with gravitons as low energy degrees of freedom, which includes the world we live in \cite{Ruhdorfer:2019qmk}. While experimentally the quantum nature of gravity is not yet accessible, assuming that it is quantized like all other dynamical degrees of freedom in nature is perhaps the most conservative assumption, whose consequences we study here.
In particular, we will devote special attention to the study of gravity's RG, whose intrinsic theoretical appeal dates back to the 70s, with the seminal works of t' Hooft, Veltman, Deser and others~\cite{tHooft:1974toh,Deser:1974cy,Barvinsky:1981rw,Deser:1974cz,Deser:1974xq,Goroff:1985sz}, with many interesting developments since then, especially in the context of supergravity (see e.g.~\cite{Bern:2018jmv,Abreu:2019rpt,Henn:2019rgj} for recent work on the ultraviolet behaviour of $\mathcal{N} = 8$ supergravity).

The interest in quantum gravitational corrections certainly extends beyond the divergent structure of gravitational theories \cite{Donoghue:1994dn}, with implications in the context of inflation, see \cite{Donoghue:2017pgk} and reference therein, modified gravity theories, e.g.~\cite{deRham:2013qqa,Noller:2019chl}, or the weak gravity conjecture~\cite{Cheung:2014ega,Bellazzini:2019xts,Charles:2019qqt,Jones:2019nev}. On-shell techniques for loop computations in gravity have also been used in~\cite{Dunbar:1995ed,Norridge:1996he,Bern:2015xsa,Bern:2017puu,Dunbar:2017szm,Abreu:2020lyk}, and have recently led to a burst of novel calculations in classical gravitational dynamics motivated by the detection of gravitational waves, with many interesting new results (see e.g.~\cite{Bern:2020gjj}).

\vspace{.2cm}

Approaching EFTs including gravitons with on-shell methods allowed us to uncover very interesting structures, both at tree-level and at one-loop. Concerning tree-level helicity amplitudes including standard $|h_i|\leq 1$ degrees of freedom \emph{and} gravitons, we see a remarkable extension of the non-trivial fact, observed in ordinary renormalizable gauge theories in \cite{Cheung:2015aba}, that almost all 4-point amplitudes constructed with marginal couplings, with only one exception, have $h\equiv\sum_i h_i=0$. We find that the natural extension of this to amplitudes including minimally coupled gravitons requires the introduction of a modified helicity, that we call $\tilde{h}$, under which gravitons carry \emph{one} unit instead of two! Modulo the same exceptional four-fermion amplitude with $|h|=2$ (absent in supersymmetry), we find that \emph{all} 4-point amplitudes in a marginal theory minimally coupled to gravity have $\tilde{h}=0$.

This fact, which is highly non obvious as we are going to see, is then successfully employed for the study of the one-loop renormalization structure of the same class of minimally coupled theories, including ${\cal O}(\Mp^{-2})$ effects whose systematic study is new. As a matter of fact, our bounds on the tree-level helicity structure allows us to simplify the computations of RG effects enormously, by \emph{a priori} drastically reducing the number of relevant counterterms. As a complementary tool we will systematically take inputs from angular momentum analysis, which provides further insight into the structure of anomalous dimensions.

As it is emerging more and more clearly, the structure of low energy EFTs which are consistent with the principles of unitarity, locality and causality is highly constrained~\cite{Adams:2006sv}, with infinitely many positivity conditions on the Wilson coefficients of the EFT expansion (see e.g.~\cite{Tolley:2020gtv,Caron-Huot:2020cmc,Bellazzini:2020cot,Arkani-Hamed:2020blm} for some of the latest results) showing up at tree-level and, even more interestingly, extendable to the loop level. In this direction, we present some results on the one-loop renormalization of certain dimension-eight operators whose Wilson coefficients are expected to be strictly positive in the absence of gravity (gravitational positivity bounds have been recently discussed in e.g.~\cite{Bellazzini:2019xts,Alberte:2020jsk,Caron-Huot:2021rmr,Bern:2021ppb}). Interestingly, we find that all the dimension-eight coefficients grow towards the infrared, i.e.~their anomalous dimensions are negative, consistently with the standard positivity constraints.

The paper is organized as follows. In Section \ref{sectree} we discuss bounds on the total helicity of tree-level amplitudes including minimally coupled gravitons, introducing the modified helicity $\tilde{h}$. In Section \ref{anomdim} we explain our method to extract UV anomalous dimensions from on-shell amplitudes, and derive new non-renormalization theorems for theories involving gravitons. In Section \ref{toy} we use our findings to derive the renormalization structure of arbitrary theories coupled to gravity, at the level of 4-point amplitudes and up to dimension-eight EFT operators. Then we present our conclusions in Section \ref{conclusions}.

In several appendices we expand on the calculations leading to our results. In Appendix \ref{ward} we show how the gravitational helicity bounds on tree-level amplitudes can be understood from supersymmetric Ward identities, and in \ref{bubble} how Hermite reduction is used to extract the massive bubble coefficients of the loop amplitudes. We discuss the (non-)renormalization of the energy momentum tensor and collinear anomalous dimensions in Appendix \ref{collinear}, non-minimal gravitational couplings of scalars and fermions in \ref{conformalgen}, and operator mixing and (loop) power counting in gravity in Appendix \ref{opMix}.
%
\section{Helicity bounds on gravity amplitudes at tree level}
\label{sectree}
%
Here and throughout the present paper we will consider scattering amplitudes for massless states with
definite helicity, whose momenta are considered as all-incoming. An $n$-point amplitude will be denoted
by $\amp (1_{\Phi_{1}},2_{\Phi_{2}},\ldots,n_{\Phi_{n}})$, where $\Phi_i$
specifies the type of particle, i.e.~$\phi$ for scalars, $\psi$ and $\bar{\psi}$ for respectively helicity $\pm 1/2$ fermions, while $V_{\pm}$ and $h_{\pm}$ will denote respectively positive and negative helicity gauge bosons and gravitons.
When some labels are not necessary, we will suppress them for simplicity, and
will often specify amplitudes according to just their number of legs, as $\amp_n$. To write down their explicit expression, we will use the spinor helicity variables, following the conventions of \cite{Baratella:2020lzz}.

The advantages that come from organizing amplitudes according to the helicity of the
external states are enormous and well known. First of all, this allows to systematically take into
account the constraints coming from the little group. Another very fruitful observation is that, in a \emph{marginal} theory (in particular, where no 3-point scalar couplings are present), almost all tree-level 4-point amplitudes $\amp (1,2,3,4)$ that are non-zero satisfy the following condition \cite{Cheung:2015aba}
\be\label{amp4}
h_1+h_2+h_3+h_4 =0\,,
\ee
where $h_i$ is the helicity of state $i$.
The only exceptions to the above rule are the four-fermion amplitude
$\amp(\psi,\psi,\psi,\psi)$ and its complex conjugate
$\amp (\bar{\psi},\bar{\psi},\bar{\psi},\bar{\psi})$,\footnote{The complex conjugated amplitude of $\amp(1,2,\ldots,n)$ is $\amp(\bar{1},\bar{2},\ldots,\bar{n})$, where the bar denotes a state with opposite quantum numbers, including helicity (but same momentum). In the spinor-helicity formalism, it is obtained by $\la \cdot\ra\leftrightarrow [\cdot ]$ and complex conjugation of all couplings.} with total helicity $h\equiv \sum_i h_i$ of 
respectively 2 and $-2$.
These amplitudes can be zero under some group theoretic condition but are in general present (e.g. in the SM).

The fact expressed by \eq{amp4} is highly non-trivial, as we are now going to explain.
In a marginal theory all $n\geq 4$ point amplitudes, with the exception of four-scalar amplitudes for which \eq{amp4} is trivially satisfied, are on-shell constructible from lower-point ones,\footnote{\label{foot:marginal}$n$-point amplitudes 
have mass dimension $[\amp_n ] = 4-n$. In a theory with only marginal couplings the negative mass dimension for $n>4$ has to be supplied by kinematic invariants. Locality implies that configurations in which these kinematic invariants vanish and the amplitude becomes singular must correspond to a factorization channel of the amplitude into lower-point amplitudes. Therefore all  $n>4$ amplitudes are on-shell constructible. At $n=4$ the only dimensionless amplitude which does not factorize is a constant, corresponding to a four-scalar amplitude. From a field theory point of view this can be phrased as the fact that, with the exception of a four-scalar operator, there are no marginal local operators comprised of $n\geq 4$ fields which could generate a contact amplitude.} and therefore their total helicity satisfies
\be \label{eq:helAddition}
h(\amp_n ) = h (\amp_m ) + h( \amp_{n-m+2} )\,,
\ee
where $\amp_m$ and $\amp_{n-m+2}$ represent any pair of lower-point amplitudes into which $\amp_{n}$
factorizes. The fundamental building blocks in this recursive relation are 3-point amplitudes, which are completely fixed by the little-group scaling of the external particles. Dimensional analysis additionally relates the total helicity $h(\amp_3 )$ to the mass dimension of the coupling constant $g_3$
\be \label{eq:3ptHel}
| h(\amp_3 ) | = 1 - [g_3 ]\,,
\ee
implying that 3-point amplitudes in a marginal theory have total helicity $| h(\amp_3 ) | = 1$. The helicity composition rule in Eq.~(\ref{eq:helAddition}) therefore fixes the helicity of 4-point amplitudes to be $| h(\amp_4 )| =0,2$ but does not explain the vanishing of most $| h(\amp_4 )| =2$ amplitudes. An explanation requires either direct computation or the use of further tools, such as supersymmetric Ward identities \cite{Azatov:2016sqh}.

Taking $| h(\amp_4 )| =0$ as an input, and using that $| h(\amp_3 ) | = 1$ in a marginal theory, Eq.~(\ref{eq:helAddition}) yields a non-trivial bound on the helicity of all $n$-point amplitudes which do not contain the exceptional four-fermion amplitude in a factorization channel
\be \label{eq:nPointTreeHelBound}
|h(\amp_n )| \leq n-4~~~~~~~~~~~~(n\geq 4)\,.
\ee
This tree-level helicity bound in marginal theories was successfully
employed in \cite{Cheung:2015aba} to derive powerful non-renormalization theorems for EFTs (especially the SM EFT). In the following, we are going to provide a very natural generalization of these results --~bound on $|h(\amp_n )|$ and non-renormalization theorems~-- to theories
involving gravitons, as well as (supersymmetric) theories with gravitons and gravitinos.
\subsection{Gravity minimally coupled to marginal theories} 
\label{eq:treeMinCoupl}
Including a minimal coupling of gravity to an otherwise marginal theory changes the above picture, as we now describe. Before presenting our results, it is important to say some words about precisely what set of amplitudes we are going to make our statements on, and what are their relevant properties. The most straightforward definition of the set of amplitudes we are considering (though somewhat outside of the spirit of the paper) is in terms of a marginal Lagrangian minimally coupled to gravity, called ${\cal L}$: we consider all tree-level on-shell amplitudes constructed from it. Very schematically, these amplitudes are of the form
\be
\amp_n = {\rm couplings}\times {\rm polarizations}\times \left(\{ {\rm poles} \}+ \{ {\rm contact} \}\right)\,,
\ee
where `polarizations' stands for a {numerator}, written with the minimal required amount of spinor-helicity variables, accounting for the external helicities; the terms in parentheses are functions of Mandelstam invariants, the first being reconstructible from lower amplitudes with factorization arguments, while the second being purely contact, i.e.~with no poles. Contrary to the analogous problem without gravity, which is reviewed in Footnote~\ref{foot:marginal}, 
the presence of the dimensionful coupling constant $1 / \Mp$ in gravity
makes the characterization of $\{$contact$\}$ somehow complicated. In most cases, like for amplitudes with at least one external graviton~\cite{ArkaniHamed:2008yf,Cheung:2008dn} or with $|h|>2,\, n \geq 4$~\cite{Cohen:2010mi}, the contact term is shown to be absent, so that the amplitude is on-shell constructible. But this is not always the case, and a subset of our amplitudes, like for example some of those with $h=0$, can come with contact term ambiguities, for any $n$.

However, on-shell constructibility --~or equivalently the absence of $\{$contact$\}$~-- is not needed for our scope, which is to inductively infer the possible range of helicities of $\amp_n$, like we did in the previous section. The only property which our set of amplitudes is required to satisfy is the following recursive fact: that the $\{$poles$\}$ part factorizes, on a given pole, to $\amp_m\times\amp_{n-m+2}$,
with $\amp_m$ and $\amp_{n-m+2}$ being lower-point amplitudes which are themselves ``minimally coupled'', i.e.~extracted from ${\cal L}$ like $\amp_n$ is. But what if $\amp_n$ is a pure contact amplitude? Once again, we let the Lagrangian guide us. Considering, as we do here, a minimal coupling of gravity to a marginal theory, and with the only exception of the four-scalar amplitude that we already discussed, pure contact terms in ${\cal L}$ {with $n \geq 4$}, that correspond to purely $\{$contact$\}$ amplitudes, are actually absent, and the inductive step can always be invoked.

\vspace{.2cm}

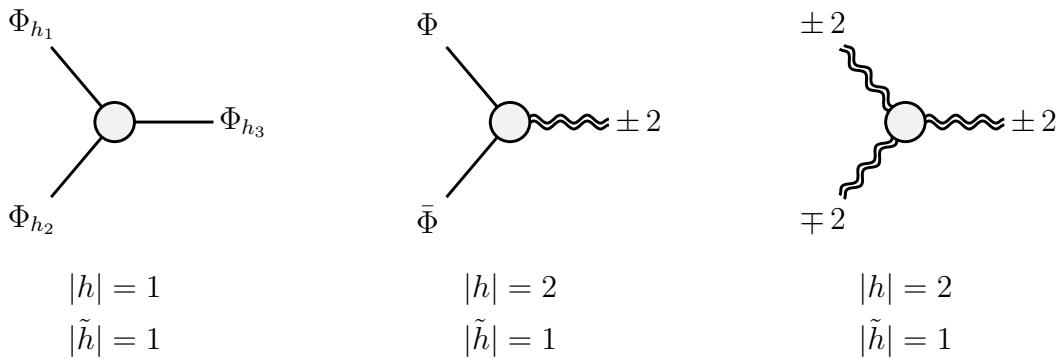
\begin{figure}[t]
\centering
\begin{tikzpicture}[line width=1.1 pt, scale=1.3, baseline=(current bounding box.center)]
	\draw[fermionnoarrow] (-130:1)--(0,0);
	\draw[fermionnoarrow] (130:1)--(0,0);
	\draw[fermionnoarrow] (0:1)--(0,0);
	\draw[fill=gray!10] (0,0) circle (.2);
	\node at (-130:1.3) {$\Phi_{h_2}$};
	\node at (130:1.3) {$\Phi_{h_1}$};
	\node at (1.3,0) {$\Phi_{h_3}$};
	\node at (0,-1.7) {$|h|= 1$};
	\node at (0,-2.2) {$|\tilde{h}| = 1$};
	\begin{scope}[shift={(4,0)}]
	\draw[fermionnoarrow] (130:1)--(0,0);
	\draw[fermionnoarrow] (-130:1)--(0,0);
	\draw[graviton] (1,0.03)--(-0.03,0.03);
	\draw[graviton] (1,-0.03)--(-0.03,-0.03);
	\draw[fill=gray!10] (0,0) circle (.2);
	\node at (-130:1.3) {$\bar{\Phi}$};
	\node at (130:1.3) {$\Phi$};
	\node at (1.3,0) {$\pm\, 2$};
	\node at (0,-1.7) {$|h|= 2$};
	\node at (0,-2.2) {$|\tilde{h}| = 1$};
	\end{scope}
	\begin{scope}[shift={(8,0)}]
	\begin{scope}[rotate = 130]
	\draw[graviton] (1,0.03)--(0,0.03);
	\draw[graviton] (1,-0.03)--(0,-0.03);
	\end{scope}
	\begin{scope}[rotate = -130]
	\draw[graviton] (1,0.03)--(0,0.03);
	\draw[graviton] (1,-0.03)--(0,-0.03);
	\end{scope}
	\draw[graviton] (1,0.03)--(0,0.03);
	\draw[graviton] (1,-0.03)--(0,-0.03);
	\draw[fill=gray!10] (0,0) circle (.2);
	\node at (-130:1.3) {$\mp\, 2$};
	\node at (130:1.3) {$\pm\, 2$};
	\node at (1.3,0) {$\pm\, 2$};
	\node at (0,-1.7) {$|h|=2$};
	\node at (0,-2.2) {$|\tilde{h}| = 1$};
	\end{scope}
 \end{tikzpicture}
 \caption{\emph{Three-point amplitudes in a marginal theory minimally coupled to gravity. Amplitudes exclusively among the matter particles $\Phi$, which can be scalars $\phi$, fermions $\psi$ or vectors $V$, carry total helicity $h= \pm 1$. Gravitational amplitudes in contrast have helicity $h = \pm 2$, the sign being determined by the graviton helicities only. Characterized in terms of a modified helicity $\tilde{h} = h -\tfrac{1}{2} h^g$, all 3-point amplitudes are on the same footing.}}
 \label{fig:modHelicity}
 \end{figure}

As we said, gravity comes with a dimensionful coupling constant, the inverse Planck mass $1 / \Mp$, which introduces a new set of 3-point amplitudes with $|h(\amp_3)| = 2$ (see Fig.~\ref{fig:modHelicity}). 
Thus a general 4-point amplitude might now have total helicity $|h(\amp_4 )|=0,1,2,3,4$: $|h|=0,2$ if it factorizes into two marginal amplitudes, $|h| = 1,3$ in case it factorizes into one marginal and one gravitational amplitude and $|h| = 0,4$ if the factorization is into two gravitational amplitudes. This seems to suggest that a non-trivial bound on the helicity requires a detailed knowledge of all factorization channels, and in particular of how many of them contain gravitational amplitudes. 

A remedy of this shortcoming of the traditional helicity counting approach requires the characterization of amplitudes according to a quantity which treats gravitational and marginal amplitudes on the same footing. Such a quantity can be found by noticing that, due to the parity conserving nature of minimally coupled gravity, all 3-point amplitudes are of the form $\amp(h_\pm,\Phi,\bar{\Phi})$ with the total helicity being completely determined by the helicity of the graviton ${h_{\pm}}$. Taking only half of the graviton's helicity would put it on the same footing as an amplitude with gauge bosons 
(that are always parity preserving and with ${h}=\pm 1$).
Motivated by this observation, we define a modified helicity $\tilde{h}$, which for all particles with $|h|\leq 1$ coincides with the regular helicity, i.e.~$\tilde{h} = h$, but under which gravitons carry $\tilde{h} = \pm 1$. For an arbitrary $n$-point amplitude the modified helicity can be defined as
\be \label{eq:modHelicity}
\tilde{h} (\amp_n ) \equiv
h(\amp_n ) - \tfrac{1}{2} h^{g} (\amp_n )\,,
\ee
where $h^{g} (\amp_n)$ is the sum of all external graviton helicities in $\amp_n$. As is shown in Fig.~\ref{fig:modHelicity}, the modified helicity puts all {$3$-point} amplitudes on the same footing, such that 
\be\label{h3pt}
|\tilde{h}(\amp_3)|=1~~~~~~~~\textrm{for \emph{all} {3-point} amplitudes}
\ee
in a minimally coupled marginal theory. The benefits of the modified helicity are not limited to $3$-point amplitudes: $\tilde{h}$ is additive with respect to factorization in the same way as the regular helicity (cf.~Eq.~(\ref{eq:helAddition})). Thus all $4$-point amplitudes, including gravitational ones, can only come with $|\tilde{h}|=0, 2$. As previously discussed all $4$-point amplitudes containing exclusively marginal couplings have $\tilde{h}=0$, with the exception of the four-fermion amplitude. The non-trivial vanishing of $|\tilde{h}|=2$ amplitudes surprisingly extends to gravitational ones. Through direct calculation, or more elegantly from supersymmetric Ward identities (see Appendix~\ref{ward}), one indeed finds that
\be \label{eq:htilde2amp}
\begin{split}
0 &= \amp(h_{+},h_{+},h_{+},h_{-})= \amp(h_{+},h_{+},\phi,{\phi})=\amp(h_{+},h_{+},V_+,V_-)=\amp(h_{+},h_{+},\bar{\psi},\psi)\\
&=\amp(h_{+},V_+,V_+,V_-)=\amp(h_{+},V_+,\phi,\phi )= \amp(h_{+},\psi,\psi,\phi)= \amp(h_{+},V_+,\bar{\psi},\psi) \,,
\end{split}
\ee
and the same holds for the complex conjugated amplitudes, that is those where particles and antiparticles are
exchanged. Hence
\be\label{h4pt}
\tilde{h}(\amp_4)= 0~~~~~~~~\textrm{for \emph{all} 4-point amplitudes}\,,
\ee
modulo the exceptional one, in a marginal theory minimally coupled to gravity. By recursively constraining the $\tilde{h}$ of $n$-point amplitudes from their factorization into lower-point amplitudes, and using \eq{h4pt} as our ``base case'', we can  straightforwardly generalize Eq.~(\ref{eq:nPointTreeHelBound}) to gravitational amplitudes, and find
%
\be \label{eq:nPointTreeModHelBound}
\big|\tilde{h}(\amp_n)\big| \leq n-4~~~~~~~~~~~~(n\geq 4)\,.
\ee
This will allow us to formulate unified non-renormalization theorems in the spirit of \cite{Cheung:2015aba} for EFTs including gravity, and in particular for the GRSMEFT \cite{Ruhdorfer:2019qmk}.

\vspace{.2cm}

Before ending this Section let us make a few comments about the modified helicity. The strength of the modified helicity bound in Eq.~(\ref{eq:nPointTreeModHelBound}) is its generality. It applies to both gravitational and marginal theories and automatically reduces to Eq.~(\ref{eq:nPointTreeHelBound}), the bound for the regular helicity, in the absence of gravity. However, this generality comes with a price: some amplitudes which are allowed by Eq.~(\ref{eq:nPointTreeModHelBound}) are actually forbidden by regular helicity composition rules. One such example is the graviton-gauge boson amplitude $\amp(h_{+},h_{+},V_{-},V_{-})$, which has $\tilde{h}= 0$ and therefore seems to be allowed by the modified helicity bound, but whose actual helicity $h=2$ is not consistent with the fact that it has to factorize into two gravitational 3-point amplitudes (allowing only $|h|=0,4$ as discussed previously). The reason for this shortcoming is the fact that gravitons are effectively treated as helicity-one particles. If they appear in amplitudes together with actual helicity-one vectors, an additional piece of information is required to tell if the amplitude is allowed or not. Such information can be provided by $h^g$, the total helicity in gravitons that the amplitude carries. At four points $|h^g (\amp_4)| \leq 2$ and it can at most increase by two units for each additional external state, implying that
\be \label{eq:bound2}
\big|h^g (\amp_n)\big| \leq 2 (n-3)\,.
\ee
This additional information forbids the $\amp(h_{+},h_{+},V_{-},V_{-})$ amplitude. Hence Eq.~(\ref{eq:nPointTreeModHelBound}) should be seen as a conservative bound which in some cases can be refined by Eq.~\eqref{eq:bound2} or an explicit study of the factorization channels.\footnote{{Equivalently, the helicity bounds Eqs.~(\ref{eq:nPointTreeModHelBound}) and (\ref{eq:bound2}) can be re-written as $|h(\amp_n)| \leq 2n-7$ for $|h| > \frac{1}{2}|h^g|$ or simply $|h(\amp_n)| \leq n-3 = \frac{1}{2}|h^g(\amp_n)|$ otherwise. In addition, for
 $h, -h^g \geq 0$ or $-h, h^g \geq 0$, then $|h(\amp_n)| \leq n-4$.}}

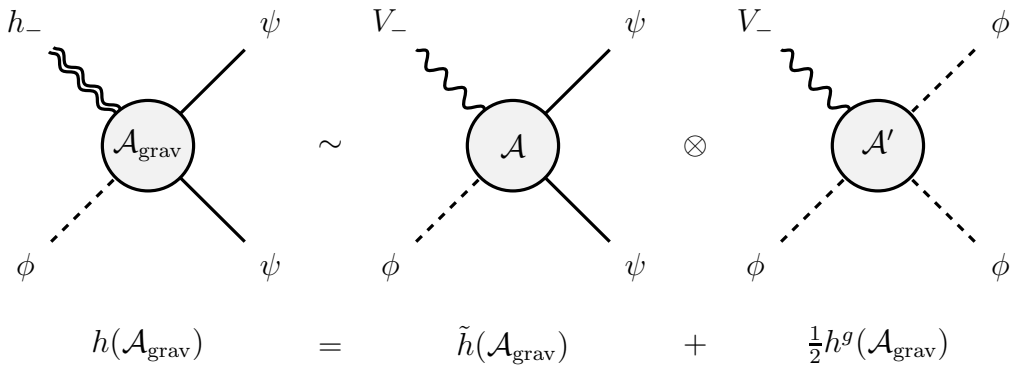
\begin{figure}[t]
\centering
\begin{tikzpicture}[line width=1.2 pt, scale=1.2, baseline=(current bounding box.center)]
	\begin{scope}[rotate = 135]
	\draw[graviton] (1.5,0.035)--(0,0.035);
	\draw[graviton] (1.5,-0.035)--(0,-0.035);
	\end{scope}
	\draw[scalarnoarrow] (-135:1.5)--(0,0);
	\draw[fermionnoarrow] (45:1.5)--(0,0);
	\draw[fermionnoarrow] (-45:1.5)--(0,0);
	\draw[fill=gray!10] (0,0) circle (.5);
	\node at (-135:1.9) {$\phi$};
	\node at (135:1.9) {$h_-$};
	\node at (45:1.9) {$\psi$};
	\node at (-45:1.9) {$\psi$};
	\node at (0,0) {$\amp_{\rm grav}$};
	\node at (2,0) {$\sim$};
	\node at (0,-2.2) {$h(\amp_{\rm grav})$};
	\node at (2,-2.25) {$=$};
	\node at (6,-2.2) {$+$};
	\begin{scope}[shift={(4.0,0)}]
	\draw[scalarnoarrow] (-135:1.5)--(0,0);
	\draw[vector] (135:1.5)--(0,0);
	\draw[fermionnoarrow] (45:1.5)--(0,0);
	\draw[fermionnoarrow] (-45:1.5)--(0,0);
	\node at (-135:1.9) {$\phi$};
	\node at (135:1.9) {$V_-$};
	\node at (45:1.9) {$\psi$};
	\node at (-45:1.9) {$\psi$};
	\draw[fill=gray!10] (0,0) circle (.5);
	\node at (0,-2.2) {$\tilde{h} (\amp_{\rm grav})$};
	\node at (0,0) {$\amp$};
	\end{scope}
	\node at (6,0) {$\otimes$};
	\begin{scope}[shift={(8.0,0)}]
	\draw[scalarnoarrow] (-135:1.5)--(0,0);
	\draw[vector] (135:1.5)--(0,0);
	\draw[scalarnoarrow] (45:1.5)--(0,0);
	\draw[scalarnoarrow] (-45:1.5)--(0,0);
	\draw[fill=gray!10] (0,0) circle (.5);
	\node at (-135:1.9) {$\phi$};
	\node at (135:1.9) {$V_-$};
	\node at (45:1.9) {$\phi$};
	\node at (-45:1.9) {$\phi$};
	\node at (0,-2.2) {$\tfrac{1}{2} h^g (\amp_{\rm grav})$};
	\node at (0,0.05) {${\amp}'$};
	\end{scope}
 \end{tikzpicture}
 \caption{\emph{Interpretation of the modified helicity bounds for the $\amp (h_-, \psi,\psi ,\phi )$ amplitude in terms of double copy relations. The modified helicity treats the graviton as a helicity-one particle and preserves the helicity of $|h|\leq 1$ matter. Thus splitting the helicity as $h=\tilde{h} + \tfrac{1}{2} h^g$ implies a factorization of the amplitude into a helicity $\tilde{h}$ gauge amplitude $\amp$ including all matter particles and a helicity $\tfrac{1}{2} h^g$ gauge amplitude ${\amp}'$ with the matter particles replaced by scalars. The modified helicity bound $|\tilde{h}|\leq n-4$ corresponds in this picture to the helicity bound for marginal theories applied to $\amp$.}}
 \label{fig:doubleCopy}
\end{figure}

The idea of using a modified helicity to study gravitational amplitudes has some resemblance with the famous KLT relations \cite{Kawai:1985xq} and the double copy structure of gravity (see \cite{Bern:2019prr} for a review), which relate gravity amplitudes to products of two gauge-theory amplitudes. If one of the gauge-theory amplitudes vanishes due to standard helicity bounds, so does the gravity amplitude. It seems natural to interpret the modified helicity bounds in a similar fashion. The modified helicity effectively treats the graviton as a helicity-one particle and preserves the helicity of the remaining matter, what in the double copy language corresponds to a factorization of gravitons into two vectors, and each matter particle into a same helicity state and a scalar, similar to the KLT construction in \cite{Bern:1999bx}. This is schematically shown in Fig.~\ref{fig:doubleCopy} for the $\amp (h_- ,\psi,\psi ,\phi )$ amplitude.\footnote{Note that here we do not try to construct explicit KLT relations for such a factorization of gauge theories. The aim of this discussion is merely to highlight the structural similarity of the modified helicity to double copy constructions.}
The first factor is a gauge amplitude with total helicity $\tilde{h}$, which 
is only non-vanishing for $|\tilde{h}| \leq n-4$ according to the helicity bounds for marginal theories. This exactly coincides with the modified helicity bound. In an explicit KLT construction we would also get a helicity bound from the second factor. However, since the second gauge theory must contain three-scalar couplings, the 4-point amplitude can have helicity $|h|=0,1$ resulting in a generally weaker bound of $|h| \leq n-3$. Note that, after we identify $h= \tfrac{1}{2} h^g$, this additional bound coincides with Eq.~\eqref{eq:bound2}.

It is important to point out that, while the modified helicity bounds are consistent with KLT and double copy constructions and might have an interpretation in terms of such relations, the derivation of the bounds does not depend on them. 

Let us finally mention that it is straightforward to extend the definition of the modified helicity to massless helicity-$3/2$ particles $\zeta$, that is gravitinos. Note, however, that a consistent theory with massless gravitinos requires superpartners for all matter particles \cite{McGady:2013sga}, i.e.~the theory must be fully supersymmetric. All of the leading gravitino 3-point amplitudes have total helicity $|h|=2$ and are of the form $\amp (h_+ , \zeta_+, \zeta_-),\, \amp (\zeta_+ , V_+, \bar\psi),\, \amp (\zeta_+ , {\psi}, \phi)$. 
These have the same helicity structure as matter amplitudes in a marginal theory if we treat the gravitino as a helicity-$1/2$ particle, i.e.~if we assign to it the modified helicity $\tilde{h} = 1/2$. With this the definition the modified helicity of any amplitude $\amp_n$ is given by
\be
\tilde{h}(\amp_n ) = h (\amp_n ) -\tfrac{1}{2} h^g (\amp_n) -\tfrac{2}{3} h^{\zeta} (\amp_n)\,,
\ee
where $h^{\zeta} (\amp_n)$ is the sum of external gravitino helicities. At tree level any 4-point amplitude satisfies $\tilde{h} (\amp_4 ) = 0$ (see Appendix~\ref{ward}),\footnote{Note that there is no exception to this rule, since the $|\tilde{h}| = 2$ four-fermion amplitude requires non-holomorphic Yukawa couplings to be non-vanishing. However, in a supersymmetric theory such couplings are absent.} implying that the bound in Eq.~\eqref{eq:modHelicity} also holds for gravitino amplitudes.

\subsection{Beyond minimal coupling}
\label{beyondmin}

So far we have been focusing on amplitudes which are constructed from either marginal interactions or minimal
coupling with gravity. Following an EFT point of view, we are now going to include all the interactions that are
allowed by the symmetries of the low energy theory. These new interactions are in one to one correspondence with higher-dimensional operators in an effective Lagrangian description.

In our approach, ``higher-dimensional amplitudes'' $\amp_i$ are defined through their expression in terms of spinor-helicity variables (in an all incoming configuration), and come with a dimensionless coupling $C_i$ that corresponds to a Wilson coefficient in the operator approach. Being new building blocks of the theory, they are  not tied to lower-point amplitudes by factorization, i.e.~they are contact amplitudes.

\begin{figure}[t]
\centering
\begin{tikzpicture}

\draw[thick,->] (0,0) -- (8,0) node[anchor=west] {$n$};
\draw[thick,->] (0,0) -- (0,6.5) node[anchor=south] {$\tilde{h}$};

\foreach \posx / \labelx in {1/3,3/4,5/5,7/6}
   \draw (\posx cm,2pt) -- (\posx cm,-2pt) node[anchor=north] {$\labelx$};

\foreach \posy / \labely in {1/0,2.5/1,4/2,5.5/3}
    \draw (2pt,\posy cm) -- (-2pt,\posy cm) node[anchor=east] {$\labely$};

\node[align=center] (n3h3) at (1,5.5) {{\small $F^3$};\\ {\small $C^3$}, {\small $C F^2$}};

\node[align=center] (n4h2) at (3,4) {{\small $F\psi^2\phi$}, {\small $\psi^4$},\\ {\small $F^2 \phi^2$}; {\small $C^2 \phi^2$}};

\node[align=center] (n4h0) at (3,1) {{\small $\psi^2\bar{\psi}^2$}, \\ {\small $\psi\bar{\psi} \phi^2 D$},\\ {\small $\phi^4 D^2$}};

\node[align=center] (n5h1) at (5,2.5) {{\small $\psi^2 \phi^3$}};

\node[align=center] (n6h0) at (7,1) {{\small $\phi^6$}};
\end{tikzpicture}
\caption{\emph{Classes of ${\cal O}(\Lambda^{-2})$ amplitudes in a generic theory including gravity, with gravitational amplitudes separated by a semicolon. Each class is characterized according to the modified helicity and number of external particles that the amplitude carries. The non-renormalization theorem in Eq.~(\ref{htildemixing}) implies that a Wilson coefficient $C_j$ can only renormalize a $C_i$ on its right, with the further condition that it lies inside the cone $\tilde{h}_i \leq \tilde{h}_{j} \pm (n_i-n_j)$. Our notation for contact amplitudes reflects the operator language, for example $\amp_{CF^2}$ is generated by an operator with a `right-handed' Weyl tensor \cite{Ruhdorfer:2019qmk} and two right-handed field strength tensors. A bar denotes the left-handed counterpart, and $D$ counts the number of derivatives that the operator carries.}}
\label{fig:helOpPlot}
\end{figure}
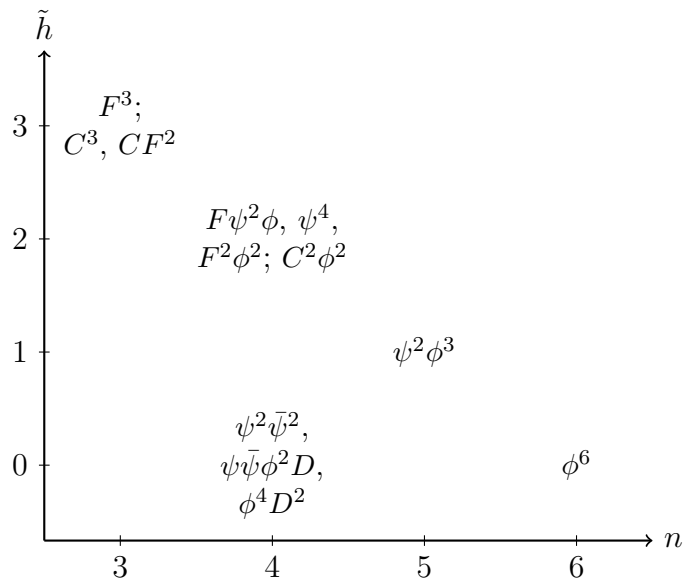

Similarly as in the operator approach, higher-dimensional amplitudes can be classified according to the inverse power of some UV-theory
scale $\Lambda$ that they carry. Notice that, when involving gravitons
as external states, amplitudes are also expected to carry a factor
$\Mp^{-1}$ for each graviton.
Beyond the $\Lambda$- and $\Mp$-scaling, as hinted by the previous analysis we also find it very convenient
to characterize amplitudes by their number of legs $n$ and their \emph{modified} helicity $\tilde{h}$. In Fig.~\ref{fig:helOpPlot}
we organize accordingly all amplitudes at $\Lambda^{-2}$ including those with gravitons.

As an example of our method, we list here all contact amplitudes at $\Lambda^{-1}$ and $\Lambda^{-2}$
that include at least one graviton.
At $\Lambda^{-1}$ we only find the following amplitude
\begin{align}
\amp_{C^2\phi}(1_{h_+},2_{h_+},3_\phi)&=C_{C^2\phi}\frac{[12]^4}{\Lambda\,\Mp^2}\,,
\end{align}
with $n=3$ and $\tilde{h}=2$. At $\Lambda^{-2}$ we have
\begin{align}
\amp_{CF^2}(1_{h_+},2_{V_+},3_{V_+})&=C_{CF^2}\frac{[12]^2[13]^2}{\Lambda^2\,\Mp}\,,\label{CF2} \\
\amp_{C^3}(1_{h_+},2_{h_+},3_{h_+})&=C_{C^3}\frac{[12]^2[23]^2[13]^2}{\Lambda^2\,\Mp^3}\,,
\end{align}
with $n=3$ and $\tilde{h}=3$, the second amplitude being constructed with only gravitons. Finally, we have
\be\label{C2phi2}
\amp_{C^2\phi^2}(1_{h_+},2_{h_+},3_\phi,4_\phi)=C_{C^2\phi^2}\frac{[12]^4}{\Lambda^2\,\Mp^2}\,,
\ee
that has $n=4$ and $\tilde{h}=2$. A systematic treatment of this problem goes beyond the scope
of this work and can be found in \cite{Ruhdorfer:2019qmk},
where the equivalent operator language is adopted. It is very important to observe that contact amplitudes suppressed by
powers of $\Lambda$ typically do not respect the $\tilde{h}$ bounds presented in the previous subsection.

As we will see in the following section, the Wilson coefficients $C_i$ of higher-dimensional amplitudes
acquire a scale dependence when loop corrections are included. This will be studied in Section~\ref{toy}, and the modified helicity will turn out to be an invaluable tool. 
%
%
\section{Anomalous dimensions from on-shell amplitudes}\label{anomdim}
%

Having presented the structure of tree-level amplitudes in a generic theory coupled to gravity,
we move on to the study of its properties at one loop, focusing in particular on the divergence structure, or
equivalently on the RG flow.

As we will see, the study of the RG is tightly connected to the properties of the theory at tree level.
This is because the UV-divergent part of a loop amplitude can be expressed in terms of products of two tree-level amplitudes integrated over some phase space. By virtue of this fact, quantitatively expressed in \eq{gamma}, we can then for example systematically translate the helicity bounds on tree amplitudes of Section~\ref{sectree} to similar bounds on amplitudes at one loop. We will see that the modified helicity $\tilde{h}$ that we have introduced plays a key role in the organization of the loop computation.

It was shown in \cite{Baratella:2020lzz} that, when there are no IR divergences, the anomalous
dimension $\gamma_i\equiv dC_i/d\ln\mu$, which encodes the RG flow of the coefficient $C_i$ in
$\amp_i$, is given by
\be\label{gamma}
\gamma_i \frac{\amp_i(1,2,\ldots,n)}{C_i}=-\frac{1}{4\pi^3}\sum_{\rm cuts}\,\sum_{\ell_{1},\ell_{2}}\sigma_{\ell_{1}\ell_{2}}\int d{\rm LIPS} ~\amp_L(\ldots,-\bar{\ell}_2,-\bar{\ell}_1)\,\amp_R(\ell_1,\ell_2,\ldots)\,,
\ee
where ${\amp}_{L,R}$ are tree-level on-shell amplitudes with at least four legs
and $\sum_{\rm cuts}$ stands for a sum
over all possible ways in which the external legs $1,2,\ldots ,n$ can be distributed in ${\amp}_{L}$
and ${\amp}_{R}$.\footnote{Notice that, in general, the cut implies a reordering of the external legs in the RHS
of \eq{gamma} with respect to the ordering 1,2,...,$n$ that is established in the LHS. When the reordering implies an odd number
of fermion permutations, the RHS of
\eq{gamma}, and similarly that of \eq{gamma2}, must include an additional minus sign.}
The second summation $\sum_{\ell_{1},\ell_{2}}$ is over all possible choices of internal legs,
to be chosen consistently in ${\amp}_{L}$ and ${\amp}_{R}$, i.e.~the internal legs of ${\amp}_{L}$ must carry opposite sign momentum, helicity and all other quantum numbers 
with respect to those of ${\amp}_{R}$. The factor $\sigma_{\ell_{1}\ell_{2}}$ is defined by
$\sigma_{\ell_{1}\ell_{2}}=(-i)^{F_{\ell_{1}\ell_{2}}}$, where $F_{\ell_{1}\ell_{2}}$ counts the number
of fermions in the list $\{\ell_{1},\ell_{2}\}$. This factor arises from the convention in which
$|-\lambda\ra=i|\lambda\ra$ and $|-\lambda]=i|\lambda]$. The integral
is over the Lorentz Invariant Phase Space (LIPS) associated with $\ell_{1}$ and $\ell_{2}$,
and it is normalized as $\int d{\rm LIPS}=\pi/2$. When $\ell_{1}$ and $\ell_{2}$
label indistinguishable particles, a symmetry factor 1/2 must be included in \eq{gamma}.

In general loop amplitudes contain IR divergences, which have to be subtracted in order to
extract the UV-divergent part we are interested in. As explained for example in \cite{Baratella:2020dvw},
IR divergences are structurally of two kinds: they can either be associated to
triangle and box integrals, or to
massless bubble integrals. The divergences contained in triangles and boxes
are purely IR and can be simply projected out,
for example along the lines of Appendix~\ref{bubble} (for other methods see e.g.~\cite{Mastrolia:2009dr}).
Massless bubbles contain instead both a UV and an IR part, which are equal and opposite and cancel.
To account for the UV part of massless bubbles, it is enough to add a ``collinear'' contribution \cite{Baratella:2020dvw}.
We then have
\be\label{gamma2}
\gamma_i \frac{\amp_i(1\ldots n)}{C_i}=-\frac{1}{4\pi^3}\sum_{{\rm cuts}\,\atop \ell_{1},\ell_{2}}\sigma_{\ell_{1}\ell_{2}}{\cal R}\int d{\rm LIPS} ~\amp_L(...-\bar{\ell}_2-\bar{\ell}_1)\,\amp_R(\ell_1\ell_2...)+\gamma_{\rm coll}\widehat\amp_i(1\ldots n).
\ee
Here ${\cal R}$ takes the rational part of the $d{\rm LIPS}$ integral (i.e.~it ``annihilates'' all the transcendental functions such as logs and dilogs), an operation that is equivalent to projecting out
triangle and boxes from the loop amplitude and keep only the bubble contribution \cite{Baratella:2020lzz}.
The last term of \eq{gamma2} accounts for collinear divergences (or equivalently massless bubbles):
$\gamma_{\rm coll}$ is given by a sum over the external particles' collinear anomalous dimensions, that is $\gamma_{\rm coll}=\sum_{j} \gamma_{\rm coll}^{(j)}$, while $\widehat\amp_i$ is a tree-level amplitude
with the same external legs and same $\Lambda$ and $\Mp$ scaling as $\amp_i$. In Section~\ref{toy} we will present
several examples of the use of \eq{gamma2}, which will also clarify the properties and role of $\widehat\amp_i$.
In Appendix~\ref{collinear} we present the computation of collinear anomalous dimensions.

A last comment on Eqs.~(\ref{gamma}) and (\ref{gamma2}) is that
in general there can be many $\amp_i$ with the same external
legs, differing by flavor or Lorentz structure. In these cases
their LHS becomes a sum over the index $i$ of all relevant amplitudes. The choice of
the $\amp_i$'s amounts to a choice of basis.

\vspace{.2cm}

The advantages that come from using \eq{gamma} instead of doing a blind loop computation are huge.
Primarily, \eq{gamma} inherits all the transparency
and compactness of the on-shell formalism, these qualities being especially manifest when dealing with massless
particles with helicity 1 or 2, that in an ordinary Lagrangian approach require the introduction of some degree of
gauge redundancy. Second, also connected to the
on-shellness of the various amplitudes entering the formula,
\eq{gamma} allows to see how properties of $\amp_L$ and $\amp_R$ are inherited by the ``counterterm''
$\amp_i$, or conversely to constrain the form of relevant amplitudes $\amp_{L,R}$ once some a priori
knowledge on the form of $\amp_i$ is available. An example of this are the `modified helicity' selection rules, 
as we now explain and will come back to with explicit examples in Section~\ref{toy}.

A first, simple but powerful condition coming from \eq{gamma} is that
\be\label{wcut}
w_i+k_i=w_L+w_R+k_L+k_R\,,
\ee
where $w(\amp)$ and $k(\amp)$ are the total powers in respectively $\Lambda$ and $\Mp$ that are carried by $\amp$, i.e.~$\amp \propto 1/\Lambda^{w}\Mp^{k}$.
A similar yet physically much richer condition holds for the modified helicity $\tilde{h}$, which is additive on the cut and imposes
\be\label{hcut}
\tilde{h}_i=\tilde{h}_L+\tilde{h}_R\,.
\ee
Its simplicity is key to a couple of very powerful results that we are going to present here and will use in
Section~\ref{toy}. The first statement is that, when both $\amp_L$ and $\amp_R$ are amplitudes constructed out
of marginal and minimal gravity couplings, we can combine \eq{hcut} with \eq{eq:nPointTreeModHelBound}
to get
\be\label{hboundloop}
|\tilde{h}_i|\leq |\tilde{h}_L|+|\tilde{h}_R|\leq (m-4)+[(n-m+4)-4]=n-4\,,
\ee
where in the first step we used the triangle inequality, while in the second we used
\eq{eq:nPointTreeModHelBound} twice, for $\amp_{L,R}$ with respectively $m$ and $(n-m+4)$ legs.
Notice that this remarkable fact is valid at any order in $\Mp^{-1}$. We have in particular
\be\label{n4htilde}
\tilde{h}_i^{\rm (1\,loop)}=0~~~~~~~~~~~~(n=4)\,,
\ee
in any minimally coupled model without trilinear scalar couplings and no exceptional
four-fermion amplitude.

The second statement, similar in nature but slightly different in purpose, concerns the way in which a
Wilson coefficient $C_j$ can enter the renormalization of another coefficient $C_i$. More specifically, we take
$\amp_L\equiv \amp_j$ to be the amplitude proportional to $C_j$, 
and $\amp_R\equiv \amp$ to be an ${n}$-point amplitude with modified helicity $\tilde{h}$ constructed out of only marginal and minimal gravity couplings.
Then, we find
\be\label{htildemixing}
|\tilde{h}_i-\tilde{h}_j|=|{\tilde{h}}|\leq {n}-4=n_i-n_j\,,
\ee
where in the first step we have used \eq{hcut} and in the second \eq{eq:nPointTreeModHelBound}, while the last step is just the topological statement that 
{$\amp$} makes $n_j$ ``jump'' to $n_i=n_j+{n}-4$. This is a natural generalization of the non-renormalization theorems of \cite{Cheung:2015aba} when gravity is included, and reduces to it for $\tilde{h}\to h$.
When applying \eq{htildemixing}, one should pay attention to the fact that, unlike for the standard scenario in which it has only
marginal couplings, {$\amp$} can in general carry also the dimensional coupling $\Mp^{-1}$. Therefore it can induce mixing
between coefficients $C_i$ associated to a different $\Lambda$-scaling. 
Indeed, since ${w} = 0$, from \eq{wcut} $w_i - w_j = k_j + {k} - k_i$, which is non-zero whenever gravitons are exchanged.
We will come back to this point in Section \ref{mixinggrav} when discussing explicit examples and in Appendix \ref{opMix}.

\section{One-loop divergence structure of theories coupled to gravity}
\label{toy}

Having analysed the general properties of \eq{gamma} and \eq{gamma2}, and with the aim of showing their
enormous effectiveness, we now move on and focus on some specific problem. Our purpose is to study
the one-loop structure of a gauge theory --~with scalars and fermions transforming under
the gauge group and interacting among themselves through Yukawas and scalar quartics~-- that is minimally coupled to gravity.

This structure incorporates the world we live in, as the SM (and any other consistent low-energy theory) is so made. Since the world is quantum, due to the presence of the negative dimensional coupling $\Mp^{-1}$ we expect to generate new interactions at one loop that are not present at the minimal-coupling level. What is the amplitude/operator spectrum
that is required by consistency of the theory at one loop? This is the first question we are going to address, providing in Sections~\ref{Ord2} and
\ref{Ord4} general formulas for divergent 4-point amplitudes at ${\cal O}(\Mp^{-2})$ and ${\cal O}(\Mp^{-4})$ respectively.

As explained in the previous section, in the setup of \ref{Ord2} and \ref{Ord4} our modified helicity constraints prove to be extremely powerful.
This is because, thanks to \eq{gamma}, we see that one-loop divergences are nothing but convolutions of two tree-level amplitudes of the minimally coupled model, which are both bounded in (modified) helicity.
At 4-point we then have $\tilde{h}_i^{\rm (1\,loop)}=0$ according to \eq{hboundloop}.

As a further introductory comment on \ref{Ord2} and \ref{Ord4}, we would like to mention that, for what concerns one-loop divergent 4-point amplitudes in a minimally coupled theory, our results are completely exhaustive. In fact, going beyond the marginal level, which is nothing but the study of renormalizable theories, divergences are found only at ${\cal O}(\Mp^{-2})$ and ${\cal O}(\Mp^{-4})$. This follows from simple power counting, see Appendix~\ref{opMix}.

In Section~\ref{mixinggrav} we are going to study a complementary case. If the previous
question regards the minimal operator content required at one loop
for consistency with the world being quantum \emph{and} gravitational, the next question concerns the leading RG effects on higher-dimensional 
amplitudes including gravitons (those of Section~\ref{beyondmin}), the effects being induced by marginal SM-like interactions. In this setup we will
be able to fruitfully employ \eq{htildemixing}, that will prove as powerful as the corresponding bound in the usual SMEFT context.

Some of our results were obtained before with alternative methods.
For the comparison with previous literature, see Section~\ref{previous}.

\vspace{.2cm}

Gravitational amplitudes for generic matter particles
relevant to the discussion that follows read
\begin{align}
\amp_{\rm min}(1_{\bar{\phi}},2_{\phi},3_{\phi'},4_{\bar{\phi}'})&=\frac{1}{\Mp^2}\left(\frac{tu}{s}-\frac{s}{6} \right) 
+\frac{1}{\Mp^2}\,\frac{s}{6}\,,\label{J2phi} \\
\amp_{\rm min}(1_{\bar{\psi}},2_{\psi},3_{\phi},4_{\bar{\phi}})&=\frac{1}{\Mp^2}\frac{t-u}{2s}\la 13 \ra [23]\,,\label{J2phipsi} \\
\amp_{\rm min}(1_{\bar{\psi}},2_{\psi},3_{\psi'},4_{\bar{\psi}'})&=\frac{1}{\Mp^2}\frac{3t-u}{4s}\la 14 \ra [23]\,,\label{J2psi}\\
\amp_{\rm min}(1_{V_-},2_{V_+},3_{\Phi_h},4_{\Phi_{-h}})&=\frac{1}{\Mp^2}(-1)^{\delta_{h,1}}\frac{\la 13 \ra^2 [23]^2}{s}\left(\frac{\la 14 \ra}{\la 13 \ra}\right)^{2h}\label{J2vec}\,,
\end{align}
where in the first equation we have split the four-scalar amplitude into its $J=2$ and $J=0$ partial wave components (in the $1, 2 \rightarrow 3, 4$ channel),
and
in the last equation we have grouped together all amplitudes with at least a pair of opposite-helicity vectors ($h=0,1/2,1$).

\subsection{Divergences in minimally coupled theories: ${\cal O}(\Mp^{-2})$}
\label{Ord2}

In this section we present our results for the renormalization of minimally coupled gauge theories at one-loop, $n=4$
and ${\cal O}(\Mp^{-2})$. These are the first effects when going beyond the renormalizable level.
In this section we will talk about $\epsilon\equiv(4-d)/2$ divergent parts of the amplitudes, called $\amp_{\rm UV}$,
instead of anomalous dimensions.
The two pieces of information are indeed equivalent, as $\sum_i\gamma_i\amp_i/C_i=-2\epsilon\amp_{\rm UV}$. The advantage of
talking about the divergent part is that we do not need to specify any basis of higher-dimensional amplitudes $\amp_i$.

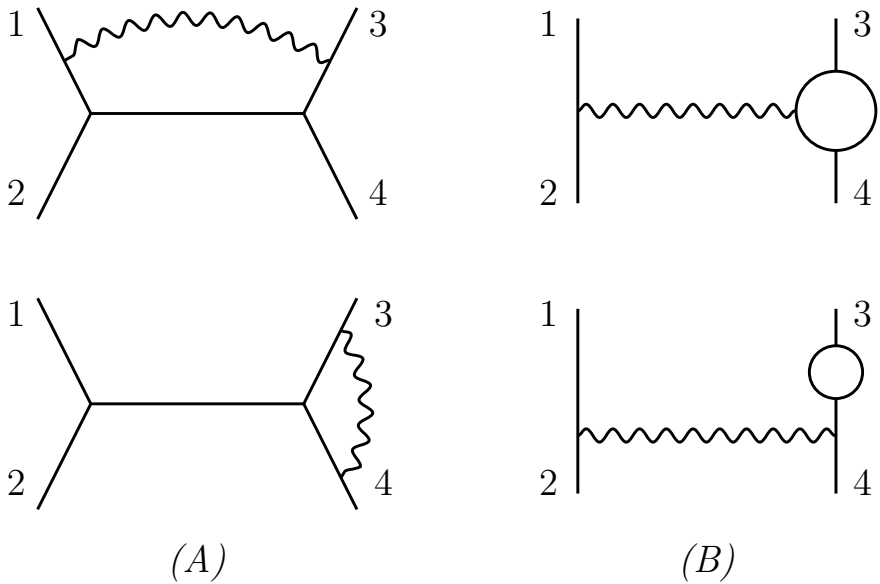
\begin{figure}[t]
\centering
\begin{tikzpicture}[line width=1.1 pt, scale=.7, baseline=(current bounding box.center)]

	\draw (-10,5) -- (-9,3) -- (-10,1) ;
	\draw (-9,3) -- (-5,3) ;
	\draw (-4,5) -- (-5,3) -- (-4,1) ;
	\draw[vector] (-4.5,4) to [bend right] (-9.5,4) ;
	\node at (-10.4,4.7) {{\large 1}};
	\node at (-10.4,1.5) {{\large 2}};
	\node at (-3.6,4.7) {{\large 3}};
	\node at (-3.6,1.5) {{\large 4}};
	
	\begin{scope}[shift={(1-1,0-1)}]
	\draw (-10,0.5) -- (-9,-1.5) -- (-10,-3.5) ;
	\draw (-9,-1.5) -- (-5,-1.5) ;
	\draw (-4,.5) -- (-5,-1.5) -- (-4,-3.5) ;
	\draw[vector] (-4.3,-0.1) to [bend left] (-4.3,-2.9) ;
	\end{scope}
	\begin{scope}[shift={(1.1-1,-4.5-1)}]
	\node at (-10.5,4.7) {{\large 1}};
	\node at (-10.5,1.5) {{\large 2}};
	\node at (-3.6,4.7) {{\large 3}};
	\node at (-3.6,1.5) {{\large 4}};
	\end{scope}
	
	\node at (-6.-1,-4.5-1) {{\large \emph{(A)}}};
	
	\node at (3.58-1,-4.5-1) {{\large \emph{(B)}}};
	
	\begin{scope}[shift={(-2.0,0.3)}]
	\draw (2.15,4.5) -- (2.15,1) ;
	\draw (7,4.5) -- (7,3.5) ;
	\draw (7,2) -- (7,1) ;
	\draw (7,2.75) circle (0.75cm) ;
	\draw[vector] (2.15,2.75) to (6.25,2.75) ;
	\end{scope}
	\begin{scope}[shift={(10.1,0)}]
	\node at (-10.5,4.7) {{\large 1}};
	\node at (-10.5,1.5) {{\large 2}};
	\node at (-4.6,4.7) {{\large 3}};
	\node at (-4.6,1.5) {{\large 4}};
	\end{scope}
	
	
	\begin{scope}[shift={(-2,0.3-1)}]
	\draw (2.15,0) -- (2.15,-3.5) ;
	\draw (7,0) -- (7,-0.7) ;
	\draw (7,-1.7) -- (7,-3.5) ;
	\draw (7,-1.2) circle (0.5cm) ;
	\draw[vector] (2.15,-2.4) to (7,-2.4) ;
	\end{scope}
	\begin{scope}[shift={(10.1-1,-4.5-1)}]
	\node at (-9.5,4.7) {{\large 1}};
	\node at (-9.5,1.5) {{\large 2}};
	\node at (-3.6,4.7) {{\large 3}};
	\node at (-3.6,1.5) {{\large 4}};
	\end{scope}

 \end{tikzpicture}
 \caption{\emph{Some of the Feynman diagrams relevant for Section~\ref{Ord2}, with continuous lines for particles with
 $|h|=0,1/2,1$ and wavy lines for gravitons. Diagrams of class {(A)}
 on the left and {(B)} on the right respectively remain connected and get disconnected when the unique graviton 
 propagator is removed. Most of class {(B)} diagrams, with well understood exceptions in the scalar sector,
 cancel among themselves as a consequence of the conservation of the energy momentum tensor $T^{\mu\nu}$.
 This is discussed in Appendix~\ref{collinear}.} }
 \label{fig:dim6}
 \end{figure}

The first guidance to the UV-divergent structure of our minimally coupled models comes from \eq{hboundloop} that,
together with the fact that divergences must be absorbable by local counterterms,
instructs us to look at the coordinates $(n=4,\tilde{h}=0)$ of Fig.~\ref{fig:helOpPlot} and, especially, forget all other counterterms.
Therefore, potentially divergent amplitudes are just
\be
\amp_{\rm UV}(1_{\phi_1},2_{\phi_2},3_{\phi_3},4_{\phi_4})\,,~~~~~\amp_{\rm UV}(1_{\bar\psi_1},2_{{\psi}_2},3_{\phi_3},4_{\phi_4})\,,~~~~~\amp_{\rm UV}(1_{\bar\psi_1},2_{{\psi}_2},3_{{\psi}_3},4_{\bar\psi_4})\,,
\ee
where $\phi_{1...4}$ and $\psi_{1...4}$ can be respectively any scalar and any positive helicity
fermion in the theory (distinguishable or not). Notice that the 
above statement is true only modulo the exceptional amplitude (and absent three-scalar couplings).
In non-holomorphic theories like the SM, we could also expect a fourth category:
\be\label{exceptionalUV}
\amp_{\rm UV}(1_{\psi_1},2_{{\psi}_2},3_{{\psi_3}},4_{\psi_4})\,.
\ee
In summary, we find that the external legs of potentially divergent amplitudes are either helicity $\pm1/2$ fermions or scalars.
The $\Mp$ dependence can then only come from one --~and only one~-- internal graviton 
(for the
sake of this discussion, it is useful to think in terms of ordinary Feynman diagrams and uncut loop amplitudes).
This fact implies that each of the loop diagrams contributing to some given $\amp_{\rm UV}$ can be divided into two categories,
characterized by the topology of the corresponding diagram: a diagram can either \emph{(A)}
remain connected or \emph{(B)} get disconnected when the graviton
propagator is removed, and this is equivalent to the propagator being part of the loop or not, respectively.
This is shown in Fig.~\ref{fig:dim6}.\footnote{We observe that diagrams in class \emph{(A)} do not contain corrections to the external legs, which are all in \emph{(B)}. This means that, when we cut the diagrams of Fig.~\ref{fig:dim6}, the contribution in \eq{gamma2} proportional to $\gamma_{\rm coll}$ will be entirely part of class \emph{(B)}.}

Our first claim, which is proven and thoroughly discussed in Appendix~\ref{collinear}, and that we also checked by direct computation, is that (almost) all contributions to class \emph{(B)} cancel among themselves. This can be seen by noticing that the diagrams in class \emph{(B)} correspond to
\begin{align}\label{ThhT}
&\left. \amp(1,2,3,4)\right|_{(B)}=\la 1,2 | T^{\mu\nu}|\Omega \ra^{(0)}\la \Omega | h_{\mu\nu}h_{\rho\sigma} |\Omega \ra ^{(0)}\la \Omega | T^{\rho\sigma}|3,4 \ra^{(1)}+...
\end{align}
where the ellipsis stand for analogous contributions with $\la 1,2 | T^{\mu\nu}|\Omega \ra^{(1)}$ and $\la\Omega | T^{\rho\sigma}|3,4 \ra^{(0)}$, the suffix denoting the loop order at which the form factor is computed. To get to this expression, we have used the fact that the graviton couples to matter through $h_{\mu\nu} T^{\mu\nu}$, and split the relevant set of one loop contributions, i.e.~those of class \emph{(B)}, thanks to Wick's theorem. As we discuss in Appendix~\ref{collinear}, the conservation of the energy-momentum tensor $T^{\mu\nu}$ implies that its \emph{traceless part} is UV-finite. Using \eq{ThhT}, this implies that, if 1,2 and 3,4 are fermions
\be\label{zeroB}
\left. \amp_{\rm UV}(1,2,3,4)\right|_{(B)}=0\,,
\ee
thanks to the property that their energy-momentum tensor is automatically traceless. The vanishing of \eq{zeroB} holds also when either 1,2 \emph{or} 3,4 are fermions, thanks to certain orthogonality property that we are going to discuss at the end of this section and in Appendix \ref{collinear} with a slightly different formalism. However, \eq{zeroB} is not necessarily true when both 1,2 and 3,4 are scalars, which admit a $T^{\mu\nu}$ with non-vanishing trace.
Leaving the fully scalar exception for 
the end, we now explore the consequences of the vanishing of class \emph{(B)} for the other amplitudes.
 
 \vspace{.2cm}
 
The key observation now is that each term in class \emph{(A)} is in one-to-one correspondence with a tree-level diagram in the marginal theory with identical external states, whose amplitude we call $\amp_{\rm tree}$, which is obtained by removing the graviton internal leg. Thanks to the helicity and flavor preserving nature
of the minimal gravitational coupling, the diagrams in \emph{(A)} and $\amp_{\rm tree}$ have the same structure,
both flavor and helicity-wise.
As we now show, this observation allows us to present our results in complete generality for any model,
without having to assume a particular flavor/color structure, which is the only source of model dependence.
Once again, we see that what governs the physics of scattering amplitudes is helicity. 

\subsubsection{Divergences of the form $\amp_{\rm UV}(1_{\bar\psi_1},2_{{\psi}_2},3_{\phi_3},4_{\phi_4})$}

We start by considering the divergent part of an amplitude with two opposite-helicity fermions and two scalars.
At tree level and with marginal couplings only, i.e.~without gravity, such an amplitude can only be of the following form
\be\label{tree1}
\amp_{\rm tree}(1_{\bar\psi_1},2_{{\psi}_2},3_{\phi_3},4_{\phi_4})=\left(\frac{T_{s}}{s}+\frac{Y_{t}}{t}+\frac{Y_{u}}{u}\right)\la 13 \ra [23]\,,
\ee
where $T$ and $Y$ are flavor/color tensors whose index structure is governed by the external legs, and we use different symbols to emphasize their gauge ($T$) or Yukawa ($Y$) nature. They also carry all the marginal coupling dependence.
By using \eq{gamma2} we find the ${\cal O}(\Mp^{-2})$ UV-divergent part of the amplitude to be
\be\label{UV1}
\boxed{~\amp_{\rm UV}(1_{\bar\psi_1},2_{{\psi}_2},3_{\phi_3},4_{\phi_4})=-\frac{7}{64\pi^2\Mp^2\epsilon}\left(3\, T_{s}+Y_{t}+Y_{u} \right)\la 13 \ra [23]\,.~}
\ee
We see that, apart from the loop factor $(4\pi)^{-2}$ and the $\Mp$ dependence, all the non-trivial information
is encoded in a couple of coefficients: 21/4 for the vector channels and 7/4 for the Yukawa channels.

Let us give some details on how to go from \eq{tree1} to \eq{UV1} using \eq{gamma2}.
We focus in particular on the contributions to the $s$-channel cut which, thanks to our main formula \eq{gamma2}, read
\begin{align}\label{exampleUV}
&\left.\amp_{\rm UV}(1_{\bar\psi_1},2_{{\psi}_2},3_{\phi_3},4_{\phi_4})\right|_{s-ch}=\frac{1}{8\pi^3\epsilon} {\cal R}\int d{\rm LIPS}_{\ell\ell'} \bigg[ \amp_{\rm tree}(1_{\bar\psi_1},2_{{\psi}_2},-\ell_{\phi_3},-\ell'_{\phi_4})\nonumber\\
&~~~~\times\amp_{\min}(\ell_{\bar{\phi}_3},\ell'_{\bar{\phi}_4},3_{\phi_3},4_{\phi_4})+
\amp_{\min}(1_{\bar\psi_1},2_{{\psi}_2},-\ell_{{\psi}_1},-\ell'_{\bar\psi_2})\,\amp_{\rm tree}(\ell_{\bar\psi_1},\ell'_{{\psi}_2},3_{\phi_3},4_{\phi_4})\nonumber\\
&~~~~+\sum_{V^a}\sum_{\pm}\amp(1_{\bar\psi_1},2_{{\psi}_2},-\ell'_{V^a_\pm},-\ell_{h_\mp})\,\amp(\ell_{h_\pm},\ell'_{V^a_\mp},3_{\phi_3},4_{\phi_4}) \bigg] \,,
\end{align}
where we stress again that the above equation is valid because we are only considering cuts of class \emph{(A)} amplitudes. One can see that the contributions are of two kinds, corresponding to the first two lines and third line respectively. Each of the first two terms is a product of $\amp_{\rm tree}$ and an amplitude that is mediated by a graviton exchange, that we have called $\amp_{\min}$. Instead, the last term is a product of two amplitudes involving three matter fields and one graviton. The index $a$, over which we sum, labels all distinct gauge bosons of the theory. Another summation is over the possible polarizations of the internal $V^a$ and $h$. The same structure governs the $t$ and $u$-channel cuts.

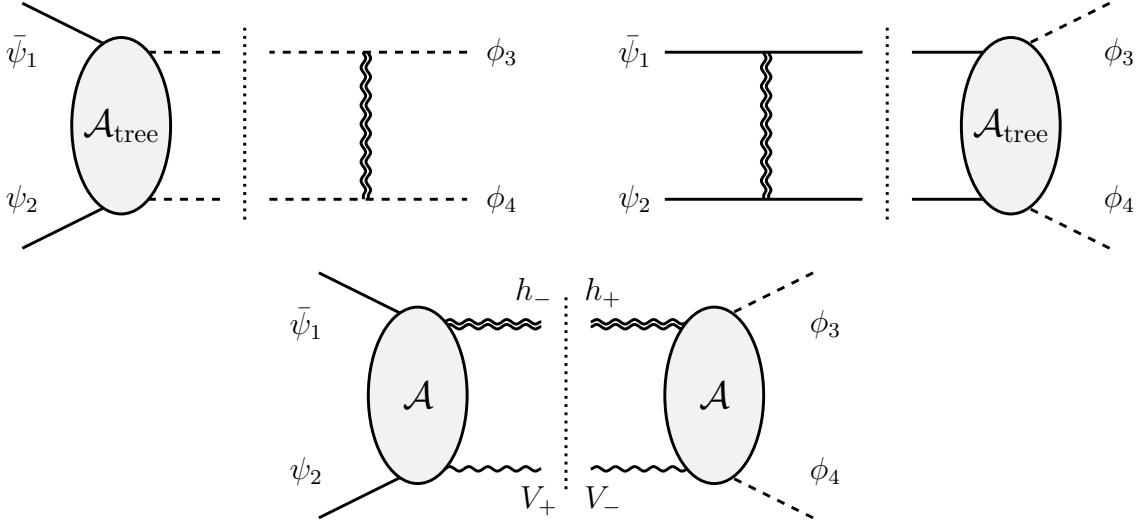
\begin{figure}[t]
\centering
\begin{tikzpicture}[line width=1.1 pt, scale=.65, baseline=(current bounding box.center)]
	\filldraw [white] (0,0) circle (2pt) 
	(-10,0) circle (2pt)
	(10,0) circle (2pt)
	(0,-7) circle (2pt)
	(0,4) circle (2pt);
	
	\draw (-11,4) -- (-9,3) -- (-9,0) -- (-11,-1) ;
	\draw[dashed] (-7,3) -- (-9,3) ;
	\draw[dashed] (-7,0) -- (-9,0) ;
	\draw[fill=gray!10] (-9,1.5) ellipse (1cm and 1.8cm) ;
	\node at (-8.93,1.5) {\large $\amp_{\rm tree}$\,};
	
	\draw[dotted] (-6.5,3.5) -- (-6.5,-0.5) ;
	
	\draw[dashed] (-2,3) -- (-6,3) ;
	\draw[dashed] (-6,0) -- (-2,0) ;
	\draw[decorate,decoration={snake,amplitude=.3mm,segment length=2.5 mm}] (-4,3) -- (-4,0) ;
	\draw[decorate,decoration={snake,amplitude=.3mm,segment length=2.5 mm}] (-4.1,3) -- (-4.1,0) ;
	
	\begin{scope}[shift={(0,0)}]
	\node at (-11.,3.) {${\bar\psi_1}$};
	\node at (-11.,-0.) {${{\psi}_2}$};
	\node at (-1.3,3.) {${\phi_3}$};
	\node at (-1.3,-0.) {${\phi_4}$};
	\end{scope}
	
	
	\draw[dashed] (11,4) -- (9,3) ;
	\draw[dashed] (11,-1) -- (9,0) ;
	\draw (7,3) -- (9,3) ;
	\draw (7,0) -- (9,0) ;
	\draw[fill=gray!10] (9,1.5) ellipse (1cm and 1.8cm) ;
	\node at (9.07,1.5) {\large $\amp_{\rm tree}$\,};
	
	\draw[dotted] (6.5,3.5) -- (6.5,-0.5) ;
	
	\draw (2,3) -- (6,3) ;
	\draw (6,0) -- (2,0) ;
	\draw[decorate,decoration={snake,amplitude=.3mm,segment length=2.5 mm}] (4,3) -- (4,0) ;
	\draw[decorate,decoration={snake,amplitude=.3mm,segment length=2.5 mm}] (4.1,3) -- (4.1,0) ;
	
	\begin{scope}[shift={(12.9,0)}]
	\node at (-11.5,3.) {${\bar\psi_1}$};
	\node at (-11.5,-0.) {${{\psi}_2}$};
	\node at (-1.7,3.) {${\phi_3}$};
	\node at (-1.7,-0.) {${\phi_4}$};
	\end{scope}
	
\begin{scope}[shift={(0,-.5)}]

	\draw (-3,-2) -- (-3,-5) ;
	\draw (-5,-1) -- (-3,-2) ;
	\draw (-5,-6) -- (-3,-5) ;
	\draw[decorate,decoration={snake,amplitude=.3mm,segment length=2.5 mm}] (-0.5,-2) -- (-3,-2) ;
	\draw[decorate,decoration={snake,amplitude=.3mm,segment length=2.5 mm}] (-0.5,-2.1) -- (-3,-2.1) ;
	\draw[decorate,decoration={snake,amplitude=.3mm,segment length=2.5 mm}] (-0.5,-5) -- (-3,-5) ;
	\draw[fill=gray!10] (-3,-3.5) ellipse (1cm and 1.8cm) ;
	\node at (-2.93,-3.5) {\large $\amp$\,};
	
	\draw[dotted] (0,-1.5) -- (0,-5.5) ;
	
	\draw[dashed] (3,-2) -- (3,-5) ;
	\draw[dashed] (5,-1) -- (3,-2) ;
	\draw[dashed] (5,-6) -- (3,-5) ;
	\draw[decorate,decoration={snake,amplitude=.3mm,segment length=2.5 mm}] (0.5,-2) -- (3,-2) ;
	\draw[decorate,decoration={snake,amplitude=.3mm,segment length=2.5 mm}] (0.5,-2.1) -- (3,-2.1) ;
	\draw[decorate,decoration={snake,amplitude=.3mm,segment length=2.5 mm}] (0.5,-5) -- (3,-5) ;
	\draw[fill=gray!10] (3,-3.5) ellipse (1cm and 1.8cm) ;
	\node at (3.07,-3.5) {\large $\amp$\,};
	
	\begin{scope}[shift={(6.55,-5.05)}]
	\node at (-11.8,3.) {${\bar\psi_1}$};
	\node at (-11.8,-0.) {${{\psi}_2}$};
	\node at (-1.3,3.) {${\phi_3}$};
	\node at (-1.3,-0.) {${\phi_4}$};
	\node at (-7.1,-.6) {${V_+}$};
	\node at (-5.8,-.6) {${V_-}$};
	\node at (-7.2,3.65) {${h_-}$};
	\node at (-5.8,3.65) {${h_+}$};
	\end{scope}
	
\end{scope}
	
	 \end{tikzpicture}
 \caption{\emph{Cut diagrams corresponding to the RHS of \eq{exampleUV}.}}
 \label{fig:dim6Cut}
 \end{figure}

We now consider in turn the three contributions to the RHS of \eq{exampleUV}, which are also represented in Fig.~\ref{fig:dim6Cut}. The first term gives
\begin{align}
&{\cal R}\int d{\rm LIPS}_{\ell\ell'}\,\amp_{\rm tree}(1_{\bar\psi_1},2_{{\psi}_2},-\ell_{\phi_3},-\ell'_{\phi_4})\amp_{\min}(\ell_{\bar{\phi}_3},\ell'_{\bar{\phi}_4},3_{\phi_3},4_{\phi_4})\nonumber\\
&={\cal R}\int d{\rm LIPS}_{\ell\ell'}\left(\frac{T_{s}}{s}+\frac{Y_{t}}{-2\ell. p_1}+\frac{Y_{u}}{-2 \ell.p_2}\right)\la 1\ell \ra [\ell 2]\times \frac{s}{\Mp^2}\frac{\ell.p_4}{\ell.p_3}= -\frac{\pi}{2}\frac{2\, T_s}{\Mp^2} \la 13\ra [23]\,,
\end{align}
where we used \eq{tree1} and \eq{J2phi}. Similarly, for the second term we find
\begin{align}
&{\cal R}\int d{\rm LIPS}_{\ell\ell'}\,\amp_{\min}(1_{\bar\psi_1},2_{{\psi}_2},-\ell_{{\psi}_1},-\ell'_{\bar\psi_2})\,\amp_{\rm tree}(\ell_{\bar\psi_1},\ell'_{{\psi}_2},3_{\phi_3},4_{\phi_4})\nonumber\\
&={\cal R}\int d{\rm LIPS}_{\ell\ell'}\,\frac{1}{\Mp^2}\frac{3s+2\ell.p_2}{{8\ell.p_1}}\la 1\ell'\ra [2\ell]\times \left(\frac{T_{s}}{s}+\frac{Y_{t}}{{2\ell. p_3}}+\frac{Y_{u}}{{ 2 \ell.p_4}}\right)\la \ell 3 \ra [\ell' 3]\\
&= -\frac{\pi}{2} \frac{19\, T_s -3(Y_t+Y_u)}{12\, \Mp^2} \la 13\ra [23]\,,\nonumber
\end{align}
using \eq{J2psi} and \eq{tree1} to go to the second line. While it is manifest that the above terms can be written in terms of the tree-level flavor structures $T_s$, $Y_{t}$ and $Y_u$, the same
is not obvious for the cuts with one internal graviton, as they do not contain a ``factor'' of $\amp_{\rm tree}$. However, we now show that the same tensors are recovered in these cuts after a couple of manipulations. We first note that all 4-point tree-level
amplitudes of a minimally coupled theory that have one graviton can be written as follows
\be\label{1graviton}
\amp(1,2,3,4_{h_-})=\frac{1}{\Mp}\amp(1,2,3)\frac{\la 4i \ra [ij] \la j4 \ra}{[4i] \la ij \ra [j4]}\,,
\ee
where all choices of $i,j=1,2,3$ with $i\neq j$ are equivalent thanks to momentum conservation, with an analogous expression when the graviton has positive helicity, specifically with angle and square brackets exchanged. It is understood that $\amp(1,2,3,4_{h_-})$ must be allowed to be non-zero by the helicity bounds of Section \ref{eq:treeMinCoupl}. The expression in \eq{1graviton}, which is reminiscent of soft behaviour but can be fixed with simple factorization arguments is,\footnote{Notice that the spinorial contraction in \eq{1graviton} can be rewritten as $-\la 4i \ra^2 [ij]^2 \la j4 \ra^2/stu$.} to our knowledge, new.
Then the last cut reads
\begin{align}
&{\cal R}\int d{\rm LIPS}_{\ell\ell'}\,\amp(1_{\bar\psi_1},2_{{\psi}_2},-\ell'_{V^a_+},-\ell_{h_-})\,\amp(\ell_{h_+},\ell'_{V^a_-},3_{\phi_3},4_{\phi_4})\nonumber\\
&={\cal R}\int d{\rm LIPS}_{\ell\ell'}\,\frac{1}{\Mp}\amp(1_{\bar\psi_1},2_{{\psi}_2},-\ell'_{V^a_+})\frac{\la \ell1 \ra [12] \la 2\ell \ra}{[\ell1] \la 12 \ra [2\ell]}\times \frac{1}{\Mp}\amp(\ell'_{V^a_-},3_{\phi_3},4_{\phi_4})\frac{[\ell3] \la 34 \ra [4\ell]}{\la \ell 3 \ra [34] \la 4 \ell \ra}\nonumber \\
&=\frac{1}{\Mp^2}{\cal R}\int d{\rm LIPS}_{\ell\ell'}\, \frac{-[2\ell']^2}{[12]}g_a t_L^{a}\,\frac{\la \ell1 \ra [12] \la 2\ell \ra}{[\ell1] \la 12 \ra [2\ell]}\times \frac{\la \ell' 3 \ra\la\ell'4 \ra}{\la 34\ra}g_a t_R^{a}\,\frac{[\ell3] \la 34 \ra [4\ell]}{\la \ell 3 \ra [34] \la 4 \ell \ra}\nonumber \\
&=\frac{\pi}{2}\frac{g_a^2 t_L^{a}t_R^{a}}{6\, \Mp^2} \la 1 3\ra [2 3]\,, \label{eq:PsiPsiPhiPhi1GravSChan}
\end{align}
plus an analogous contribution with ${h_+},{V_-} \leftrightarrow {h_-},{V_+}$, which gives the same result. In~\eq{eq:PsiPsiPhiPhi1GravSChan} we introduced the generators $t_{L,R}^{a}$ that govern the interaction of $V^a$ with respectively $\psi_1\bar{\psi}_2$
and $\phi_3\phi_4$, and $g_a$ which fixes the strength of the coupling. After summing upon all the vectors of the theory
as dictated by \eq{exampleUV}, the flavor/color part becomes
\be
\sum_a g_a^2\, t_L^{a}t_R^{a}= T_s\,,
\ee
and we recover the same 
tensor that enters in \eq{tree1}. The fact that the above correspondence must hold in general can be seen explicitly by factorization of \eq{tree1}. In fact, factorization in the $s$-channel imposes
\begin{align}
T_s\la 13 \ra[23]&=\lim_{\la 12\ra \to 0} s \amp_{\rm tree}(1_{\bar\psi_1},2_{{\psi}_2},3_{\phi_3},4_{\phi_4})\nonumber \\
&= -\sum_a \amp(1_{\bar\psi_1},2_{{\psi}_2},-\ell_{V^a_+})\times\amp(\ell_{V^a_-},3_{\phi_3},4_{\phi_4})\nonumber \\
&=-\sum_a g_a t_L^{a}\frac{-[2\ell]^2}{[12]}\times g_a t_R^{a}\frac{\la \ell 3 \ra\la\ell4 \ra}{\la 34\ra}=\sum_a g_a^2\, t_L^{a}t_R^{a} \la 13 \ra[23]\,,
\end{align}
where we used the fact that $\ell=p_1+p_2=-p_3-p_4$. When we finally put everything together, we get that the $s$-channel
contribution to \eq{UV1} is given by
\be
\left.\amp_{\rm UV}(1_{\bar\psi_1},2_{{\psi}_2},3_{\phi_3},4_{\phi_4})\right|_{s-ch}=\frac{\la 13 \ra [23]}{64\pi^2\Mp^2\epsilon}(Y_t+Y_u-13T_s)\,.
\ee
Analogously we can proceed for the $t$-channel cut, whose result we quote directly after giving the expression for the
phase space integral:
\begin{align}\label{exampleUVt}
&\left.\amp_{\rm UV}(1_{\bar\psi_1},2_{{\psi}_2},3_{\phi_3},4_{\phi_4})\right|_{t-ch}=\frac{-i}{8\pi^3\epsilon} {\cal R}\int d{\rm LIPS}_{\ell\ell'} \bigg[ \amp_{\rm tree}(1_{\bar\psi_1},3_{\phi_3},-\ell_{{\psi}_2},-\ell'_{\phi_4}) \nonumber\\
&~~~\times\amp_{\min}(\ell_{\bar\psi_2},\ell'_{\bar{\phi}_4},2_{{\psi}_2},4_{\phi_4})+
\amp_{\min}(1_{\bar\psi_1},3_{\phi_3},-\ell_{{\psi}_1},-\ell'_{\bar{\phi}_3})\,\amp_{\rm tree}(\ell_{\bar\psi_1},\ell'_{\phi_3},2_{{\psi}_2},4_{\phi_4})\nonumber\\
&~~~+\sum_{\bar\xi_i}\amp(1_{\bar\psi_1},3_{\phi_3},-\ell'_{\bar\xi_i},-\ell_{h_+})\,\amp(\ell_{h_-},\ell'_{{\xi}_i},2_{{\psi}_2},4_{\phi_4}) \bigg] = \frac{\la 13 \ra [23]}{16\pi^2\Mp^2\epsilon}(Y_u-3Y_t-T_s)\,.
\end{align}
The sum that appears in the last term is over all possible fermions $\xi_i$ in the theory that can mediate a Yukawa interaction between
$\bar\psi_1\phi_3$ on one side and ${\psi}_2\phi_4$ on the other.
By summing over $\xi_i$ one reconstructs the tensor $Y_t$, similarly
as for $T_s$ in the $s$-channel. The helicity structure of the $u$-channel is exactly the same as that of the $t$-channel, and
we get
\be
\left.\amp_{\rm UV}(1_{\bar\psi_1},2_{{\psi}_2},3_{\phi_3},4_{\phi_4})\right|_{u-ch}=\frac{\la 13 \ra [23]}{16\pi^2\Mp^2\epsilon}(Y_t-3Y_u-T_s)\,.
\ee
By summing over all channels, we then obtain \eq{UV1}.

\subsubsection{Fully fermionic divergences}

We now consider divergences in the fully fermionic sector, starting with the ``exceptional'' $h=-2$ amplitudes, i.e.~those of the form of \eq{exceptionalUV}. In non-holomorphic theories this amplitude can have a contribution at the marginal level, with the tree-level amplitude being given by
\be \label{eq:exceptionalTree}
\amp_{\rm tree}(1_{\psi_1},2_{{\psi}_2},3_{{\psi}_3},4_{\psi_4})=\left(Y_s+Y_u+Y_t \right)\frac{[ 12 ]}{\la 34\ra}\,.
\ee
This structure may look surprising at first sight due to the apparent absence of $t$- and $u$-channel poles, which is not expected in general since scalars can be exchanged in all channels. However, using kinematic identities like $[ 12 ] \la 24\ra+[13] \la 34\ra=0$, one can show for example that a $t$-channel pole like $[13]/\la 24\ra$ can be rewritten as $-[12]/\la 34\ra$.
This is why \eq{eq:exceptionalTree} turns out to be crossing symmetric.

If present at the marginal level, this amplitude could generate a non-vanishing contribution to the divergence of $\amp_{\rm UV}(1_{\psi_1},2_{{\psi}_2},3_{{\psi}_3},4_{\psi_4})$ at $\mathcal{O}(1/\Mp^2)$. A divergence in this $|\tilde{h} |= 2$ amplitude is the only allowed exception to the helicity non-renormalization rule in~\eq{n4htilde}. However, as we now show, the one-loop contributions from~\eq{eq:exceptionalTree} to~\eq{exceptionalUV} add up to zero, such that the rule~\eq{n4htilde} is exact in any marginal theory minimally coupled to gravity.

We proceed by giving the expression for the $s$-channel cuts contributing to \eq{exceptionalUV}. We note that a contribution with one cut graviton is absent here, because one would need a tree-level amplitude with $\tilde{h}\neq 0$ containing one graviton,
which is not allowed according to the tree-level helicity rules of Section~\ref{sectree}. We therefore have
\begin{align}\label{exampleUVt}
&\left.\amp_{\rm UV}(1_{\psi_1},2_{{\psi}_2},3_{\psi_3},4_{\psi_4})\right|_{s-ch}=\frac{1}{8\pi^3\epsilon} {\cal R}\int d{\rm LIPS}_{\ell\ell'} \bigg[ \amp_{\rm tree}(1_{\psi_1},2_{{\psi}_2},-\ell_{{\psi}_3},-\ell'_{\psi_4}) \nonumber\\
&~~~\times\amp_{\min}(\ell_{\bar{\psi}_3},\ell'_{\bar{\psi}_4},3_{\psi_3},4_{\psi_4})+
\amp_{\min}(1_{\psi_1},2_{{\psi}_2},-\ell_{\bar{\psi}_1},-\ell'_{\bar{\psi}_2})\,\amp_{\rm tree}(\ell_{\psi_1},\ell'_{\psi_2},3_{\psi_3},4_{\psi_4}) \bigg] \,.
\end{align}
The two terms are structurally equivalent and they give the same result. Let us quote here the result of the relevant phase space
integration:
\begin{align}\label{excepzero}
\allowdisplaybreaks
&{\cal R}\!\int d{\rm LIPS}_{\ell\ell'} \amp_{\min}(1_{\psi_1},2_{{\psi}_2},-\ell_{\bar{\psi}_1},-\ell'_{\bar{\psi}_2})\,\amp_{\rm tree}(\ell_{\psi_1},\ell'_{\psi_2},3_{\psi_3},4_{\psi_4})\nonumber \\
&={\cal R}\!\int d{\rm LIPS}_{\ell\ell'} \,\frac{1}{\Mp^2}\frac{s+6\ell.p_2}{-8\ell.p_1}[12] \la\ell\ell'\ra\times{ (Y_s+Y_u+Y_t)}\frac{[34]}{\la\ell\ell'\ra}\nonumber \\
&=\frac{Y_s+Y_u+Y_t}{\Mp^2}\,[12][34]~{\cal R}\!\!\int d{\rm LIPS}_{\ell\ell'}\left(\frac{3}{4}{-\frac{s}{2\ell.p_1}}
\right)
=\frac{3\pi(Y_s+Y_u+Y_t)}{8\Mp^2}\,[12][34]\,.
\end{align}
In the first step we just gave the explicit expression of the phase space integrand. In the second step we reduced the integrand
to the sum of a pure bubble and a pure triangle, which is a baby example of the Passarino-Veltman integral reduction. Thanks
to the properties of ${\cal R}$, we could then project out the triangle, what leaves us with a trivial phase space integral.
A reduction like this is always possible but often complicated, so that most of the times we let ${\cal R}$ do it automatically as explained in Appendix~\ref{bubble}.
Due to the identical helicity structure, the $t$- and $u$-channel cuts can be obtained in a similar way. Paying attention to the signs, one can see that sum over the three cuts is proportional to the crossing invariant combination
\be
\amp_{\rm UV}(1_{\psi_1},2_{{\psi}_2},3_{{\psi}_3},4_{\psi_4})\propto [12][34] +[14][23]+[13][42]\,,
\ee
which is \emph{zero} thanks to the Schouten identity. This result can be traced back to the crossing symmetry of \eq{eq:exceptionalTree}, which minimal gravity does not spoil being color/flavor blind. All in all we find that
\be
\boxed{{\color{white} \frac{1}{1}}\amp_{\rm UV}\left(1_{\psi_1},2_{{\psi}_2},3_{{\psi}_3},4_{\psi_4}\right)=0\,.{\color{white} \frac{1}{1}}}
\ee
While the all-minus fermionic amplitude is not renormalized, the combination with total helicity $h=0$ is renormalized. Since the procedure
should be clear by now, we just quote in turn the tree-level amplitude and the corresponding UV-divergent amplitude.
At tree level, the most general amplitude for two positive and two negative helicity fermions is given by
\be\label{tree2}
\amp_{\rm tree}(1_{\bar\psi_1},2_{{\psi}_2},3_{{\psi}_3},4_{\bar\psi_4})=\left(\frac{T_s}{s}+\frac{T_t}{t}+\frac{Y_u}{u}\right)\la 14 \ra [23]\,,
\ee
while the minimal gravity induced UV-divergence is found to be
\be\label{UV2}
\boxed{~\amp_{\rm UV}(1_{\bar\psi_1},2_{{\psi}_2},3_{{\psi}_3},4_{\bar\psi_4})=-\frac{55}{192\pi^2\Mp^2\epsilon}\left(T_s+T_t\right)\la 14 \ra [23]\,,~}
\ee
the relevant coefficients being 55/12 for the vector channels and, perhaps more interestingly, \emph{zero} for the Yukawa channel.
Our methods cannot explain why the Yukawa channel, which is mediated by a scalar, is not associated to any UV divergence.

\subsubsection{Fully scalar divergences}

Let us now present our results for the divergences of purely scalar amplitudes
$\amp(1_{\phi_1},2_{\phi_2},3_{\phi_3},4_{\phi_4})$. The topological splitting into class \emph{(A)} and \emph{(B)} in Fig.~\ref{fig:dim6} is still valid, but there is a novelty here with respect to the previous cases, as class \emph{(B)} does not vanish completely.

We first consider \emph{(A)}, and start as usual by quoting the relevant tree-level amplitude in the marginal theory, which on general grounds takes the
form
\be\label{tree3}
\amp_{\rm tree}(1_{\phi_1},2_{\phi_2},3_{\phi_3},4_{\phi_4})=C+\frac{t-u}{s}\,T_s+\frac{u-s}{t}\,T_t+\frac{s-t}{u}\,T_u\,,
\ee
where $C$ is a contact term coming from quartic scalar couplings. By using \eq{gamma2}, we can then compute
the loop divergent part of the amplitude at ${\cal O}(\Mp^{-2})$, which is found to be
\be\label{UV3A}
\boxed{~\left.\amp_{\rm UV}(1_{\phi_1},2_{\phi_2},3_{\phi_3},4_{\phi_4})\right|_{(A)}=-\frac{5}{12\pi^2\Mp^2\epsilon}\left[(t-u)\,T_s+(u-s)\,T_t+(s-t)\,T_u\right]\,.~}
\ee
We see that the quartic term is not associated to any UV divergence, as one can show by a simple readaptation of \eq{excepzero}.
In this case, the term proportional to $C$ vanishes thanks to $s+t+u=0$ after summing over all cuts.

\begin{figure}[t]
\centering
\begin{tikzpicture}[line width=1.1 pt, scale=.7, baseline=(current bounding box.center)]
	\filldraw [white] (0,0) circle (2pt) 
	(-11.6,0) circle (2pt)
	(12,0) circle (2pt)
	(0,4) circle (2pt);
	
	\draw[dashed] (-10.9,3.5) -- (-9.9,2) -- (-10.9,0.5) ;
	\draw[graviton] (-9.9,2.05) -- (-8,2.05) ;
	\draw[graviton] (-9.9,1.95) -- (-8,1.95) ;
	\draw[dashed] (-7,3.5) -- (-8,2) -- (-7,0.5) ;
	\draw[decorate,decoration={snake,amplitude=.4mm,segment length=1.2 mm}] (-7.5,2.75) to [bend left] (-7.5,1.25) ;
	\node at (-8.9,2.5) {\footnotesize $J=0$};
	\node at (-11.15,3.5) {$\phi$};
	\node at (-11.15,0.5) {$\bar{\phi}$};
	\node at (-6.7,3.5) {$\phi'$};
	\node at (-6.8,0.5) {$\bar{\phi}'$};

	\draw[dashed] (-4.9,3.5) -- (-3.9,2) -- (-4.9,0.5) ;
	\draw[graviton] (-3.9,2.05) -- (-2,2.05) ;
	\draw[graviton] (-3.9,1.95) -- (-2,1.95) ;
	\draw[dashed] (-1,3.5) -- (-2,2) -- (-1,0.5) ;
	\draw[decorate,decoration={snake,amplitude=.4mm,segment length=1.2 mm}] (-1.2,3.2) to [bend left=80] (-1.8,2.3) ;
	\node at (-2.9,2.5) {\footnotesize $J=0$};
	\begin{scope}[shift={(6,0)}]
	\node at (-11.15,3.5) {$\phi$};
	\node at (-11.15,0.5) {$\bar{\phi}$};
	\node at (-6.7,3.5) {$\phi'$};
	\node at (-6.8,0.5) {$\bar{\phi}'$};
	\end{scope}
	
	\draw[dashed] (5,3.5) -- (4,2) -- (5,0.5) ;
	\draw[graviton] (4,2.05) -- (2.1,2.05) ;
	\draw[graviton] (4,1.95) -- (2.1,1.95) ;
	\draw[dashed] (1.1,3.5) -- (2.1,2) -- (1.1,0.5) ;
	\draw[fill=white] (4.5,2.75) circle (0.3cm) ;
	\node at (3,2.5) {\footnotesize $J=0$};
	\begin{scope}[shift={(12,0)}]
	\node at (-11.15,3.5) {$\phi$};
	\node at (-11.15,0.5) {$\bar{\phi}$};
	\node at (-6.7,3.5) {$\phi'$};
	\node at (-6.8,0.5) {$\bar{\phi}'$};
	\end{scope}
	
	\draw[dashed] (11,3.5) -- (10,2) -- (11,0.5) ;
	\draw[graviton] (9.6,2.05) -- (8,2.05) ;
	\draw[graviton] (9.6,1.95) -- (8,1.95) ;
	\draw[dashed] (7,3.5) -- (8,2) -- (7,0.5) ;
	\draw[dashed,fill=white] (9.5,2) circle (0.48cm) ;
	\node at (8.47,2.4) {\scriptsize $J=0$};
	\begin{scope}[shift={(17.95,0)}]
	\node at (-11.25,3.5) {$\phi$};
	\node at (-11.15,0.5) {$\bar{\phi}$};
	\node at (-6.65,3.5) {$\phi'$};
	\node at (-6.8,0.5) {$\bar{\phi}'$};
	\end{scope}
	
 \end{tikzpicture}
 \caption{\emph{Feynman diagrams of class (B) that ``survive'' the cancellation. They are all proportional
 to the $J=0$ component of the gravity mediated amplitude $\amp_{\min}(1_\phi,2_{\bar{\phi}},3_{\bar{\phi}'},4_{\phi'})$.
 The analogous terms proportional to the $J=2$ component of $\amp_{\min}$ cancel among themselves, as we explain
 in Appendix~\ref{collinear}.}}
 \label{fig:scalar6}
 \end{figure}
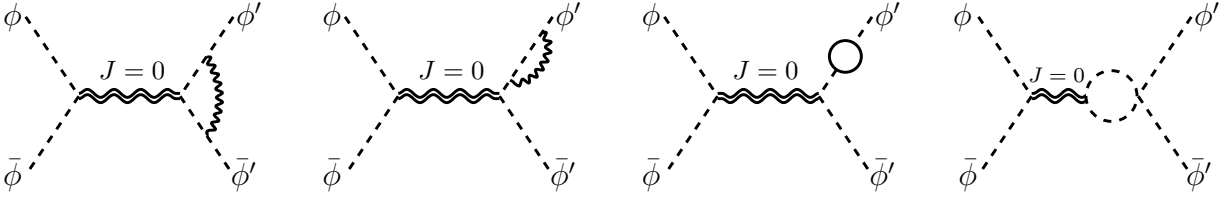
 
\vspace{.2cm}

We move now to class \emph{(B)}, elaborating on \eq{ThhT}. As a preliminary observation, we note that, because of the flavor/color-blindness of gravity, particles 1,2,3,4 must come in particle antiparticle pairs to possibly contribute to \emph{(B)}. We will therefore take 1,2 and 3,4 to be particle antiparticle pairs, respectively $\phi,\bar{\phi}$ and $\bar{\phi}',\phi'$. We will also take 3 and 4 \emph{not} to be 1's antiparticle.

Like we said previously, the crucial point is that scalars couple to gravity through a stress tensor $T^{\mu\nu}$ with non-vanishing trace. We therefore split $T^{\mu\nu}=T^{\mu\nu}_0+T^{\mu\nu}_2$, where $T_{0,2}$ are ``purely trace'' and traceless, both conserved (see Appendix~\ref{collinear} for the precise definitions). In particular we have at tree level
\be
\la 1,2| T^{\mu\nu}|\Omega \ra^{(0)}=\la 1,2 | T_0^{\mu\nu}|\Omega \ra^{(0)}+\la 1,2 | T_2^{\mu\nu}|\Omega \ra^{(0)}\,.
\ee
As a consequence of the conservation of $T^{\mu\nu}$, we show in Appendix~\ref{collinear} that the UV-divergent part of $\la \Omega | T^{\rho\sigma}|3,4 \ra^{(1)}$, in which we are interested as it enters the UV-divergent part of \emph{(B)}, reads
\be
\left.\la \Omega | T^{\rho\sigma}|3,4 \ra^{(1)}\right|_{\rm UV}=-\frac{\gamma_0^{(\phi')}}{2\epsilon}\la \Omega | T_0^{\rho\sigma}|3,4 \ra^{(0)}\,,
\ee
where $\gamma_0^{(\phi')}$ is the UV anomalous dimensions of $T_0$ in the $\phi'$ sector. Notice that only $T_0$ enters the RHS. By using the harmonic gauge, we can then compute
\be
\la \Omega | h_{\mu\nu}h_{\rho\sigma} |\Omega \ra ^{(0)}=\frac{\eta_{\mu\rho}\eta_{\nu\sigma}+\eta_{\mu\sigma}\eta_{\nu\rho}-\eta_{\mu\nu}\eta_{\rho\sigma}}{2(p_1+p_2)^2}\,.
\ee
When we put things together as dictated by \eq{ThhT}, we get after dropping the $(0)$ suffix and using $P\equiv p_1+p_2$
\begin{align}
&\la 1,2 | T_0^{\mu\nu}+T_2^{\mu\nu}|\Omega \ra\,\frac{2\eta_{\mu\rho}\eta_{\nu\sigma}-\eta_{\mu\nu}\eta_{\rho\sigma}}{2P^2}\,\frac{\gamma_0^{(\phi')}}{2\epsilon}\la \Omega | T_0^{\rho\sigma}|3,4 \ra\nonumber \\
&=\frac{\gamma_0^{(\phi')}}{2\epsilon P^2}\left( \la 1,2 | (T_0)^{\mu\nu}+(T_2)^{\mu\nu}|\Omega \ra\la \Omega | (T_0)_{\mu\nu}|3,4 \ra-\frac{1}{2}\la 1,2 | (T_0)_{\mu}^{\mu}|\Omega \ra\la \Omega | (T_0)_{\sigma}^{\sigma}|3,4 \ra\right)\nonumber \\
&=\frac{\gamma_0^{(\phi')}}{2\epsilon P^2}\left( \la 1,2 | (T_0)^{\mu\nu}|\Omega \ra\la \Omega | (T_0)_{\mu\nu}|3,4 \ra-\frac{1}{2}\la 1,2 | (T_0)_{\mu}^{\mu}|\Omega \ra\la \Omega | (T_0)_{\sigma}^{\sigma}|3,4 \ra\right)\nonumber \\
&=\frac{\gamma_0^{(\phi')}}{2\epsilon}\amp_{\rm min}^{(J=0)}(1_{\phi},2_{\bar{\phi}},3_{\bar{\phi}'},4_{{\phi}'})=\frac{\gamma_0^{(\phi')}}{2\epsilon}\,\frac{s}{6\Mp^2}\,.
\end{align}
This result requires several explanations. From the algebraic point of view, we used $(T_2)^\mu_\mu=0$ and that $\la 1,2 |(T_2)^{\mu\nu}|\Omega \ra\la \Omega | (T_0)_{\mu\nu}|3,4 \ra=0$. This last orthogonality property comes
from the fact that $T_0$ and $T_2$ transform differently under rotations, in a spin 0 and a spin 2 representation, respectively (cf.~Appendix~\ref{collinear}).
In the final step we recovered the $J=0$ component of \eq{J2phi}. In fact, with similar considerations to those that lead to \eq{ThhT}, we have
\be
\amp_{\rm min}(1_{\phi},2_{\bar{\phi}},3_{\bar{\phi}'},4_{{\phi}'})=\la 1,2 | T_0^{\mu\nu}+T_2^{\mu\nu}|\Omega \ra\,\frac{2\eta_{\mu\rho}\eta_{\nu\sigma}-\eta_{\mu\nu}\eta_{\rho\sigma}}{2P^2}\,\la \Omega | T_0^{\rho\sigma}+T_2^{\rho\sigma}|3,4 \ra\,,
\ee
where all quantities are tree-level. The above expression can be decomposed as a term with $T_2\times T_2$ and a term with $T_0 \times T_0$, corresponding precisely to the $J=2,0$ components of \eq{J2phi}.

When we add the terms coming from loops inserted on the 1,2 side instead of the 3,4 side (see Fig.~\ref{fig:dim6}), we finally get
\be\label{UV3B}
\boxed{~\left.\amp_{\rm UV}(1_{\phi},2_{\bar{\phi}},3_{\bar{\phi}'},4_{\phi'})\right|_{(B)}=-\frac{s}{12\Mp^2\epsilon}( \gamma_0^{(\phi)}+\gamma_0^{(\phi')})\,.~}
\ee

The above, very general, derivation tracks the class \emph{(B)} UV-divergent contributions of the fully scalar amplitude to the UV anomalous dimension of $T_0$ in the $\phi$ and $\phi'$ sector. A more direct derivation goes through the computation of diagrams like those of Fig.~\ref{fig:scalar6}.\footnote{We note here that the amplitude $\amp_{C^2\phi^2}$, that we study in Section~\ref{mixinggrav}, undergoes a self renormalization which has many similarities to Fig.~\ref{fig:scalar6}.}

We close this section by noticing that \eq{UV3B} becomes \emph{zero} when the scalar is coupled to gravity \emph{conformally} instead of minimally, 
since the $J=0$ piece of \eq{J2phi} vanishes in this case. 
We stress however that, for conformally coupled scalars, also the other divergences that we computed get modified. In Appendix~\ref{conformalgen} we extend all our results to a conformal coupling of gravity to scalars (as well as to non-vanishing torsion, in the form of a contact four-fermion gravitational interaction).

\subsection{Divergences and positivity in minimally coupled theories: ${\cal O}(\Mp^{-4})$}
\label{Ord4}

In this section we continue our analysis of the one-loop divergence structure of a generic minimally coupled theory, moving to
${\cal O}(\Mp^{-4})$ and focusing on 4-point amplitudes. At this level, only the gravitational interactions matter, so that
marginal couplings do not play any role. Like in the previous section, \eq{n4htilde} plays a leading role by leaving only a handful
of amplitude renormalizations to compute, all with $\tilde{h}=0$. The flavor structure is also extremely constrained,
allowing only amplitudes involving particle-antiparticle pairs, that is of the form $\amp(1_{\bar\Phi},2_{{\Phi}},3_{{\Phi}'},4_{\bar\Phi'})$, where $\Phi,\Phi'$ have helicity $0,1/2,1$.
We will fix $h(\Phi)\geq h(\Phi')$ and assume that $1$ and $4$ are distinguishable particles, as the identical case is simply recovered with crossing arguments.

It is convenient to parameterize the relevant contact amplitudes in terms of an amplitude basis with definite angular momentum in the $1, 2 \rightarrow 3, 4$ channel, i.e.~in the so-called angular momentum basis (see e.g. \cite{Jiang:2021tqo}). We have
\begin{align}
\amp (1_{\bar\phi},2_{{\phi}},3_{{\phi}'},4_{\bar\phi'})&=\frac{1}{\Mp^4}\left[ C^{(2)}_{\phi\phi'}\, \frac{s^2-6\,tu}{6} + C^{(1)}_{\phi\phi'}\, \frac{s (t-u) }{2}+ C^{(0)}_{\phi\phi'}\, s^2 \right]\,, \label{eq:scalarDim8}\\
\amp (1_{\bar\psi},2_{{\psi}},3_{{\phi}},4_{\bar\phi})&=-\frac{1}{\Mp^4}\left[ C^{(2)}_{\psi\phi}\, \frac{t-u}{2} + C^{(1)}_{\psi\phi}\, s\right]\la 13 \ra [23]\,, \label{eq:fermionScalarDim8}\\
\amp(1_{\bar\psi},2_{{\psi}},3_{{\psi}'},4_{\bar\psi'})&=-\frac{1}{\Mp^4}\left[ C^{(2)}_{\psi\psi'}\, \frac{3t-u}{4} + C^{(1)}_{\psi\psi'}\, s \right]\la 14 \ra [23]\,,\label{eq:fermionDim8}\\
\amp (1_{V_-},2_{V_+},3_{{\Phi}},4_{\bar\Phi})&=-\frac{C^{(2)}_{V\Phi}}{\Mp^4}(-1)^{\delta_{1,h_\Phi}}\la 13 \ra^2 [23]^2
\left( \frac{\la 14 \ra}{\la 13 \ra} \right)^{2h_\Phi}\,, \label{eq:vectorDim8}
\end{align}
where the part of the amplitude proportional to $C^{(J)}$, that we call $\amp^{(J)}$, has angular momentum $J$ in the $1, 2 \rightarrow 3, 4$ channel.\footnote{The coefficients of the polynomials in Eqs.~(\ref{eq:scalarDim8}--\ref{eq:vectorDim8}) are fixed as follows. Considering for example $J=2$, we want $\amp^{(2)}$ to match the corresponding amplitude among Eqs.~(\ref{J2phi}--\ref{J2vec})
under the substitution $C^{(2)}\to -\Mp^2/s$
(with the $J = 0$ component of the four-scalar amplitude set to zero, or $\alpha=0$ in \eq{eq:ScalarNonMin}).
Similarly for the $\amp^{(1)}$'s and the corresponding amplitudes mediated by annihilation into a vector boson, that are obtained by $C^{(1)}\to -\Mp^4g^2/s^2$.}

\begin{figure}[t]
\centering
\begin{tikzpicture}[line width=1.1 pt, scale=.7, baseline=(current bounding box.center)]

	\draw[decorate,decoration={snake,amplitude=.55mm,segment length=1.8 mm}] (-1.5,.5) to [bend left] (1.5,.5) ;
	\draw[decorate,decoration={snake,amplitude=.55mm,segment length=1.8 mm}] (-1.5,-.5) to [bend right] (1.5,-.5) ;
	\draw (-2.5,1) -- (-1.5,.5) ;
	\draw (-2.5,-1) -- (-1.5,-.5) ;
	\draw (2.5,1) -- (1.5,.5) ;
	\draw (2.5,-1) -- (1.5,-.5) ;
	\draw[dotted] (0,1.4) -- (0,-1.4) ;
	\draw[fill=gray!15] (-1.5,0) ellipse (.5 cm and .9 cm) ;
	\draw[fill=gray!15] (1.5,0) ellipse (.5 cm and .9 cm) ;
	\node at (-2.8,1) { 1} ;
	\node at (-2.8,-1) { $\bar{1}$} ;
	\node at (2.8,1) { 2} ;
	\node at (2.8,-1) { $\bar{2}$} ;
	\node at (-.5,1.5) {\small $+$} ;
	\node at (-.5,-1.5) {\small $-$} ;
	\node at (.7,1.5) {\small $-$} ;
	\node at (.7,-1.5) {\small $+$} ;
	\node at (4.2,2.2) {\large $s{\rm -cut}$} ;

	\begin{scope}[shift={(8.0,0)}]
	\draw (-2.2,1) -- (-1.7,0) -- (-2.2,-1) ;
	\draw (2.2,1) -- (1.7,0) -- (2.2,-1) ;
	\draw[decorate,decoration={snake,amplitude=.55mm,segment length=1.8 mm}] (-1.7,0) -- (0,0) ;
	\draw[decorate,decoration={snake,amplitude=.55mm,segment length=1.8 mm}] (1.7,0) -- (0,0) ;
	\draw[fill=white] (0,0) circle (0.65cm) ;
	\draw[dotted] (0,1.2) -- (0,-1.2) ;
	\node at (-2.6,1) { 1} ;
	\node at (-2.6,-1) { $\bar{1}$} ;
	\node at (2.6,1) { 2} ;
	\node at (2.6,-1) { $\bar{2}$} ;
	\node at (-.5,1.) {\small $3$} ;
	\node at (-.5,-1.) {\small $\bar{3}$} ;
	\node at (.5,1.) {\small $\bar{3}$} ;
	\node at (.5,-1.) {\small $3$} ;
	\node at (-4,0) {\LARGE ${,}$} ;
	\end{scope}

	\begin{scope}[shift={(8.0,-4)}]
	\draw (-2.2,1) -- (-1.7,0) -- (-2.2,-1) ;
	\draw (2.2,1) -- (1.7,0) -- (2.2,-1) ;
	\draw[decorate,decoration={snake,amplitude=.55mm,segment length=1.8 mm}] (-1.7,0) -- (0,0) ;
	\draw[decorate,decoration={snake,amplitude=.55mm,segment length=1.8 mm}] (1.7,0) -- (0,0) ;
	\draw[fill=white] (0,0) circle (0.65cm) ;
	\draw[dotted] (0,1.2) -- (0,-1.2) ;
	\node at (-2.6,1) { 1} ;
	\node at (-2.6,-1) { $\bar{1}$} ;
	\node at (2.6,1) { 2} ;
	\node at (2.6,-1) { $\bar{2}$} ;
	\node at (-.5,1.) {\small $1$} ;
	\node at (-.5,-1.) {\small $\bar{1}$} ;
	\node at (.5,1.) {\small $\bar{1}$} ;
	\node at (.5,-1.) {\small $1$} ;
	\node at (-4,0) {\LARGE ${+}$} ;
	\end{scope}

	\begin{scope}[shift={(8.0,-8)}]
	\draw (-2.2,1) -- (-1.7,0) -- (-2.2,-1) ;
	\draw (2.2,1) -- (1.7,0) -- (2.2,-1) ;
	\draw[decorate,decoration={snake,amplitude=.55mm,segment length=1.8 mm}] (-1.7,0) -- (0,0) ;
	\draw[decorate,decoration={snake,amplitude=.55mm,segment length=1.8 mm}] (1.7,0) -- (0,0) ;
	\draw[fill=white] (0,0) circle (0.65cm) ;
	\draw[dotted] (0,1.2) -- (0,-1.2) ;
	\node at (-2.6,1) { 1} ;
	\node at (-2.6,-1) { $\bar{1}$} ;
	\node at (2.6,1) { 2} ;
	\node at (2.6,-1) { $\bar{2}$} ;
	\node at (-.5,1.) {\small $2$} ;
	\node at (-.5,-1.) {\small $\bar{2}$} ;
	\node at (.5,1.) {\small $\bar{2}$} ;
	\node at (.5,-1.) {\small $2$} ;
	\node at (-4,0) {\LARGE ${+}$} ;
	\end{scope}

	\begin{scope}[shift={(0.0,-8)}]
	\draw (-2.3,.8) -- (-1.8,0) -- (-2.3,-.8) ;
	\draw (-.1,.8) -- (-.6,0) -- (-.1,-.8) ;
	\draw[decorate,decoration={snake,amplitude=.55mm,segment length=1.8 mm}] (-1.8,0) -- (-.6,0) ;
	\draw (0.1,.8) -- (2.2,.8) ;
	\draw (0.1,-.8) -- (2.2,-.8) ;
	\draw[decorate,decoration={snake,amplitude=.55mm,segment length=1.8 mm}] (1.1,.8) -- (1.1,-.8) ;
	\draw[dotted] (0,1.2) -- (0,-1.2) ;
	\node at (-2.6,1) { 1} ;
	\node at (-2.6,-1) { $\bar{1}$} ;
	\node at (2.6,1) { 2} ;
	\node at (2.6,-1) { $\bar{2}$} ;
	\node at (-.4,1.2) {\small ${2}$} ;
	\node at (-.4,-1.2) {\small $\bar{2}$} ;
	\node at (.4,1.2) {\small $\bar{2}$} ;
	\node at (.4,-1.2) {\small ${2}$} ;
	\end{scope}

	\begin{scope}[shift={(0.0,-4)}]
	\draw (2.3,.8) -- (1.8,0) -- (2.3,-.8) ;
	\draw (.1,.8) -- (.6,0) -- (.1,-.8) ;
	\draw[decorate,decoration={snake,amplitude=.55mm,segment length=1.8 mm}] (1.8,0) -- (.6,0) ;
	\draw (-0.1,.8) -- (-2.2,.8) ;
	\draw (-0.1,-.8) -- (-2.2,-.8) ;
	\draw[decorate,decoration={snake,amplitude=.55mm,segment length=1.8 mm}] (-1.1,.8) -- (-1.1,-.8) ;
	\draw[dotted] (0,1.2) -- (0,-1.2) ;
	\node at (-2.6,1) { 1} ;
	\node at (-2.6,-1) { $\bar{1}$} ;
	\node at (2.6,1) { 2} ;
	\node at (2.6,-1) { $\bar{2}$} ;
	\node at (-.4,1.2) {\small $\bar{1}$} ;
	\node at (-.4,-1.2) {\small ${1}$} ;
	\node at (.5,1.2) {\small ${1}$} ;
	\node at (.5,-1.2) {\small $\bar{1}$} ;
	\end{scope}
	
	\draw[thin]  (12,1) -- (12,-9) ;
	\draw[thin]  (12,-4) -- (19,-4) ;

	\begin{scope}[shift={(16.0,-1.5)}]
	\draw (-2.2,.9) -- (2.2,.9) ;
	\draw (-2.2,-.95) -- (2.2,-.95) ;
	\draw[dotted] (0,1.2) -- (0,-1.2) ;
	\draw[decorate,decoration={snake,amplitude=.55mm,segment length=1.8 mm}] (1.1,.9) -- (1.1,-.95) ;
	\draw[decorate,decoration={snake,amplitude=.55mm,segment length=1.8 mm}] (-1.1,.9) -- (-1.1,-.95) ;
	\node at (-2.6,1) { 1} ;
	\node at (-2.6,-1) { 2} ;
	\node at (2.6,1) { $\bar{1}$} ;
	\node at (2.6,-1) { $\bar{2}$} ;
	\node at (0,2.) {\large $t{\rm -cut}$} ;
	\end{scope}

	\begin{scope}[shift={(16.0,-7.)}]
	\draw (-2.2,.9) -- (2.2,.9) ;
	\draw (-2.2,-.95) -- (2.2,-.95) ;
	\draw[dotted] (0,1.2) -- (0,-1.2) ;
	\draw[decorate,decoration={snake,amplitude=.55mm,segment length=1.8 mm}] (1.1,.9) -- (1.1,-.95) ;
	\draw[decorate,decoration={snake,amplitude=.55mm,segment length=1.8 mm}] (-1.1,.9) -- (-1.1,-.95) ;
	\node at (-2.6,1) { 1} ;
	\node at (-2.6,-1) {$\bar{2}$} ;
	\node at (2.6,1) { $\bar{1}$} ;
	\node at (2.6,-1) {2 } ;
	\node at (0,2.) {\large $u{\rm -cut}$} ;
	\end{scope}
	
	\filldraw [white] (0,-10) circle (2pt) ;
		
 \end{tikzpicture}
 \caption{\emph{Cut diagrams entering the computation of the anomalous dimensions at ${\cal O}(\Mp^{-4})$ in \eq{eq:gammaDim8}, with solid lines for matter and wavy lines for gravitons. To avoid clutter, we label $1\equiv \Phi$, $2\equiv \Phi'$ and $3\equiv \Phi''$. Diagrams are
 divided into $s$-, $t$- and $u$-cut. The right column of the $s$-cut panel corresponds to the universal contribution
 in Eqs.~(\ref{eq:anomDim8Scalar}-\ref{eq:anomDim8Vector}), proportional to the factor $K$ defined in \eq{Kfactor}.}}
 \label{fig:dim8}
 \end{figure}
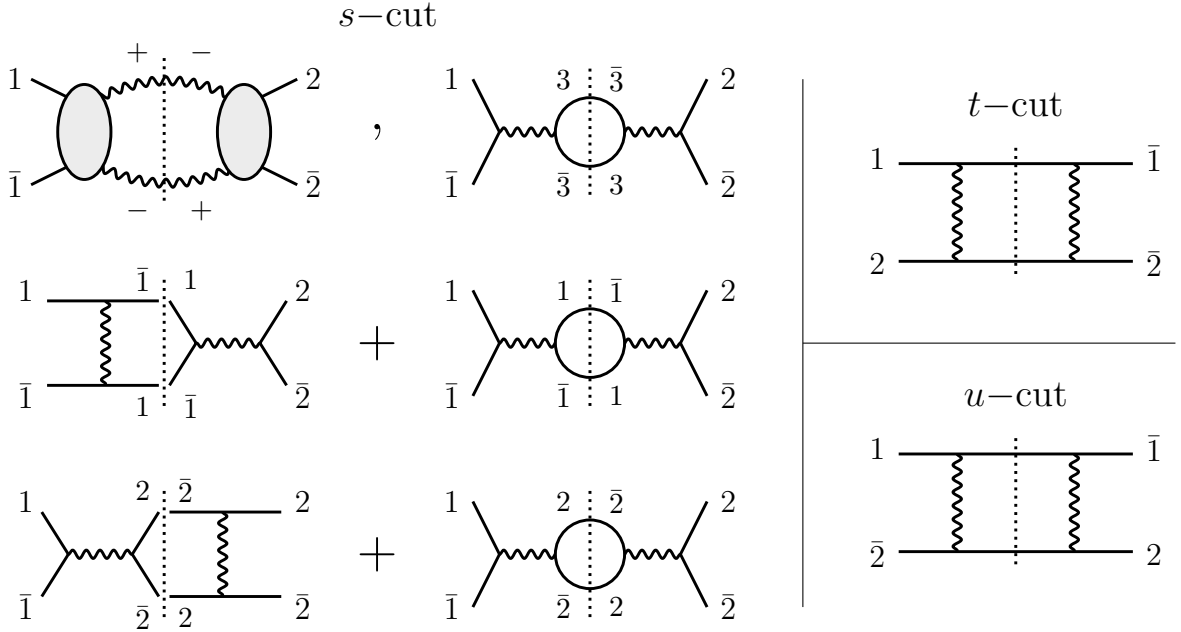

Let us now present how to obtain the one-loop anomalous dimensions $\gamma_{\Phi\Phi'}^{(J)} = dC_{\Phi\Phi'}^{(J)}/d\ln\mu$ from~\eq{gamma2}. Taking into account that gravitational interactions do not generate collinear anomalous dimensions, i.e.~$\gamma_{\rm coll}=0$,~\eq{gamma2} reduces to the sum over the cuts in the $s$-, $t$- and $u$-channels
\be \label{eq:gammaDim8}
\sum_J \gamma_{\Phi\Phi'}^{(J)} \frac{\amp_{\Phi\Phi'}^{(J)}(1_{\bar\Phi}, 2_{{\Phi}},3_{{\Phi}'}, 4_{\bar\Phi'})}{C_{\Phi\Phi'}^{(J)}}=-\frac{1}{4\pi^3}\,{\cal R}\!\!\int d{\rm LIPS}\big[ (\text{$s$-cut}) + (\text{$t$-cut}) + (\text{$u$-cut})\big].
\ee
We first consider the $s$-channel cut, whose many contributions are depicted in the leftmost panel of Fig.~\ref{fig:dim8}.
The first contribution is given by the graviton cut, that reads
\be \label{eq:scuth}
(\text{$s$-cut})_h = \amp(1_{\bar\Phi}, 2_{{\Phi}},-\ell'_{h_+},-\ell_{h_-})\,\amp(\ell_{h_+},\ell'_{h_-},3_{{\Phi}'}, 4_{\bar\Phi'})\,,
\ee
where the relevant two-graviton amplitudes are given by
\be
\amp(1_{h_-},2_{h_+},3_{{\Phi}}, 4_{\bar\Phi})=-\frac{1}{\Mp^2}\frac{(\la 14 \ra [23])^{2h_\Phi}(\la 13 \ra [32])^{4-2h_\Phi}}{stu}\,,
\ee
which can be derived as a special case of \eq{1graviton}. The second contribution, that we dub $(\text{$s$-cut})_{\Phi''}$ and corresponds to the second diagram in the first row of
Fig.~\ref{fig:dim8}, is of the following form
\be \label{eq:scut1}
(\text{$s$-cut})_{\Phi''} = \sum_{\Phi''\neq \Phi,\Phi'}\sigma_{\Phi''\Phi''} \amp_{\rm min}(1_{\bar\Phi}, 2_{{\Phi}},-\ell'_{\Phi''},-\ell_{\bar{\Phi}''})\,\amp_{\rm min}(\ell_{\Phi''},\ell'_{\bar{\Phi}''},3_{{\Phi}'}, 4_{\bar\Phi'})\,,
\ee
where the sum runs over all particle-antiparticle pairs that are different from both $\Phi\bar{\Phi}$ and $\Phi'\bar{\Phi}'$.
Finally, we have the contribution corresponding to the second and third line of the $s$-cut panel of Fig.~\ref{fig:dim8},
that reads
\begin{align} \label{eq:scut2}
(\text{$s$-cut})_{\Phi,\Phi'} &= \sigma_{\Phi\Phi} \amp_{\rm min}(1_{\bar\Phi}, 2_{{\Phi}},-\ell'_{\Phi},-\ell_{\bar{\Phi}})\,\amp_{\rm min}(\ell_{\Phi},\ell'_{\bar{\Phi}},3_{{\Phi}'}, 4_{\bar\Phi'})\nonumber \\
&+\sigma_{\Phi'\Phi'} \amp_{\rm min}(1_{\bar\Phi}, 2_{{\Phi}},-\ell'_{\Phi'},-\ell_{\bar{\Phi}'})\,\amp_{\rm min}(\ell_{\Phi'},\ell'_{\bar{\Phi}'},3_{{\Phi}'}, 4_{\bar\Phi'})\,.
\end{align}
As depicted in Fig.~\ref{fig:dim8}, each of these last two contributions can be further split into two terms corresponding
to diagrams with different topology,
respectively in the left and right side of the second and third line of the $s$-cut panel. It should be clear from the picture
that \eq{eq:scut1} and the two contributions of \eq{eq:scut2} corresponding to the rightmost diagrams in the $s$-cut panel of Fig.~\ref{fig:dim8}
(that is, collectively, the whole second column of Fig.~\ref{fig:dim8}) can be grouped together, as they are all corrections
to the graviton propagator. Indeed, in the Lagrangian formulation they result in a renormalization of operators with two Ricci scalars or Ricci tensors of the form $R^2$ and $R_{\mu\nu}^2$. Applying then the equations of motion to these operators gives a universal contribution to on-shell 4-point scattering amplitudes of matter particles. This universal contribution, which is sensitive
to the full matter content of the theory, is known (see e.g.~\cite{Goon:2016mil,Alonso:2019mok}) and will provide a non-trivial partial check of our results.

Also note that the tree amplitudes corresponding to the second column of Fig.~\ref{fig:dim8} proceed through $s$-channel graviton exchange which, except for the minimally coupled four-scalar amplitude, carries angular momentum $J=2$. Thus in general only the renormalization of the $J=2$ Wilson coefficients $C_i^{(2)}$ is sensitive to the full particle spectrum of the theory, the exception being $C_{\phi\phi'}^{(0)}$ which gets renormalized by all scalars due to the $J=0$ component of the minimally coupled four-scalar amplitude (see Appendix~\ref{collinear}).

The first column of Fig.~\ref{fig:dim8} corresponds instead to contributions that 
only depend on the external legs.
As can be seen from the rightmost panels of Fig.~\ref{fig:dim8}, the same is true for the $t$- and $u$-channel cuts, given by
\begin{align}
(\text{$t$-cut}) &= \sigma_{\Phi\Phi'}\amp_{\rm min}(1_{\bar\Phi}, 3_{{\Phi}'},-\ell'_{\bar{\Phi}},-\ell_{\Phi'})\,\amp_{\rm min}(\ell_{\bar{\Phi}'},\ell'_{\Phi},2_{{\Phi}}, 4_{\bar\Phi'})\,,\\
(\text{$u$-cut}) &= \sigma_{\Phi\Phi'}\amp_{\rm min}(1_{\bar\Phi}, 4_{\bar\Phi'},-\ell'_{{\Phi}},-\ell_{{\Phi}'})\,\amp_{\rm min}(\ell_{\bar{\Phi}'},\ell'_{\bar\Phi},2_{{\Phi}}, 3_{{\Phi}'})\,,
\end{align}
with additional minus signs to account for Fermi statistics in amplitudes involving fermions, as the external legs are not
ordered as 1,2,3,4.

With these ingredients, we 
then obtain the anomalous dimensions
\begin{align}
\gamma_{\phi\phi'}^{(2)} &= -\frac{1}{8\pi^2} \bigg( K + \frac{67}{10}\bigg)\,,\qquad~~~ \gamma_{\phi\phi'}^{(1)}=0\,,\qquad \gamma_{\phi\phi'}^{(0)} = -\frac{1}{48\pi^2} \bigg(\frac{N_\phi}{6} + \frac{7}{2}\bigg)\,, \label{eq:anomDim8Scalar}\\
\gamma_{\psi\phi}^{(2)} &= -\frac{1}{8\pi^2}\bigg(K + \frac{129}{20}\bigg)\,,\qquad~~ \gamma_{\psi\phi}^{(1)} = 0\,,\label{eq:anomDim8FermionScalar}\\
\gamma_{\psi\psi'}^{(2)} &= -\frac{1}{8\pi^2}\bigg(K + \frac{181}{30}\bigg)\,,\qquad~\, \gamma_{\psi\psi'}^{(1)} = -\frac{1}{16\pi^2}\frac{25}{24}\,,\label{eq:anomDim8Fermion}\\
\gamma_{V\Phi}^{(2)} &= -\frac{1}{8\pi^2}\bigg(K + \frac{131}{30}\bigg)\,,\label{eq:anomDim8Vector}
\end{align}
%
where $K$ parameterizes the full matter
content of the theory, and is given by
\be\label{Kfactor}
K=\frac{N_\phi}{30}+\frac{N_\psi}{20}+\frac{N_V}{5}\,.
\ee
$N_\Phi$ counts the number of complex scalar, Weyl fermion and vector degrees of freedom for $\Phi=\phi,\psi,V$ respectively.\footnote{For real scalars $\varphi$ one has to make the replacement $N_\phi \rightarrow N_\varphi / 2$.} Interestingly, we see that all $\gamma_{V\Phi}$ turn out to be equal, for any $\Phi$.

Before moving on let us comment on these results. We have chosen the amplitude basis such that all non-vanishing anomalous dimensions are negative. This causes the Wilson coefficients to become increasingly positive due to RG running from a high scale $\Lambda$ to an IR scale $\mu \ll \Lambda$
\be
C^{(J)}_{\Phi\Phi'} (\mu) = C^{(J)}_{\Phi\Phi'} (\Lambda) - \gamma^{(J)}_{\Phi\Phi'} \ln\frac{\Lambda}{\mu}\,.
\ee
If the scale separation is large enough the logarithmic running will dominate over the initial value $C^{(J)}_{\Phi\Phi'} (\Lambda)$, what implies that regardless of the UV completion the Wilson coefficients will be positive if evaluated deep in the IR. Note that this conclusion is independent of the matter content, as all particle species contribute to $K$ with the same sign. Similar arguments have recently been used to argue that, due to quantum corrections, the mild form of the weak gravity conjecture is asymptotically satisfied in a large class of non-supersymmetric low-energy EFTs containing a photon and a graviton~\cite{Cheung:2014ega,Bellazzini:2019xts,Charles:2019qqt,Jones:2019nev}.

More generally, the negativity of the anomalous dimensions is interesting when contrasted with the
positivity bounds on Wilson coefficients obtained from dispersion relations, in particular from the analyticity properties of elastic $2\rightarrow 2$ scattering amplitudes, see e.g.~\cite{Adams:2006sv}. In the presence of gravity it is yet unclear how robust these bounds are, as 
the $t$-channel massless graviton pole in tree-level amplitudes (or in general the presence of singular terms in the $t \to 0$ limit) 
prevents one from taking the forward limit of the elastic scattering amplitude. 
For this reason, it is important to check if the positivity of Wilson coefficients due to gravitational quantum effects aligns with the consistency conditions which are derived neglecting the $t$-channel graviton pole. For the amplitudes in Eqs.~\eqref{eq:scalarDim8} to~\eqref{eq:vectorDim8}, these positivity bounds take the form
\begin{equation} \label{eq:posBounds}
\begin{split}
C_{\phi\phi'}^{(2)} > 0\,,\quad &2\,C_{\phi\phi'}^{(2)} +3\, C_{\phi\phi'}^{(0)} > 0\,,\qquad C_{\phi\phi'}^{(2)} + C_{\phi\phi'}^{(1)} > 0\,, \quad C_{\psi\phi}^{(2)} > 0\,,\\ 
&C_{\psi\psi'}^{(2)} > 0\,,\quad C_{\psi\psi'}^{(2)} + C_{\psi\psi'}^{(1)} > 0\,,\quad C_{V\Phi}^{(2)} > 0\,,
\end{split}
\end{equation}
which have been obtained by requiring that the $s^2$-term 
of the elastic forward amplitude $\amp (ab\rightarrow ab)|_{t\rightarrow 0}$ is positive
for $a,b= \phi,\phi',\psi,\psi',V$ and their antiparticles.\footnote{Note that in order to obtain all the bounds for the scalar Wilson coefficients one has to scatter states which are superpositions of $\phi,\bar{\phi},\phi',\bar{\phi}'$ (see e.g. \cite{Remmen:2019cyz}).} 
\eq{eq:posBounds} coincides with the bounds 
found in \cite{Remmen:2019cyz} after converting to their operator basis.

It therefore appears that all positivity bounds in~\eq{eq:posBounds} are indeed not spoiled 
at low energy scales, on the contrary 
the gravitational RG running 
makes the Wilson coefficients more 
positive.
In this regard, we can understand why some of the cuts in Fig.~\ref{fig:dim8} are negative from the fact that they can be related to a cross section. Specifically, the $s$-channel cuts in the first row of Fig.~\ref{fig:dim8}, for $1$ and $2$ of the same helicity, can be simply derived from \eq{gamma} after taking its forward limit. For instance, \eq{eq:scut1} leads to
\be \label{eq:scutsigma}
(\text{$s$-cut})_{\Phi''}|_{t\rightarrow 0} = - \frac{2 s}{\pi} \sigma(\Phi \bar{\Phi} \to \Phi'' \bar{\Phi}'')\,,
\ee
and similarly for the graviton cut \eq{eq:scuth}.
This is regardless of $\Phi$ and $\Phi'$ being distinguishable species, i.e.~of different flavor/color (yet same helicity), since they couple to gravity in the same way.
Despite the simplicity of this argument, let us stress that it does not hold for the rest of the cuts in Fig.~\ref{fig:dim8}.
In light of this, it would be interesting to gain further understanding on the negativity of the anomalous dimensions in Eqs.~(\ref{eq:anomDim8Scalar}--\ref{eq:anomDim8Vector}).

\subsection{RG mixing of operators including gravitons}
\label{mixinggrav}
Now we would like to 
show how our methods allow to efficiently study the running of higher-dimensional amplitudes that include gravitons,
which have been presented in Section~\ref{beyondmin}. The relevant question here is the following: given a ``spectrum'' of higher-dimensional amplitudes (including gravitons), what are the leading contributions to the RG flow of their Wilson coefficients?

Our analysis starts from \eq{wcut}. As one can see, the powers of $\Lambda$ and $\Mp$ in general do not have to
match separately in the left and right hand sides of \eq{gamma}. This is what we saw in the previous sections, with
minimally coupled amplitudes --~suppressed by powers of $\Mp$~-- contributing to the running of $\Lambda^{-2}$ and
$\Lambda^{-4}$ amplitudes. However, when we focus on the leading contributions to the running of a given higher-dimensional
amplitude, assuming a large separation between the EFT and Planck scale, i.e.~$\Lambda\ll\Mp$, we should only keep
the contributions on the RHS of \eq{gamma} with the smallest $k_L+k_R$, that is those which are less $\Mp$ suppressed.

In order to characterize these leading contributions, we first observe that all tree-level 
amplitudes carry \emph{at least} one power of $\Mp^{-1}$ for each external graviton,\footnote{The only exception are amplitudes
with just gravitons, like $\amp(1_{h_-},2_{h_-},3_{h_+},4_{h_+})\propto \Mp^{-2}$. However our conclusions do not change after we include
these particular cases.}
while, as we pointed out in Section~\ref{beyondmin}, higher-dimensional amplitudes 
come with \emph{exactly} as many powers of $\Mp^{-1}$ as external gravitons.
Take then $\amp_i$ in the LHS of \eq{gamma} with a given $k_i$, i.e.~with exactly $k_i$ external gravitons.
The legs of $\amp_L$ and $\amp_R$ are either external
or inside the cut. The external ones must overall match those of $\amp_i$, and in particular there will be exactly $k_i$ gravitons.
The cut legs can be gravitons or $|h|\leq 1$ particles. If some of them are gravitons, we will have necessarily
$k_L+k_R>k_i$. On the contrary, if all the cut particles have $|h|\leq 1$, then it is possible to satisfy $k_L+k_R=k_i$,
which is also the minimum possible $k_L+k_R$.\footnote{It is possible but not automatic, as $\amp_L$ and $\amp_R$ could also carry additional powers of $\Mp^{-1}$,
not due to external gravitons but instead to internal ones, like for example in \eq{J2phi}. This is discussed in Appendix~\ref{opMix}.}
In summary, the dominant contributions to the renormalization of $\amp_i$ in the limit
$\Lambda\ll\Mp$ will be those \emph{without cut gravitons}, in which case the following equality can be satisfied
\be\label{wrule}
w_i=w_L+w_R\,.
\ee
As an example, 
we provide now all the anomalous dimensions involving 3- and 4-point higher-dimensional amplitudes
that include at least one graviton, at order $\Lambda^{-2}$ (see Fig.~\ref{fig:helOpPlot}). 
At leading order in $\Mp$, 
following \eq{wrule} we could have
either $w_L=w_R=1$, which turns out to be empty,
or $w_L=2,~w_R=0$, on which we focus from now on. This corresponds to
mixings among amplitudes in Fig.~\ref{fig:helOpPlot}.

In this setup, we can state very simple renormalization rules
that 
reduce 
the computational
work to the minimum. First, we have \eq{htildemixing}, which is the natural extension of the non-renormalization
theorems of \cite{Cheung:2015aba} to our enlarged playground. An amplitude $\amp_j$ can renormalize $\amp_i$
only if the latter is in the 
left ``light cone'' of the former. Second, since gravitons can only be external, their number $n^{(h)}$ can only
increase, meaning that
\be
n^{(h)}_i\geq n^{(h)}_j\,.
\ee
This leaves us with only a few renormalizations to compute: $F^3\to CF^2$, $CF^2\to CF^2$, $CF^2\to C^2\phi^2$
and $C^2\phi^2\to C^2\phi^2$. The self-renormalization $C^3\to C^3$ is zero because some gravitons are necessarily in the cut.

\vspace{.2cm}

To specify the problem, we need to fix the field content that enters the above mixings. We take from now on a simple non-abelian gauge group $G$ with vector bosons $V^a$, and a complex scalar multiplet $\phi$ that transforms irreducibly under it, and comment on generalizations later on.

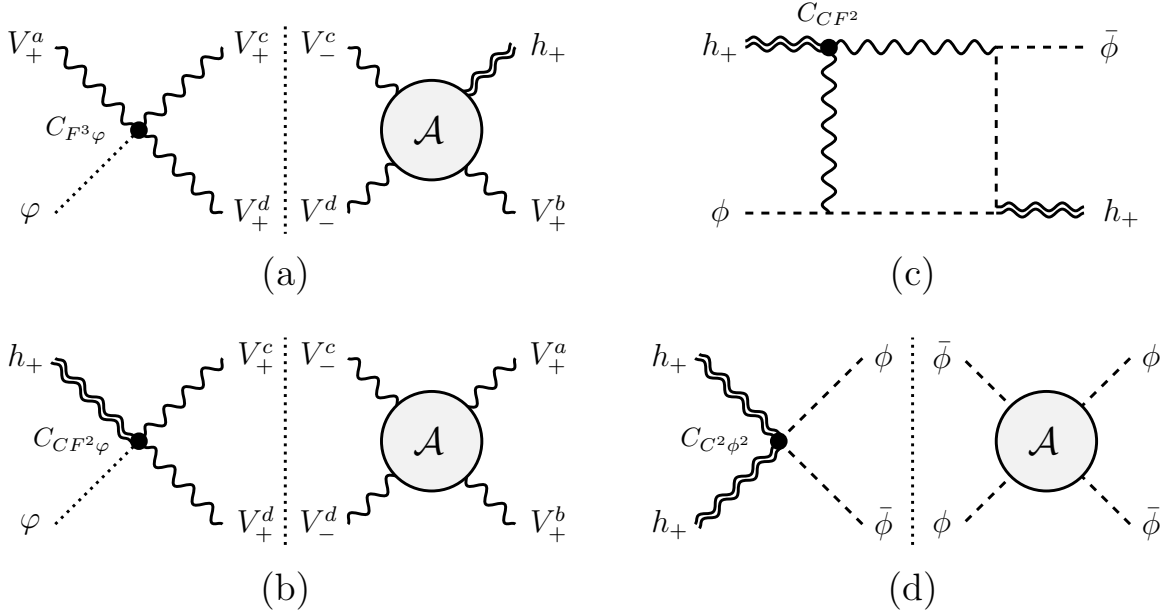
\begin{figure}[t]
\centering
\begin{tikzpicture}[line width=1.1 pt, scale=.55, baseline=(current bounding box.center)]
	
	\begin{scope}[shift={(-1,0)}]
	\begin{scope}[shift={(-.5,0)}]
	\draw[vector] (-12,5) -- (-10,3) ;
	\draw[vector] (-8,5) -- (-10,3) ;
	\draw[dotted] (-12,1) -- (-10,3) ;
	\draw[vector] (-8,1) -- (-10,3) ;
	\filldraw [black] (-10,3) circle (5pt) ;
	\node at (-12.7,5.) {${V_+^a}$};
	\node at (-12.6,1.) {$\varphi$};
	\node at (-11.5,3) {\footnotesize $C_{F^3\varphi}$};
	\end{scope}
	
	\draw[dotted] (-7,5.5) -- (-7,.5) ;
	\node at (-7,-0.5) {\large (a)};
	
	\begin{scope}[shift={(.5,0)}]
	\draw[vector] (-6,5) -- (-4,3) ;
	\draw[graviton] (-2,5) -- (-4,3) ;
	\draw[graviton] (-2.1,5.1) -- (-4.1,3.1) ;
	\draw[vector] (-6,1) -- (-4,3) ;
	\draw[vector] (-2,1) -- (-4,3) ;
	\draw[fill=gray!10] (-4,3) circle (1.2cm) ;
	\node at (-1.2,5.) {${h_+}$};
	\node at (-1.2,1.) {${V_+^b}$};
	\node at (-4,3) {\large $\amp$\,};
	\end{scope}

	\node at (-6.2,5) {${V_-^c}$};
	\node at (-6.2,1.) {${V_-^d}$};
	\node at (-7.8,5) {${V_+^c}$};
	\node at (-7.8,1.) {${V_+^d}$};
	\end{scope}


	\begin{scope}[shift={(0.0,-1.5)}]
	
	\begin{scope}[shift={(-.5,0)}]
	\draw[dotted] (-13,-5) -- (-11,-3) ;
	\draw[vector] (-9,-5) -- (-11,-3) ;
	\draw[graviton] (-13,-1) -- (-11,-3) ;
	\draw[graviton] (-13.1,-1.1) -- (-11.1,-3.1) ;
	\draw[vector] (-9,-1) -- (-11,-3) ;
	\filldraw [black] (-11,-3) circle (5pt) ;
	\node at (-13.7,-1.) {${h_+}$};
	\node at (-13.6,-5.) {$\varphi$};
	\node at (-12.6,-3) {\footnotesize $C_{CF^2\varphi}$};
	\end{scope}
	
	\draw[dotted] (-8,-5.5) -- (-8,-.5) ;
	
	\begin{scope}[shift={(.5,0)}]
	\draw[vector] (-7,-5) -- (-5,-3) ;
	\draw[vector] (-3,-5) -- (-5,-3) ;
	\draw[vector] (-7,-1) -- (-5,-3) ;
	\draw[vector] (-3,-1) -- (-5,-3) ;
	\draw[fill=gray!10] (-5,-3) circle (1.2cm) ;
	\node at (-5,-3) {\large $\amp$\,};
	\node at (-2.2,-1.) {${V_+^a}$};
	\node at (-2.2,-5.) {${V_+^b}$};
	\end{scope}
	
	\node at (-8,-6.6) {\large (b)};
	\node at (-7.2,-1) {${V_-^c}$};
	\node at (-7.2,-5) {${V_-^d}$};
	\node at (-8.7,-1) {${V_+^c}$};
	\node at (-8.7,-5) {${V_+^d}$};
	\end{scope}
	
	
	\begin{scope}[shift={(0,-1.5)}]
	
	\begin{scope}[shift={(.2,0)}]
	\draw[dashed] (12,-5) -- (10,-3) -- (8,-5) ;
	\draw[dashed] (12,-1) -- (10,-3) -- (8,-1) ;
	\draw[fill=gray!10] (10,-3) circle (1.2cm) ;
	\node at (10,-3) {\large $\amp$\,};
	\node at (12.5,-5) {$\bar{\phi}$};
	\node at (12.5,-1) {$\phi$};
	\node at (7.5,-5) {$\phi$};
	\node at (7.5,-1) {$\bar{\phi}$};
	\end{scope}
	
	\draw[dotted] (7,-5.5) -- (7,-.5) ;
	\node at (7,-6.6) {\large (d)};
	
	\begin{scope}[shift={(-.2,0)}]
	\draw[dashed] (6,-5) -- (4,-3) ;
	\draw[graviton] (2.1,-5.1) -- (4.1,-3.1) ;
	\draw[graviton] (2,-5) -- (4,-3) ;
	\draw[dashed] (6,-1) -- (4,-3) ;
	\draw[graviton] (2,-1) -- (4,-3) ;
	\draw[graviton] (2.1,-0.9) -- (4.1,-2.9) ;
	\filldraw [black] (4,-3) circle (5pt) ;
	\node at (2.5,-3) {\footnotesize $C_{C^2\phi^2}$};
	\node at (6.5,-1) {${\phi}$};
	\node at (6.5,-5) {$\bar{\phi}$};
	\node at (1.4,-1) {$h_+$};
	\node at (1.4,-5) {$h_+$};
	\end{scope}
	
	\end{scope}
	
	
	\draw[graviton] (3,5) -- (5,5) ;
	\draw[graviton] (3,5.15) -- (5,5.15) ;
	\draw[vector] (5,1) -- (5,5) ;
	\draw[vector] (9,5) -- (5,5) ;
	\filldraw [black] (5,4.99) circle (5pt) ;
	\draw[dashed] (3,1) -- (9,1) -- (9,5) -- (11.2,5) ;
	\draw[graviton] (11.1,1) -- (9,1) ;
	\draw[graviton] (11.1,1.15) -- (9,1.15) ;
	\node at (5,5.8) {\footnotesize $C_{CF^2}$};
	\node at (7,-0.5) {\large (c)};
	\node at (2.4,1) {$\phi$};
	\node at (12,1) {$h_+$};
	\node at (2.4,5) {$h_+$};
	\node at (11.7,5) {$\bar{\phi}$};
		
 \end{tikzpicture}
 \caption{\emph{Diagrams relevant for mixings among operators containing at least one graviton at ${\cal O}(\Lambda^{-2})$.}}
 \label{fig:dim6mix}
 \end{figure}

The relevant amplitudes are given in \eq{CF2}, augmented with color indices 
$2_{V_+}\to 2_{V^a_+}$, $3_{V_+}\to 3_{V^b_+}$ and $C_{CF^2} \to C_{CF^2}\delta^{ab}$, and \eq{C2phi2}. We need as well the amplitude
\be
\amp_{F^3}(1_{V^a_+},2_{V^b_+},3_{V^c_+})=C_{F^3}if^{abc}\frac{[12][23][13]}{\Lambda^2}\,,
\ee
where $f^{abc}$ are the structure constants of $G$.
In all cases, there will be only one relevant cut up to crossing, making the extraction of the anomalous dimensions
quite simple (see Fig.~\ref{fig:dim6mix}).

We first consider the $3\to 3$ renormalizations, which involve gauge bosons and gravitons. The relevant diagrams are
(a) and (b) of Fig.~\ref{fig:dim6mix}. 
To lift the degeneracies associated to the 3-point kinematics, we use a trick
that consists in studying the renormalization of the following $\Lambda^{-3}$ amplitudes
\begin{align}
\amp_{F^3\varphi}(1_{V^a_+},2_{V^b_+},3_{V^c_+},4_\varphi)&=C_{F^3\varphi}if^{abc}\frac{[12][23][13]}{\Lambda^3}\,,\\
\amp_{CF^2\varphi}(1_{h_+},2_{V^a_+},3_{V^b_+},4_\varphi)&=C_{CF^2\varphi}\delta^{ab}\frac{[12]^2[13]^2}{\Lambda^3\,\Mp}\,,
\end{align}
with an additional scalar $\varphi$ that has no tree-level couplings other than the two above. Thanks to this last property,
it behaves as a spectator in the renormalizations among $\amp_{F^3\varphi}$ and $\amp_{CF^2\varphi}$, implying in particular that
$\gamma_{CF^2}=\gamma_{CF^2\varphi}$.

We start with (a) and the $V^a\leftrightarrow V^b$ crossed topology, which have no IR divergences since they describe non-diagonal mixing. The contribution from the cut (a) shown in Fig.~\ref{fig:dim6mix} to the anomalous dimension of $C_{CF^2\varphi}$ is given by
\begin{align}
\left.\gamma_{CF^2\varphi}\right|_{\rm (a)}\,&\frac{[14]^2[34]^2}{\Lambda^3\,\Mp}\delta^{ab}=-\frac{1}{4\pi^3}\frac{1}{2}\int d{\rm LIPS}_{\ell\ell'}~\amp_{F^3\varphi}(1_{V^a_+},2_\varphi,-\ell_{V^c_+},-\ell'_{V^d_+}){\amp}(\ell_{V^c_-},\ell'_{V^d_-},3_{V^b_+},4_{h_+})\nonumber \\
&=-\frac{1}{8\pi^3}\int d{\rm LIPS}_{\ell\ell'}~iC_{F^3\varphi}f^{acd}\frac{[1\ell][1\ell'][\ell\ell']}{\Lambda^3}\times \frac{ig}{\Mp}f^{cdb}\frac{\la \ell\ell'\ra^3}{\la\ell 3\ra\la \ell' 3\ra}\frac{[43]\la 3\ell \ra[\ell 4]}{\la 43\ra[3\ell]\la\ell 4 \ra}\nonumber \\
&=\frac{g\, C_{{F^3\varphi}}}{16\pi^2}\frac{[14]^2[34]^2}{\Lambda^3\,\Mp}f^{acd}f^{cdb}=C_{{F^3\varphi}}\,\frac{g\,{\cal C}_A}{16\pi^2}\,\frac{[14]^2[34]^2}{\Lambda^3\,\Mp}\delta^{ab}\,,
\end{align}
where the factor of $1/2$ in the first line accounts for the indistinguishable
particles in the cut, and we made use of \eq{fig:dim6mix}. We also used $f^{acd}f^{cdb}={\cal C}_A\delta^{ab}$, ${\cal C}_A$ being the Casimir of the adjoint. The $t$-channel contribution can be obtained by replacing $1\leftrightarrow 3$, what gives the same result. The full anomalous dimension from the (a) topology is given by the sum of both contributions.

In a similar fashion, but now paying attention to IR divergences as it is a self-renormalization, we get for the contribution to $\gamma_{CF^2\varphi}$ from diagram (b) of Fig.~\ref{fig:dim6mix}:
\begin{align}
&\left(\left.\gamma_{CF^2\varphi}\right|_{\rm (b)} -\gamma_{\rm coll}\, C_{CF^2\varphi}\right)\,\frac{[13]^2[14]^2}{\Lambda^3\,\Mp}\delta^{ab}\nonumber\\
&=-\frac{1}{4\pi^3}\frac{1}{2}{\cal R}\!\!\int d{\rm LIPS}_{\ell\ell'}~\amp_{CF^2\varphi}(1_{h_+},2_\varphi,-\ell_{V^c_+},-\ell'_{V^d_+}){\amp_{\rm tree}}(\ell_{V^c_-},\ell'_{V^d_-},3_{V^a_+},4_{V^b_+})\nonumber \\
&=\frac{1}{8\pi^3}{\cal R}\!\!\int d{\rm LIPS}_{\ell\ell'}~C_{CF^2\varphi}\,\frac{[1\ell]^2[1\ell']^2}{\Lambda^3\,\Mp}\times g^2 {\cal C}_A\frac{[34]^4}{[\ell 3][3 \ell'][\ell' 4][4\ell]}\delta^{ab}\nonumber \\
&=C_{{CF^2\varphi}}\,\frac{3g^2{\cal C}_A}{8\pi^2}\,\frac{[13]^2[14]^2}{\Lambda^3\,\Mp}\delta^{ab}\,,
\end{align}
where we 
directly used the contracted expression
\be
\delta^{cd}{\amp_{\rm tree}}(1_{V^c_-},2_{V^d_-},3_{V^a_+},4_{V^b_+})=g^2 {\cal C}_A\frac{[34]^4}{[13][32][24][41]}\delta^{ab}
\ee
in the second line.
The amplitude $\amp_{CF^2\varphi}$ also receives divergent contributions from scaleless bubbles on the external legs, these divergences
being accounted for
by the collinear term in the above equation. In this particular case, $\gamma_{\rm coll}=2\gamma^{(V)}$, where $\gamma^{(V)}=(3-22 \,{\cal C}_A)g^2/192\pi^2$. Putting everything together (and getting rid of the spectator particle $\varphi$), we finally get
\be
\boxed{~\gamma_{CF^2}=\frac{g\,{\cal C}_A}{8\pi^2}C_{F^3}+\frac{(3+14\,{\cal C}_A)g^2}{96\pi^2}C_{CF^2}\,.~}
\ee
We have assumed a simple gauge group $G$ for the above derivation. In general, the gauge group will be $\prod_i G_i \times \prod_j U(1)_j$. 
Bose symmetry and gauge invariance dictate that $\amp_{F^3}$ can either involve vectors in the same
non-abelian factor $G_i$, or vectors in three distinct $U(1)$ factors. In the latter case $C_{CF^2}$ is however not renormalized
by $C_{F^3}$, since $\amp(V_-,V_-, V_+,h_+)$ is non-zero only when all three vectors belong to the same $G_i$. Similarly,
$\amp_{CF^2}$ has either two vectors in the same $G_i$ or two vectors belonging to two $U(1)$ factors, not necessarily distinct.
Nevertheless, also in this case only those $C_{CF^2}$ with vectors in the same $G_i$ get a non-zero anomalous dimension,
because $\amp(V_-,V_-, V_+,V_+)$ is non-zero only when all four vectors belong to the same $G_i$.
This means that our results are exhaustive and can be easily promoted to a generic gauge theory.

\vspace{.2cm}

Let us now move to the renormalization $C_{CF^2}\to C_{C^2\phi^2}$, which is a 3$\,\to\,$4 process that involves the box
depicted in Fig.~\ref{fig:dim6mix} (c). Due to the flavor preserving nature of gravity and gauge interactions, one can see that
the external scalars
must be particle-antiparticle pairs, i.e.~$\phi$ and $\bar{\phi}$.

The interaction between $\phi$ and gauge bosons is fixed by $\amp(1_{V_+^a},2_{\bar{\phi}},3_{\phi})=
g t^a[12][31]/[23]$. The diagrams which are relevant for $C_{CF^2}\to C_{C^2\phi^2}$ are then
all proportional to $g^2 t^at^a=g^2 {\cal C}_\phi$, where ${\cal C}_\phi$ is the Casimir of the $G$-representation
that is carried by $\phi$. Having fixed the group theory structure,
the remaining task --~which is the non trivial part~-- is to determine the overall coefficient, that depends
just on the kinematic structure. Thanks to \eq{gamma}, we can extract the contribution to $\gamma_{C^2\phi^2}$ coming from $C_{CF^2}$ by evaluating
\begin{align}
&\gamma_{C^2\phi^2} \frac{[14]^4}{\Lambda^2 \Mp^2}=-\frac{1}{4\pi^3}\int d{\rm LIPS}_{\ell\ell'}~\widehat{\amp}_{CF^2}(1_{h_+},2_\phi,-\ell_{V^a_+},-\ell'_{\bar{\phi}}){\amp}(\ell_{V^a_-},\ell'_{\phi},3_{\bar{\phi}},4_{h_+})\nonumber \\&-\frac{1}{4\pi^3}\int d{\rm LIPS}_{\ell\ell'}~\widehat{\amp}_{CF^2}(1_{h_+},3_{\bar{\phi}},-\ell_{V^a_+},-\ell'_{{\phi}}){\amp}(\ell_{V^a_-},\ell'_{\bar{\phi}},2_{{\phi}},4_{h_+})=\frac{g^2 {\cal C}_\phi}{8\pi^2}   \frac{[14]^4}{\Lambda^2 \Mp^2} C_{CF^2}\,,
\end{align}
the two phase space integrals being the same up to crossing. The necessary amplitudes are provided by \eq{1graviton} and by the following expression
\be
\widehat{\amp}_{CF^2}(1_{h_+},2_{V^a_+},3_\phi,4_{\bar{\phi}})=\frac{g \,t^a \,C_{CF^2}}{\Lambda^2\Mp}\frac{[12]^2[13][14]\la 34\ra}{s}\,,
\ee
which is completely fixed by factorization into \eq{CF2} and $\amp({V_-^a},{\phi},{\bar{\phi}})$, whose complex conjugated amplitude was defined above.

We finally get to the self-renormalization of $C_{C^2\phi^2}$, whose relevant cut is depicted in Fig.~\ref{fig:dim6mix} (d), and gives
\begin{align}
&\gamma_{C^2\phi^2}|_{\rm (d)} -\gamma_{\rm coll}\, C_{C^2\phi^2}
=\frac{-\Lambda^2\Mp^2}{4\pi^3[12]^4\delta_j^i}\,\,{\cal R}\!\int \!d{\rm LIPS}_{\ell\ell'}\,\amp_{C^2\phi^2}(1_{h_+},2_{h_+},-\ell_{\phi^l},-\ell'_{\bar{\phi}_k}){\amp_{\rm tree}}(\ell_{\bar{\phi}_l},\ell'_{\phi^k},3_{\phi^i},4_{\bar{\phi}_j})\nonumber \\
&=-\frac{C_{C^2\phi^2}}{4\pi^3\delta_j^i}\,{\cal R}\!\int \!d{\rm LIPS}_{\ell\ell'}\,\delta^l_k \amp_{\rm tree}(\ell_{\bar{\phi}_l},\ell'_{\phi^k},3_{\phi^i},4_{\bar{\phi}_j})=\frac{C_{C^2\phi^2}}{4\pi^3\delta_j^i}\,{\cal R}\!\int \!d{\rm LIPS}_{\ell\ell'}\,\delta^l_k 
\nonumber \\
&\times\left[ \lambda(\delta_l^k\delta_j^i+\delta_j^k\delta_l^i) -g^2(t^a)_l^k(t^a)_j^i\frac{u-t}{2s}-g^2(t^a)_j^k(t^a)_l^i\frac{u-s}{2t} \right]=\frac{2\lambda(N_\phi+1)+g^2{\cal C}_{\phi}}{16\pi^2}\,C_{C^2\phi^2}\,.
\end{align}
Note that $\amp_{\rm tree}$ is parametrized slightly differently than in \eq{tree3}, according to the simplified assumptions on the group-theory structure of $\phi$ that we specified before. When we put everything together, we find $\gamma_{C^2\phi^2}$ to be
\be
\boxed{\gamma_{C^2\phi^2}=\frac{g^2{\cal C}_\phi}{8\pi^2}  \,C_{CF^2} +\left( \frac{2\lambda(N_\phi+1)-3g^2{\cal C}_{\phi}}{16\pi^2}+2\gamma^{(\phi)}|_{y^2}\right)C_{C^2\phi^2}\,,}
\ee
where we used $\gamma_{\rm coll}=2\gamma^{(\phi)}$, with $\gamma^{(\phi)}=\gamma^{(\phi)}|_{y^2}-g^2{\cal C}_\phi/8\pi^2$, the $|_{y^2}$ term arising from Yukawa interactions that we do not specify.

The above result can be generalized to more complicated group structures by making the replacement $g^2 {\cal C}_\phi \rightarrow \sum_i g_i^2\, {\cal C}_\phi^{i}$. Let us for example take $\phi$
to be the SM Higgs $H$ and $G=G_{\rm SM}$. One should be careful in defining the coupling and generator normalizations. We take generators $t^a$ with $\mathrm{tr}(t^at^b)=\delta^{ab}/2$, and the gauge coupling $g_i$ to enter the covariant derivative as $D=\partial-igt^aA^a/\sqrt{2}$, as is standard with amplitude methods. Finally, the quartic coupling is fixed by ${\cal L}=-\lambda_H |H|^2/2$. We then get
\begin{align}
\gamma_{C^2H^2}&=\sum_i \frac{g_i^2{\cal C}_H^i}{8\pi^2}\left( C_{CF_i^2}-\frac{3}{2}C_{C^2H^2}\right)+\frac{\lambda_H(N_H+1)}{8\pi^2}C_{C^2H^2}+\sum_f \frac{y_f^2}{8\pi^2}C_{C^2H^2}\nonumber \\
&\simeq \frac{1}{32\pi^2}\left(g_1^2C_{CB^2} +3g_2^2\, C_{CW^2}\right)+\frac{1}{8\pi^2}\left( y_t^2+3\lambda_H-\frac{3}{8}g_1^2-\frac{9}{8}g_2^2 \right)C_{C^2H^2}
\end{align}
where we used that the Casimir of the $SU(2)$ fundamental is 3/4 and that the generalization
of ${\cal C}_{\phi}$ to an abelian factor is simply the charge squared, and in this case $Y_H^2=1/4$. We also used $N_H=2$, and $\gamma^{(H)}|_{y^2}=\sum_f {y_f^2}/16\pi^2\simeq y_t^2/16\pi^2$ to very good approximation.

As a final comment, we would like to stress that, even though the results in this section do not cover all the possible flavor/color structures that a generic model can have, we nevertheless covered all possible \emph{helicity} (or kinematic) structures that are allowed at this order. In a more general situation, one should first `strip off' the index structure of the relevant amplitudes, then recycle our integrals and finally plug back in the proper tensors that carry the flavor indices.

\subsection{Comparison with the literature}
\label{previous}
Some of the results presented in Section~\ref{Ord4} are not new. 
The one-loop UV divergences of (only) scalars~\cite{tHooft:1974toh}, fermions~\cite{Deser:1974cy,Barvinsky:1981rw} and vectors~\cite{Deser:1974cz,Deser:1974xq} minimally coupled to gravity were systematically studied in the 1970s 
employing background-field methods and later revisited from an amplitude point of view in~\cite{Dunbar:1995ed,Norridge:1996he}. Our results generalize these findings to an arbitrary particle content consisting of $N_\phi$ complex scalars, $N_\psi$ Weyl fermions and $N_V$ vectors, and in particular include amplitudes with multiple particle species. Here we shortly demonstrate that these classic results are contained as special cases in Eqs.~\eqref{eq:scalarDim8} to~\eqref{eq:vectorDim8}.

The only one-loop divergent non-factorizable amplitude in a theory of one massless real scalar minimally coupled to gravity, i.e.~$N_\phi = 1/2$ and $N_\psi = N_V =0$, is ${\cal A} (1_\phi , 2_\phi, 3_\phi, 4_\phi)$~\cite{tHooft:1974toh}. The corresponding UV divergence can be extracted from Eq.~\eqref{eq:scalarDim8}  and Eq.~\eqref{eq:anomDim8Scalar} after adding the crossed channels $p_2 \leftrightarrow p_3$ and $p_2 \leftrightarrow p_4$ to account for $\phi' = \phi$ and $\bar{\phi}=\phi$
%
\be \label{eq:realScalarDim8}
\amp_{\rm UV}(1_{\phi},2_{\phi},3_{\phi},4_{\phi})=\frac{1}{16\pi^2\Mp^4\epsilon}\frac{203}{40}\left( s^2 + t^2 + u^2 \right)\,.
\ee
This is exactly canceled by the amplitude induced by the counterterm in~\cite{tHooft:1974toh} after applying the equations of motion and also agrees with the result in~\cite{Dunbar:1995ed}.\footnote{Note that~\cite{tHooft:1974toh,Dunbar:1995ed} work in units where $\Mp=\sqrt{2}$ and use the abbreviation $1/\varepsilon \equiv 1/(8\pi^2 (d-4)) = -1/(16\pi^2\epsilon)$ for $d=4-2\epsilon$. After reinstating $\Mp$ by dimensional analysis the results in~\cite{tHooft:1974toh,Dunbar:1995ed} agree with Eq.~\eqref{eq:realScalarDim8}.} If the spectrum is extended to $N>1$ distinguishable real scalars the numerical prefactor in Eq.~\eqref{eq:realScalarDim8} is changed as 
$203/40\rightarrow (202+N)/40$ and there are further divergences in amplitudes of the form $\amp(1_{\phi},2_{\phi},3_{\phi'},4_{\phi'})$, which are directly given by Eq.~\eqref{eq:scalarDim8} with $N_\phi =N/2,\, N_\psi = N_V =0$. We have checked that this is consistent with the corresponding counterterm Lagrangian in~\cite{Dunbar:1995ed}.

Minimally coupling a Yang-Mills theory of a simple gauge group to gravity generates UV divergences at one loop in amplitudes of the form $\amp (1_{V_-^a}, 2_{V_+^b}, 3_{V_+^c}, 4_{V_-^d})$. The UV divergence can be obtained from Eq.~\eqref{eq:vectorDim8} and Eq.~\eqref{eq:anomDim8Vector}  by dressing the amplitude with flavor indices and adding the $p_2\leftrightarrow p_3$ crossed channel, which for $N_\phi = N_\psi= 0$ takes the form
\be
\begin{split}
\amp_{\rm UV} (1_{V_-^a}, 2_{V_+^b}, 3_{V_+^c}, 4_{V_-^d}) 
=\frac{1}{16\pi^2\Mp^4\epsilon} \frac{137 + 6 (N_V -1)}{30} \left( \delta^{ab} \delta^{cd} + \delta^{ac} \delta^{bd}\right) \la 1 4\ra^2 [23]^2\,.
\end{split}
\ee
This is consistent with the counterterm for Einstein-Yang-Mills theory in~\cite{Deser:1974xq}, and for $a=b=c=d$ and $N_V=1$ it reduces to the one for Einstein-Maxwell theory as detailed in~\cite{Deser:1974cz}. In this limit the UV-divergent part of the amplitude also agrees with~\cite{Norridge:1996he}.

Last but not least we consider a single Weyl fermion minimally coupled to gravity, i.e.~$N_\psi =1,\, N_\phi = N_V =0$. In this case the one-loop UV divergence in the four-fermion amplitude is extracted from the sum of Eq.~\eqref{eq:fermionDim8} and the $p_2 \leftrightarrow p_3$ crossed amplitude, which comes with a minus sign due to Fermi statistics, and the anomalous dimensions in Eq.~\eqref{eq:anomDim8Fermion} . The resulting divergent part of the amplitude takes the form
\be \label{eq:SingleFermionDim8}
\amp_{\rm UV}(1_{\bar\psi},2_{{\psi}},3_{{\psi}},4_{\bar\psi})=\frac{1}{16\pi^2\Mp^4\epsilon}\frac{65}{8} u\, \la 1 4\ra [2 3]\,.
\ee
The corresponding counterterm was derived in~\cite{Barvinsky:1981rw} using functional methods and is consistent with the divergence in Eq.~\eqref{eq:SingleFermionDim8}. Note that in~\cite{Norridge:1996he} an independent computation applying amplitude methods was performed and arrived at a different result with a prefactor of $59/8$ instead of the $65/8$ in Eq.~\eqref{eq:SingleFermionDim8}. Since our computation, which also uses amplitude methods, exactly agrees with the completely unrelated functional approach in~\cite{Barvinsky:1981rw}, we are confident that the result in Eq.~\eqref{eq:SingleFermionDim8} is correct.

Let us stress again that our findings in Section~\ref{Ord4} do not only reproduce these classic results, but also hold in more general setups with an arbitrary number of scalars, fermions and vectors simultaneously coupled to gravity. In particular, also amplitudes among different particle species are UV-divergent in such a general setup. Eqs.~\eqref{eq:scalarDim8} to~\eqref{eq:vectorDim8} represent the exhaustive set of one-loop UV divergences in 4-point amplitudes at $\mathcal{O}(\Mp^{-4})$ of theories with $N_\phi$ complex scalars, $N_\psi$ Weyl fermions and $N_V$ real vectors minimally coupled to gravity.
Besides, the results presented in Sections \ref{Ord2} and \ref{mixinggrav} are completely new to our knowledge.

\section{Conclusions}\label{conclusions}
%

The analysis presented here 
has shown how on-shell methods provide extremely efficient tools for studying loop effects within theories involving gravitons. 
We 
systematically explored the RG structure of minimally coupled theories, computing all corrections 
arising at 4-points 
and at any order in inverse powers of $\Mp$. 
In particular, we considered the leading effects, arising at order 
$\Mp^{-2}$, where previous literature mostly focused on theories which are free in the limit $\Mp\to\infty$, in which case the first effects arise at order $\Mp^{-4}$.
We also considered the renormalization of higher-dimensional operators involving gravitons in generic EFTs, in particular the RG of all such dimension-six operators. 
All of these results are completely new, in contrast to previous on-shell analyses of the SM EFT at one loop.

To organize our computations, we found the `modified helicity' $\tilde{h}$, whose use and definition we introduced in this work, to be an invaluable tool. This quantity has a number of important properties: (1) it reduces to the standard total helicity $h$ in the absence of gravitons, (2) it is \emph{zero} for all tree-level 4-point amplitudes of any minimally coupled marginal theory (except for the notorious $\psi^4$ `exceptional' amplitude) and (3) it provides, together with the usual $\Lambda^{-1}$ power counting, the most convenient ``coordinates'' in the space of higher-dimensional amplitudes (this is illustrated in Fig.~\ref{fig:helOpPlot}). These coordinates are in fact the most transparent in order to state the new non-renormalization results that we provided here.
In addition, we pointed out the similarity between the modified helicity and the KLT relations, a connection that would be interesting to explore further.

With respect to the ordinary Feynman-Dyson approach, our method has several advantages. Instead of dealing with hundreds of Feynman diagrams, our extraction of one-loop anomalous dimensions always reduces to the computation of a handful of cuts (often just one). These are nothing but phase space integrals of a product of tree-level amplitudes, that can be reduced to simple $(\theta,\phi$) angle integrals which can be easily automatized or even computed with pen and paper in half a page in many 
cases.

The methods we employed here are extremely versatile and can be extended to more complicated topologies (i.e.~more legs or more loops), to higher orders in the $\Lambda$ and $\Mp$ expansion, and to more general theories, like supergravity 
(the method is in principle exactly the same, with the gravitinos carrying modified helicity $\tilde{h}_{\zeta_\pm}=\pm1/2$).

Our results can be considered as a step towards a better quantitative knowledge of UV physics from a purely IR perspective. 
In this regard, we also showed the non-trivial compatibility between the IR gravitational running and the positivity constraints on EFT Wilson coefficients derived from very general assumptions on the UV dynamics.
We hope to report more in the future on this 
fascinating UV-IR connection. 

\section*{Acknowledgments}
P.B. thanks Alex Pomarol, Marco Serone and Elena Venturini for useful discussions. J.S. thanks Brando Bellazzini for illuminating discussions. This research has been partially supported by the DFG Cluster of Excellence 2094 ORIGINS and the Collaborative Research Center SFB1258. MR is supported by the Studienstiftung des deutschen Volkes. All authors warmly thank the Munich Institute for Astro- and Particle Physics (MIAPP) for hospitality, which is funded by the Deutsche Forschungsgemeinschaft (DFG, German Research Foundation) under Germany's Excellence Strategy - EXC-2094 - 390783311.

\appendix 

\section{Selection rules from supersymmetry}\label{ward}
%
A crucial ingredient for the tree-level modified helicity bounds in section~\ref{sectree} is the fact that all $|\tilde{h}|=2$ 4-point amplitudes, with the exception of the four-fermion amplitude $\amp(\psi,\psi,\psi,\psi)$ and its complex conjugate, vanish on-shell. This non-trivial statement has been shown with a combination of direct computation, supersymmetric Ward identities (SWI) and KLT relations to hold for marginal theories~\cite{Cheung:2015aba,Azatov:2016sqh}, pure gravity~\cite{Grisaru:1976vm,Grisaru:1977px} and minimally coupled gravity with two external gravitons~\cite{Bjerrum-Bohr:2013bxa}. Here we complete the proof along the lines of~\cite{Azatov:2016sqh} using SWIs and show that all $|\tilde{h}|=2$ 4-point amplitudes with one external graviton vanish, i.e.
\be \label{eq:h3amp}
0 = \amp(h_{+},V_+,V_+,V_-) = \amp(h_{+},V_+,\phi,\phi^\dagger) = \amp(h_{+},{\psi},{\psi},\phi) = \amp(h_{+},V_+,\bar{\psi},\psi)\,.
\ee
The proof is based on the observation that a marginal theory minimally coupled to gravity that has holomorphic Yukawa couplings can be embedded in a ${\cal N} = 1$ supergravity theory with an $R$-parity, the particles of the original theory being even under it.\footnote{For the SM this is the case in the limit in which either all up-type or down-type Yukawa couplings vanish.} Amplitudes in the supersymmetric theory that have only the original particles (including the gravitons) as external states coincide with the amplitudes of the original minimally coupled theory, since the $R$-parity prevents superpartners, which are odd under $R$, to appear in internal lines at tree level. Global invariance under supersymmetry (SUSY) transformations yields then non-trivial relations between amplitudes via SWIs (see e.g. \cite{Dixon:1996wi,Mangano:1990by,Dixon:2004za}).

Consider an operator $\mathcal{O}=\Phi_1\cdots \Phi_n$ containing a product of arbitrary fields $\Phi_i$ and define $Q(\xi) = \hat{\xi}^\alpha Q_\alpha$ as the supercharge multiplied by a Grassmann spinor parameter $\hat{\xi}$. Then the SWI for $\mathcal{O}$ takes the form
\begin{equation}
0 = \langle 0 | [Q(\xi),\mathcal{O}] | 0\rangle = \sum_i \langle 0 | \Phi_1\cdots [Q(\xi),\Phi_i]\cdots \Phi_n| 0\rangle\,.
\end{equation}
The SUSY transformations of the fields are given by
\begin{align}
[Q(\xi),\phi^\dagger (k)] &= \theta\, \langle k \xi \rangle\, {\psi}(k)\,,\quad &[Q(\xi),\bar\psi (k)] &= \theta\, \langle \xi k\rangle\, \phi (k)\,,\\
[Q(\xi),{\lambda} (k)] &= \theta\, \langle k \xi\rangle\, V_+(k)\,,\quad &[Q(\xi),V_- (k)] &= \theta\, \langle \xi k\rangle\, \bar\lambda (k)\,,\\
[Q(\xi),\zeta_+ (k)] &= \theta\, \langle k \xi\rangle\, h_+(k)\,,\quad &[Q(\xi),h_- (k)] &= \theta\, \langle \xi k\rangle\, \zeta_- (k)\,,
\end{align}
where particle pairs in the same chiral or vector supermultiplet are denoted respectively by $\phi,\psi$ and $\lambda,V$, while $\zeta$ and $h$ are the gravitino and graviton. We also split the Grassmann spinor parameter $\hat{\xi}^\alpha = \theta\, \xi^\alpha$ into a Grassmann parameter $\theta$ and a spinor variable $\xi^\alpha$, corresponding physically to an arbitrary massless vector $\xi^\mu$. The commutators for the remaining fields are obtained by inverting the helicities and substituting $\langle \xi k\rangle \rightarrow [k\xi]$.

In order to prove Eq.~\eqref{eq:h3amp}, we consider in turn the two operators $\mathcal{O}_1 =h_+ V_+ \lambda V_-$ and $\mathcal{O}_2 = h_+ {\lambda} \phi \phi^\dagger$. At the level of scattering amplitudes, the SWI for the first operator yields
\be \label{eq:O1SWI}
\begin{split}
0 = &-[\xi 1 ] \amp(1_{\zeta_+}, 2_{V_+}, 3_{{\lambda}}, 4_{V_-}) - [\xi 2] \amp (1_{h_+}, 2_{{\lambda}}, 3_{{\lambda}}, 4_{V_-})\\
&- \langle \xi 3\rangle \amp (1_{h_+}, 2_{V_+}, 3_{V_+}, 4_{V_-}) + \langle \xi 4 \rangle \amp (1_{h_+}, 2_{V_+}, 3_{{\lambda}}, 4_{\bar\lambda})\,.
\end{split}
\ee
In a supersymmetrized version of a marginal theory minimally coupled to gravity, there are two classes of 3-point amplitudes: graviton and gravitino 3-point amplitudes with $|h(\amp_3 )|=2$ (see e.g.~\cite{McGady:2013sga}), and marginal 3-point amplitudes with $|h(\amp_3 )|=1$. This implies that the first two amplitudes in Eq.~\eqref{eq:O1SWI}, which have total helicity $h=2$, do not factorize into 3-point amplitudes and can only be contact amplitudes, corresponding to a Lorentz invariant higher-dimensional effective operator. However, as we assume minimal coupling of gravity to a marginal theory, there are no higher{-}dimensional operators which could give rise to these amplitudes. Hence they must vanish. Now we have the freedom to choose the kinematic configurations $\xi=k_3$ or $\xi = k_4$, where respectively $\langle \xi 3 \rangle$ and $\langle \xi 4 \rangle$ in \eq{eq:O1SWI} vanish. Then it must be
\be \label{eq:SWIres1}
\amp (1_{h_+}, 2_{V_+}, 3_{V_+}, 4_{V_-})=0\,,\qquad \amp (1_{h_+}, 2_{V_+}, 3_\lambda, 4_{\bar{\lambda}})=0\,.
\ee
Similarly, the Ward identity for $\mathcal{O}_2$
\be
\begin{split}
0 = &- [\xi 1 ] \amp (1_{\zeta_+} ,2_{{\lambda}}, 3_\phi, 4_{\phi^\dagger}) - \langle \xi 2 \rangle \amp (1_{h_+}, 2_{V_+}, 3_\phi,4_{\phi^\dagger})\\
&+ [\xi 3 ] \amp (1_{h_+}, 2_{{\lambda}}, 3_{\bar\psi}, 4_{\phi^\dagger}) - \langle \xi 4 \rangle \amp (1_{h_+}, 2_{{\lambda}}, 3_\phi ,4_{{\psi}})
\end{split}
\end{equation}
contains two amplitudes with $h=2$, i.e.~$\amp (1_{\zeta_+}, 2_{{\lambda}}, 3_\phi, 4_{\phi^\dagger})$ and $\amp (1_{h_+}, 2_{{\lambda}}, 3_{\bar\psi}, 4_{\phi^\dagger})$, which trivially vanish since there are no interactions that mediate it. Taking $\xi=k_2$ or $\xi = k_4$ we then find that also the other two amplitudes vanish
\be \label{eq:SWIres2}
\amp (1_{h_+}, 2_{V_+}, 3_\phi,4_{\phi^\dagger}) = 0\,,\qquad \amp (1_{h_+}, 2_{{\lambda}}, 3_\phi ,4_{{\psi}})=0\,.
\ee
In order to complete the proof we note that helicity amplitudes factorize into color and Lorentz structure. This, in combination with the fact that gauge groups commute with the SUSY algebra, implies that Eqs.~\eqref{eq:SWIres1} and~\eqref{eq:SWIres2} do not only hold for gauginos in the adjoint or matter fermions in the fundamental representation, but for fermions in general representations \cite{Azatov:2016sqh}. This completes the proof that all amplitudes in Eq.~(\ref{eq:h3amp}) vanish.
\subsection{Gravitino amplitudes}

In section~\ref{eq:treeMinCoupl} we commented on the definition of a modified helicity in the presence of gravitinos. If we define $\tilde{h} = h -\tfrac{1}{2} h_g -\tfrac{2}{3} h_{\zeta}$ all 3-point amplitudes satisfy $|\tilde{h}(\amp_3 )|=1$, allowing for $|\tilde{h}(\amp_4 )|=0,2$. In the previous discussion we showed that $|\tilde{h}(\amp_4 )|=2$ amplitudes with matter or graviton external states vanish on shell. The same is true for $|\tilde{h}(\amp_4 )|=2$ amplitudes with external gravitinos, such that $|\tilde{h}(\amp_4 )|=0$ for arbitrary amplitudes.\footnote{A consistent theory with gravitinos must be completely supersymmetric. This forbids non-holomorphic Yukawa couplings, thus there is no exceptional four-fermion amplitude that does not respect the helicity bound.} A simple proof uses SWI to relate $|\tilde{h}(\amp_4 )|=2$ gravitino amplitudes to vanishing graviton amplitudes.

As an explicit example consider amplitudes of the form $\amp (\zeta_+ ,X, Y, Z)$ with $X,Y,Z$ being scalars, fermions or vectors with a combined helicity of ${h}=3/2$. These amplitudes could naively be constructed, yet do not satisfy $|\tilde{h}(\amp_4 )|=0$. Using the operator $\mathcal{O} = h_+ X Y Z$, one obtains the Ward identity
\begin{equation}
0 = -[\xi 1] \amp (1_{\zeta_+}, 2_X, 3_Y, 4_Z) + \ldots\,,
\label{eq:oneGravitinoWard}
\end{equation}
where the ellipsis stand for amplitudes of the form $\amp (1_{h_+}, \ldots)$ with one graviton plus three matter particles with combined helicity ${h}=1,2$, such that the modified helicity of the full amplitudes is $\tilde{h} = 2,3$, which we have shown to vanish in the previous discussion. Thus Eq.~(\ref{eq:oneGravitinoWard}) implies that amplitudes of the form $\amp (1_{\zeta_+}, 2_X, 3_Y, 4_Z)$ with $h(X,Y,Z)=3/2$ vanish on-shell. The remaining $|\tilde{h}(\amp_4 )|=2$ amplitudes with additional external gravitinos can be shown to vanish in a similar fashion.
%

\section{Cut computation with Hermite reduction}
\label{bubble}
%
In this appendix we present the method we used in this work to calculate the rational part
of the $d$LIPS integrals in \eq{gamma2},
an operation that corresponds to extracting the massive bubble coefficients of the loop amplitude (see e.g.~\cite{Baratella:2020lzz,Baratella:2020dvw}).
This method was first reported in~\cite{Mastrolia:2009dr}. Here we present a slightly modified version which employs a different parameterization of the two-particle phase space.

The main point is that the action of ${\cal R}$ in \eq{gamma2} can be algorithmically implemented at the level of the phase space \emph{integrand} using the Hermite Polynomial Reduction method (see~\cite{Mastrolia:2009dr}), in a way that we now
explain.

For internal massless momenta $\ell_{1}$ and $\ell_{2}$ satisfying $\ell_{1} + \ell_{2} = p + q$, $p$ and $q$ being massless external momenta (or combinations of external momenta with $p^2=q^2=0$),
we can parameterize the two-particle phase space by relating $| \ell_{1} \rangle$ and $| \ell_{2} \rangle$ to $| p \rangle$ and $| q \rangle$ as
\be \label{eq:PsParam}
|\ell_1\rangle = \cos\theta\, |p\rangle - \sin\theta\, e^{i\phi} |q\rangle\,,\qquad |\ell_2\rangle = \sin\theta\, e^{-i\phi} |p\rangle + \cos\theta |q\rangle\,,
\ee
with the square brackets given by the complex conjugate of these expressions. The phase space integral in this parameterization is then given by~\cite{Caron-Huot:2016cwu}
\be \label{eq:PsMeasure}
\int d\text{LIPS} =\frac{\pi}{2} \int_0^{2\pi} \frac{d\phi}{2\pi}\int_0^{\pi/2} d\theta\, 2\sin\theta\, \cos\theta\,.
\ee
Substituting $z\equiv e^{i\phi}$ and $t\equiv \tan\theta$, we can then formally re-express the $d$LIPS integral as
\be
\int d{\rm LIPS} \, \amp_L(\ldots,-\bar{\ell}_2,-\bar{\ell}_1)\,\amp_R(\ell_1,\ell_2,\ldots) = \frac{\pi}{2} \,\oint_{|z|=1} \frac{dz}{2\pi i} \int_0^\infty  \frac{2\, t\,dt}{(1+t^2)^2} f(z,t)\,,
\ee
where we defined
\be
f(z,t) = \frac{1}{z}\, \amp_L(\ldots,-\bar{\ell}_2,-\bar{\ell}_1)\,\amp_R(\ell_1,\ell_2,\ldots)\,,
\ee
with $\ell_1$ and $\ell_2$ expressed in terms of $p$ and $q$ according to Eq.~\eqref{eq:PsParam}.

We first perform the $z$ integral along the contour $|z|=1$, which yields a sum over the residues of all simple poles $z_i (t)$ of $f(z,t)$ that are within the unit circle
\be
\oint_{|z|=1} \frac{dz}{2\pi i} f(z,t) = \sum_i\, \Theta( |z_i (t)| -1)\, \text{Res}_{z=z_i (t)} f(z,t)\,,
\ee
with the Heaviside step function enforcing the condition that the pole $z_i (t)$, whose position in the complex plane is a function of $t$, lies in the region $|z|\leq 1$.

We are now left with the $dt$ integration. Because both $\amp_L$ and $\amp_R$ are tree-level amplitudes, $f(z,t)$ and therefore also $\text{Res}_{z=z_i (t)} f(z,t)$ are \emph{rational functions} of $t$. The key fact now is that the primitive of any rational function has a well defined decomposition into a rational and a logarithmic part~\cite{Bronstein}, so that in our case we have
\begin{equation}\label{rhoL}
\int dt \,\frac{2\, t}{(1+t^2)^2}\, \text{Res}_{z=z_i (t)} f(z,t) = \rho_i(t) + L_i(t) \,,
\end{equation}
where by $\rho_i(t)$ and $L_i(t)$ we denote respectively the purely rational and purely log parts of the \emph{indefinite} integral. The Hermite Polynomial Reduction method comes in at this point to efficiently obtain $\rho_i(t)$ without having to solve the full integral~\cite{Mastrolia:2009dr,Bronstein}. For this work, we have adopted a Mathematica implementation of the algorithm that can be found at~\cite{HPR}.

Once we have the rational part of \eq{rhoL}, as a final step we must plug in the correct integration boundaries $t_{\rm max}$ and $t_{\rm min}$, which are dictated by the Heaviside functions, to finally obtain
\be
{\cal R}\int d{\rm LIPS} \, \amp_L(\ldots,-\bar{\ell}_2,-\bar{\ell}_1)\,\amp_R(\ell_1,\ell_2,\ldots) = \frac{\pi}{2} \, \sum_i \big[ \rho_i(t_{{\rm max}}^{(i)}) - \rho_i(t_{{\rm min}}^{(i)}) \big]\,.
\ee

\section{Renormalization of $T^{\mu\nu}$ and collinear anomalous\\dimensions}\label{collinear}
The aim of this appendix is manifold.
The main objective is to give a proof of the cancellation {of UV divergences} among class \emph{(B)} diagrams of
Section~\ref{Ord2}, that relies on the conservation of the energy momentum tensor $T^{\mu\nu}$. In the construction
of the proof, we will however touch on other topics which could be interesting \emph{per se}.
Specifically, we will define a partial wave expansion for
Lorentz-covariant operators' form factors, and we will also provide a simple formula to express collinear anomalous dimensions
in an arbitrary theory in terms of certain partial wave coefficients with $J=2$.

The cancellation of divergences among the diagrams in class \emph{(B)} has a very well known analogue in terms of global symmetries,
where the conservation of the symmetry current $j^\mu$ implies, via Ward identities, the combined
cancellation of vertex and propagator corrections.
This phenomenon can be stated as the `non-renormalization' of the current operator $j^\mu$, that is $\gamma_j=0$.
However, strictly speaking the operator that does not get renormalized is the \emph{divergence} of the current, i.e.~$\partial_\mu j^\mu$ (see appendix A13 of \cite{ZinnJustin:2002ru} for a nice review of the quantum consequences of the
conservation of $j^\mu$
and $T^{\mu\nu}$). This distinction turns out to be crucial in QED where the vector current mixes with the divergence of the field strength $\partial_\nu F^{\nu\mu}$ and therefore acquires a non-zero anomalous dimension in the presence of a photon (see e.g.~\cite{Collins:2005nj}). However, it is possible to define an improved current out of the vector current $j^\mu$ and $\partial_\nu F^{\nu\mu}$ which does not get renormalized.\footnote{Similar comments should apply to $T^{\mu\nu}$ and the operator $\partial_\rho \partial_\sigma C^{\mu\rho\nu\sigma}$ constructed out of the Weyl tensor $C^{\mu\rho\nu\sigma}$ (see \cite{Erdmenger:1996yc} for a related discussion). However this mixing will not be relevant for the following discussion, since power counting dictates that it is a two-loop effect. We refrain from further studying this problem here.}

This observation is also important for our analysis of the renormalization of $T^{\mu\nu}$. In fact in a generic scalar theory,
while the operator $\partial_\mu T^{\mu\nu}$ has zero anomalous
dimension, one cannot exclude that $T^{\mu\nu}$ undergoes a renormalization like \cite{ZinnJustin:2002ru,Callan:1970ze}
\be\label{ZinnJ}
\left. T^{\mu\nu}\right|_{\rm ren}=T^{\mu\nu}+\frac{A}{\epsilon}\left(\eta^{\mu\nu}\partial^2-\partial^\mu\partial^\nu \right) \phi^2\,.
\ee
By taking the $\partial_\mu$ of the above, we get as expected
\be
\left. \partial_\mu T^{\mu\nu}\right|_{\rm ren}=\partial_\mu T^{\mu\nu}\,,
\ee
i.e.~$\partial_\mu T^{\mu\nu}$ is not renormalized. However, the very existence of a Lorentz-invariant divergence-free
combination like $\eta^{\mu\nu}\partial^2-\partial^\mu\partial^\nu$ does not allow us to extend the non-renormalization
property to the full $T^{\mu\nu}$. This incomplete $T^{\mu\nu}$ non-renormalization is related, as we will see, to the
fact that cancellation of class \emph{(B)} diagrams in the scalar sector is similarly not complete
(when scalars are minimally coupled).

With these generalities in mind, we move now to their concrete implementation. Our central object, whose renormalization
we will study, is the form factor
\be
T^{\mu\nu}(1_\Phi,2_{\bar{\Phi}})\equiv \la 1_\Phi,2_{\bar{\Phi}} |T^{\mu\nu}| \Omega \ra \,,
\ee
where $\Phi\bar{\Phi}$ is a particle-antiparticle pair, that for later convenience we take as outgoing. We have for example for vectors
\be\label{Tvec}
(\sigma_\mu)^{\alpha\dot{\alpha}}(\sigma_\nu)^{\beta\dot{\beta}}T^{\mu\nu}( 1_{V_-}, 2_{V_+})\equiv T^{\alpha\beta,\dot{\alpha}\dot{\beta}} ( 1_{V_-}, 2_{V_+}) = 2\, \lambda_2^\alpha \lambda_2^\beta \tilde{\lambda}_1^{\dot{\alpha}}\tilde{\lambda}_1^{\dot{\beta}}\,,
\ee
where $\lambda_2^\alpha=\la 2$ and $\tilde{\lambda}_1^{\dot{\alpha}}=1]$. Similarly we find for fermions
\be\label{Tpsi}
T^{\alpha\beta,\dot{\alpha}\dot{\beta}} ( 1_{{\psi}}, 2_{\bar\psi}) = \frac{1}{2}  \big( \lambda_1^\alpha \lambda_1^\beta \tilde{\lambda}_1^{\dot{\alpha}}\tilde{\lambda}_2^{\dot{\beta}} + \lambda_1^\alpha \lambda_1^\beta \tilde{\lambda}_1^{\dot{\beta}}\tilde{\lambda}_2^{\dot{\alpha}} - \lambda_1^\alpha \lambda_2^\beta \tilde{\lambda}_2^{\dot{\alpha}}\tilde{\lambda}_2^{\dot{\beta}} - \lambda_1^\beta \lambda_2^\alpha \tilde{\lambda}_2^{\dot{\alpha}}\tilde{\lambda}_2^{\dot{\beta}}\big)\,.
\ee
Notice that both \eq{Tvec} and \eq{Tpsi} are \emph{symmetric} in the dotted and undotted $SL(2,\mathbb{C})$ indices, which
is equivalent to the tracelessness of $T^{\mu\nu}$, that is $T^\mu_{~\mu}=0$. In both ``languages'', this means
that the energy-momentum tensor transforms irreducibly under the Lorentz group.
It is also important to stress that both of the above
expressions are uniquely fixed by little group scaling, the requirement that $p_\mu T^{\mu\nu}=0$ and the requirement that the $T^{\mu\nu}$ operator should give back the
momentum of the state \cite{Caron-Huot:2016cwu}: meaning that the stress tensor is automatically traceless in the fermion
and vector sectors.

The same is not true for scalars. For a minimally coupled scalar, the energy momentum that couples to gravity is of the form
\be\label{scalarT}
T^{\mu\nu}(x)=\partial^\mu\phi\partial^\nu \phi-\frac{1}{2}\eta^{\mu\nu}\partial_\alpha\phi\partial^\alpha\phi+\ldots\,,
\ee
where the ellipsis stand for terms with more fields, and one gets
\be\label{Tphimin}
T^{\alpha\beta,\dot{\alpha}\dot{\beta}} ( 1_{\phi}, 2_{\bar{\phi}})=-p_1^{\alpha\dot{\beta}}p_2^{\beta\dot{\alpha}}-p_1^{\beta\dot{\alpha}}p_2^{\alpha\dot{\beta}}\,,
\ee
which is \emph{not} symmetric in the $SL(2,\mathbb{C})$ indices and therefore not transforming irreducibly.
However, by coupling the scalar conformally instead of minimally, one gets a stress tensor that reads
\be\label{scalarTtilde}
\tilde{T}^{\mu\nu}(x)=\partial^\mu\phi\partial^\nu \phi-\frac{1}{2}\eta^{\mu\nu}\partial_\alpha\phi\partial^\alpha\phi+\frac{1}{6}\left(\eta^{\mu\nu}\partial^2-\partial^\mu\partial^\nu \right) \phi^2+\ldots\,,
\ee
which is traceless. Notice that the `pure trace' that we are effectively subtracting from \eq{scalarT} by
coupling conformally instead of minimally has precisely the form of the divergent term in \eq{ZinnJ}. The two-scalar form factor of $\tilde{T}^{\mu\nu}(x)$ reads
\begin{align}
\tilde{T}^{\alpha\beta,\dot{\alpha}\dot{\beta}} ( 1_{\phi}, 2_{\bar{\phi}})= \frac{1}{3} \big( p_1^{\alpha\dot{\alpha}} p_1^{\beta\dot{\beta}} + p_2^{\alpha\dot{\alpha}} p_2^{\beta\dot{\beta}} - p_1^{\alpha\dot{\alpha}} p_2^{\beta\dot{\beta}} - p_2^{\alpha\dot{\alpha}} p_1^{\beta\dot{\beta}} - p_1^{\alpha\dot{\beta}} p_2^{\beta\dot{\alpha}} - p_1^{\beta\dot{\alpha}} p_2^{\alpha\dot{\beta}}\big)\,
\end{align}
which is symmetric in the $SL(2,\mathbb{C})$ indices as expected
(remember that $p^{\alpha\dot{\alpha}}=\lambda^\alpha\tilde{\lambda}^{\dot{\alpha}}$).

Having become familiar with the relevant ${T}^{\mu\nu}$ form factors, we are finally ready to study their one-loop renormalization.
We will focus in particular on the renormalization of \eq{Tphimin}.
In light of the previous discussion, and anticipating future notation, we split $T=T_0+T_2$, where
$T_0=-(\eta^{\mu\nu}\partial^2-\partial^\mu\partial^\nu)\phi^2/6$ and $T_2\equiv \tilde{T}$. With a natural generalization of
\eq{gamma2} that allows to include form factor renormalizations (see \cite{Caron-Huot:2016cwu} for details), we get at one loop
\begin{align}\label{gammacol}
\gamma_2 \,T_2^{\mu\nu}(1_{\phi}, 2_{\bar{\phi}})&+\gamma_0 \,T_0^{\mu\nu}(1_{\phi}, 2_{\bar{\phi}})=\gamma_{\rm coll}\,T^{\mu\nu}(1_{\phi}, 2_{\bar{\phi}})\nonumber \\
&-\frac{1}{4\pi^3}\sum_{1',2'}\,{\cal R}\!\int d{\rm LIPS} ~T^{\mu\nu}(1',2')\,\amp_{\rm tree}(1',2'\to 1_{\phi}, 2_{\bar{\phi}})\,,
\end{align}
where the RHS is summed over all relevant particle pairs $1',2'$ as in \eq{gamma2}. Notice that in this particular case $1',2'$ are always particle-antiparticle pairs.
Crucially, we split the LHS into two pieces as we expect the `pure trace' $T_0$ to undergo a separate renormalization with
respect to the traceless part $T_2$, as we see from \eq{ZinnJ}. Actually, the conservation of $T^{\mu\nu}$ implies the
stronger statement that
\be\label{gam2}
\gamma_2=0\,.
\ee
As we are going to see, the components $T_0$ and $T_2$ do not mix under the $d$LIPS convolution, and
\eq{gammacol} can be equivalently rewritten as the following system of two equations
\begin{align}
\gamma_{\rm coll}\,T_2^{\mu\nu}(1_{\phi}, 2_{\bar{\phi}})&=+\frac{1}{4\pi^3}\sum_{1',2'}{\cal R}\!\!\int d{\rm LIPS} ~T_2^{\mu\nu}(1',2')\,\amp_{\rm tree}(1',2'\to 1_{\phi}, 2_{\bar{\phi}})\,,\label{T2coll}\\
(\gamma_0-\gamma_{\rm coll})\,T_0^{\mu\nu}(1_{\phi}, 2_{\bar{\phi}})&=-\frac{1}{4\pi^3}\sum_{1',2'}{\cal R}\!\!\int d{\rm LIPS} ~T_0^{\mu\nu}(1',2')\,\amp_{\rm tree}(1',2'\to 1_{\phi}, 2_{\bar{\phi}})\,,\label{T0UV}
\end{align}
where we have also enforced \eq{gam2}.
The possibility of breaking \eq{gammacol} in this way will be rigorously
proven through a partial wave analysis of $T^{\mu\nu}$ (and of the phase space integral),
that will also explain the choice of suffix for its irreducible components.
Before going into the details of this, we notice that \eq{T2coll} provides a formula to compute the collinear anomalous dimensions
(that we are going to further massage later), while $\gamma_0$ in \eq{T0UV}, which will turn out to be non-vanishing,
directly translates into the coefficient of \eq{UV3B}.

\vspace{.2cm}

The partial wave expansion of a generic amplitude $\amp$ is defined for example in \cite{Baratella:2020dvw}, whose notation
and definitions we use here. 
In the following we present the analogous steps to define an angular decomposition for Lorentz-covariant operators like $T^{\mu\nu}$,
which to our knowledge is new.

We define it operatively. Take an expression written in terms of spinors like \eq{Tvec}, and rewrite the spinors
in terms of some reference $\zeta^\alpha$ (Zenith) and $\nu^\alpha$ (Nadir), as follows
\be
\begin{array}{l}
 |1\ra=  c_{\theta/2}| \zeta\ra -s_{\theta/2} e^{-i\phi}|\nu\ra\\
|2\ra=  s_{\theta/2} e^{i\phi}|\zeta\ra+c_{\theta/2}|\nu\ra
\end{array} \quad \ \
\begin{array}{l}
|1]=c_{\theta/2}| \zeta] -s_{\theta/2} e^{i\phi}|\nu]\\
|2]=s_{\theta/2} e^{-i\phi}|\zeta]+c_{\theta/2}|\nu]
\end{array}\,.
\label{nuzeta}
\ee
The expression that one obtains can then be expanded into angular functions
\be\label{pwdT}
T^{\alpha\beta,\dot{\alpha}\dot{\beta}} ( 1, 2)=e^{-i (h_{1}-h_2)\phi}\sum_J c_{(1,2)}^J\times\sum_M e^{iM\phi } d^J_{M,h_{1}-h_2}(\theta)\,\tau_{J,M}^{\alpha\beta,\dot{\alpha}\dot{\beta}}\,,
\ee
where $d^j_{mm'}$ are Wigner functions and $c$, $\tau$ are defined by the expression itself (we do not bother to fix the normalization of $\tau$, as only its `group-theoretic' properties will enter here). For example, \eq{Tvec} can be expanded as
\begin{align}\label{Tvecexp}
T^{\alpha\beta,\dot{\alpha}\dot{\beta}} ( 1_{V_-}, 2_{V_+})=2\left( e^{4i\phi}s_{\theta/2}^4 \zeta^\alpha \zeta^\beta \tilde{\nu}^{\dot{\alpha}}\tilde{\nu}^{\dot{\beta}}+\ldots+ c_{\theta/2}^4 \nu^\alpha \nu^\beta \tilde{\zeta}^{\dot{\alpha}}\tilde{\zeta}^{\dot{\beta}}\right)\,,
\end{align}
where we quoted only the simplest terms, that have $M=2$ and $M=-2$ respectively, the ellipsis standing for those with $M=-1,0,1$.
We see that indeed $s_{\theta/2}^4\propto d^2_{2,-2}(\theta)$, while $c_{\theta/2}^4\propto d^2_{-2,-2}(\theta)$, both with
$J=2$! The reader is invited to check that the presence of only $d^{2}$ functions
extends to all the terms of \eq{Tvecexp} that we did not write,
as well as to \eq{Tpsi} when it is expanded according to \eq{nuzeta}.

Perhaps unsurprisingly at this point, the story of the scalar sector is different, and a $J=0$ component is in general present. This
is actually the case for $T^{\mu\nu}$ in \eq{scalarT} --~but \emph{not} for $\tilde{T}^{\mu\nu}$ in \eq{scalarTtilde}~--. One has in fact
\begin{align}
&{T}^{\alpha\beta,\dot{\alpha}\dot{\beta}} ( 1_{\phi}, 2_{\bar{\phi}})=\frac{1}{3}[\,\nu^\alpha \nu^\beta \tilde{\nu}^{\dot{\alpha}}\tilde{\nu}^{\dot{\beta}}+\zeta^\alpha \zeta^\beta \tilde{\zeta}^{\dot{\alpha}}\tilde{\zeta}^{\dot{\beta}}-\zeta^\alpha \nu^\beta \tilde{\zeta}^{\dot{\alpha}}\tilde{\nu}^{\dot{\beta}}-\nu^\alpha \zeta^\beta \tilde{\nu}^{\dot{\alpha}}\tilde{\zeta}^{\dot{\beta}}+2\,\nu^\alpha \zeta^\beta \tilde{\zeta}^{\dot{\alpha}}\tilde{\nu}^{\dot{\beta}}+2\,\zeta^\alpha \nu^\beta \tilde{\nu}^{\dot{\alpha}}\tilde{\zeta}^{\dot{\beta}}\,]\nonumber \\
&+\frac{1}{3}[\, -\nu^\alpha \nu^\beta \tilde{\nu}^{\dot{\alpha}}\tilde{\nu}^{\dot{\beta}}-\zeta^\alpha \zeta^\beta \tilde{\zeta}^{\dot{\alpha}}\tilde{\zeta}^{\dot{\beta}}+\nu^\alpha \zeta^\beta \tilde{\zeta}^{\dot{\alpha}}\tilde{\nu}^{\dot{\beta}}+\nu^\alpha \zeta^\beta \tilde{\nu}^{\dot{\alpha}}\tilde{\zeta}^{\dot{\beta}}+\zeta^\alpha \nu^\beta \tilde{\zeta}^{\dot{\alpha}}\tilde{\nu}^{\dot{\beta}}+\zeta^\alpha \nu^\beta \tilde{\nu}^{\dot{\alpha}}\tilde{\zeta}^{\dot{\beta}}\,]\frac{3 \,c^2_\theta -1}{2}+...\nonumber \\
&=\frac{1}{3}\tau_{0,0}^{\alpha\beta,\dot{\alpha}\dot{\beta}}d^0_{0,0}(\theta)+\frac{1}{3}\tau_{2,0}^{\alpha\beta,\dot{\alpha}\dot{\beta}}d^2_{0,0}(\theta)+\ldots
\end{align}
which has both a $J=2$ and a $J=0$ component, while $\tilde{T}$ has only the second term proportional to $d^2_{0,0}(\theta)$ (with
the same coefficient). The ellipsis stand for terms with $M=\pm 1,\pm 2$. As another important point, let us mention that $\tau_{0,0}$ and $\tau_{2,0}$ are orthogonal. Indeed, as one can easily check,
\be
(\tau_{0,0})^{\alpha\beta,\dot{\alpha}\dot{\beta}}(\tau_{2,0})_{\alpha\beta,\dot{\alpha}\dot{\beta}}=0\,.
\ee
The
decomposition in \eq{pwdT} into components with different $J$ (the different $M$ components of a given $J$
are fixed by Lorentz symmetry) is equivalent to the decomposition of the conserved
stress tensor into a traceless part and and a `conserved trace', and it can be used in a more general setup
to automatize the familiar splitting of covariant tensors into irreducible pieces. 

\vspace{.2cm}

Now, similarly to what was done in \cite{Baratella:2020dvw} for a $d$LIPS integral of two amplitudes, we can combine in
\eq{gammacol}
the angular decomposition defined by \eq{pwdT} with the partial wave decomposition of $\amp_{\rm tree}$. Thanks to the orthogonality
properties of the Wigner $d$-functions and of the $\phi$ exponentials, and using $d$LIPS$=d\phi\, d\theta s_\theta/8$,
\eq{gammacol} reduces to
\begin{align}
&\sum_{J\in \{0,2\}} c_{(\phi,\bar{\phi})}^J\,\gamma_J \sum_{M=-J}^Je^{iM\phi } d^J_{M,0}(\theta)\tau_{J,M}^{\alpha\beta,\dot{\alpha}\dot{\beta}}\nonumber \\
&=\sum_{J\in \{0,2\}}  \left(c_{(\phi,\bar{\phi})}^J\gamma_{\rm coll}-\frac{1}{8\pi^2} \sum_{1',2'}c_{(1',2')}^J\left. a^J_{( {1}',{2}'\to {\phi}, {\bar{\phi}})}\right|_{\rm reg}  \right)\sum_{M=-J}^Je^{iM\phi } d^J_{M,0}(\theta)\tau_{J,M}^{\alpha\beta,\dot{\alpha}\dot{\beta}}\,,
\end{align}
where the IR-regularized partial wave coefficients are defined in \cite{Baratella:2020dvw}, the necessary IR subtraction being
provided in \eq{gammacol} by the `projector' ${\cal R}$. Now, for the above equation to be valid for any $\phi$ and $\theta$,
it must be
\begin{align}
\gamma_{\rm coll}&=2\gamma^{(\phi)}=\frac{1}{8\pi^2}\,\left( {c_{(\phi,\bar{\phi})}^2}\right)^{-1}\,\sum_{1',2'}\,c_{(1',2')}^2\left. a^2_{( {1}',{2}'\to {\phi}, {\bar{\phi}})}\right|_{\rm reg}\,, \label{formulacoll} \\
 \gamma_0&=\gamma_{\rm coll}-\frac{1}{8\pi^2} \,\sum_{1',2'}\,\left. a^0_{({1}',{2}'\to {\phi},{\bar{\phi}})}\right|_{\rm reg}  \,,\label{formulazero}
\end{align}
where we used \eq{gam2} in the first equality. The relation between $\gamma_{\rm coll}$ and $J=2$
partial wave coefficients provided by \eq{formulacoll} is practically useful and, to our knowledge, new. The value of the coefficients
$c_{(1',2')}^2$ -- where we remind the reader that $2'$ is $1'$s antiparticle -- in \eq{formulacoll} will be provided in \eq{cfromf}.
We stress that, while we focused on the richer scalar sector, all the analysis can be extended to fermions and vectors as well, with obvious
modifications in the formulas that we provided. Let us also point out that we simplified the second equation after considering that $1',2'$ can only be scalars
in order for $a^0$ to be non-zero, and that $c_{(1',2')}^0$ depends just on helicities and not on other quantum numbers. We see that the splitting into a $J=2$
and a $J=0$ equation, which is mirrored by \eq{T2coll} and \eq{T0UV}, crucially relies in the conservation of $J$ across the cut, analogously to what was studied in \cite{Baratella:2020dvw}.

\vspace{.2cm}

We finally come back to proving the claims of Section~\ref{Ord2}. We do this in the partial wave language, as we believe it is the most informative although perhaps not the most direct. We prove the cancellation of class \emph{(B)} diagrams in the conformal limit, and at the same time we deduce \eq{UV3B}.

The partial wave coefficients of the gravitational amplitudes in Eqs.~(\ref{J2phi}--\ref{J2vec}) are
given by
\be
a^2_{({\Phi},{\bar{\Phi}}\,\to\,{\Phi'},{\bar{\Phi}'})}={f_{h_\Phi} f_{h_{\Phi'}}}\,,~~~~~~~
f_{0}=\frac{1}{\sqrt{30}}\,,~~f_{1/2}=\frac{1}{\sqrt{20}}\,,~~f_1=-\frac{1}{\sqrt{5}}\,,
\ee
while $a^0_{({\phi},{\bar{\phi}}\,\to\,{\phi'},{\bar{\phi}'})}=1/6$. 

Consider then the ${\cal O}(\Mp^{-2})$ renormalization of the scalar amplitude $\amp(1_{\phi},2_{\bar{\phi}},3_{\bar{\phi}'},4_{\phi'})$, and focus on those diagrams
of class \emph{(B)} that, as in Fig.~\ref{fig:dim6}, have matter loops inserted in the ${\phi}',\bar{\phi'}$ line. Performing
a partial wave decomposition of the corresponding terms in \eq{gamma2}, as explained for example in \cite{Baratella:2020dvw},
one gets
\begin{align}
&\amp_{\rm UV}^{(B)}(1_{\phi},2_{\bar{\phi}}\to 3_{{\phi}'},4_{\bar{\phi'}})=-\frac{\alpha s}{12\epsilon \Mp^2}\left(\gamma_{\rm coll}-\frac{1}{8\pi^2} \sum_{1',2'}\left. a^0_{({1}',{2}'\,\to \,{\phi'}, {\bar{\phi}'})}\right|_{\rm reg} \right)\,  d^{\,0}_{0,0}(\theta) \nonumber \\
&-\frac{s}{2\epsilon\Mp^2}f_0\left( f_0 \,\gamma_{\rm coll}-\frac{1}{8\pi^2} \sum_{1',2'} f' \left. a^2_{({1}',{2}'\,\to \,{\phi'}, {\bar{\phi}'})}\right|_{\rm reg}\right)\,5\, d^{\,2}_{0,0}(\theta)+\ldots=\frac{-\alpha \gamma^{(\phi')}_0 s}{12\epsilon \Mp^2} +\ldots\,,
\end{align}
where the ellipsis stand for the analogous terms where the roles of $\phi$ and $\phi'$ are exchanged. The parameter $\alpha$ is defined in \eq{eq:ScalarNonMin}, $f'\equiv f_{h_{1'}}=f_{h_{2'}}$, and in the last step we used \eq{formulacoll} and \eq{formulazero} together with the fact that
\be\label{cfromf}
c^2_{(\Phi,\bar{\Phi})}= \sqrt{\frac{10}{3}}f_{h_\Phi}\,,
\ee
as the reader can check. For all other class \emph{(B)} amplitudes, i.e.~with not just scalars as external legs, there is no $d^{\,0}$ component and the amplitude completely vanishes.

\section{Generalization to non-minimal coupling}\label{conformalgen}
%
The tree-level gravitational 4-point amplitudes of matter particles must reduce to products of 3-point amplitudes in all factorization channels. While this is a powerful consistency condition it does not fix all leading order amplitudes completely. In particular the four-scalar and four-fermion amplitude, which at $\mathcal{O}(\Mp^{-2})$ in full generality have the form
\begin{align}
\amp(1_{\bar\phi},2_{{\phi}},3_{{\phi}'},4_{\bar\phi'})&=\frac{1}{\Mp^2}\left(\frac{tu}{s}-\frac{s}{6} \right)+\frac{\alpha}{\Mp^2}\,\frac{s}{6}\,, \label{eq:ScalarNonMin}\\
\amp(1_{\bar\psi},2_{{\psi}},3_{{\psi}'},4_{\bar\psi'})&=\frac{1}{\Mp^2}\frac{3t-u}{4s}\la 14 \ra [23] +\frac{1}{\Mp^2}\frac{\beta}{4}\la 14 \ra [23] \,, \label{eq:FermionNonMin}
\end{align}
depend on two free parameters $\alpha$ and $\beta$ whose contribution to the amplitude vanishes in all factorization channels. They can be fixed e.g.~by matching to a Feynman diagrammatic computation starting from a specific action. For minimal coupling to gravity one finds $\alpha = 1$ and $\beta= 0$, for which the amplitudes reduce to Eq.~\eqref{J2phi} and~\eqref{J2psi}. Different values of $\alpha$ originate from a non-minimal coupling of the scalar field to gravity of the form $\sim R \phi^\dagger\phi$, where $R$ is the Ricci scalar. A particularly interesting scenario is $\alpha= 0$, the so-called conformal coupling, which makes the scalar energy-momentum tensor traceless (cf.~Appendix~\ref{collinear}). Deviations from $\beta=0$ can be interpreted as an effect of non-vanishing torsion which can be captured in an effective four-fermion operator (see e.g. \cite{Alonso:2019ptb} for the exact relation to torsion). Note that splitting the amplitudes as in Eq.~\eqref{eq:ScalarNonMin} and~\eqref{eq:FermionNonMin} allows for an interpretation in terms of angular momentum $J$ in the $1,2 \rightarrow 3,4$ channel. The first term in both expressions corresponds to the $J=2$ component of the amplitude what one would expect from the exchange of an on-shell graviton. $\alpha, \beta \neq 0$ induce a $J=0$ and $J=1$ component for the scalar and fermion amplitude, respectively.\footnote{This explains why $\alpha$ and $\beta$ cannot be fixed by factorization arguments. On the factorization channel the internal graviton goes on-shell and can, due to angular momentum conservation, only contribute to the $J\geq 2$ part of the amplitude and the $J=0$ and $J=1$ components must vanish.}

All previous results were obtained assuming minimal coupling to gravity, i.e.~$\alpha=1,\beta=0$. In this appendix we generalize the main results of Section~\ref{Ord2} and~\ref{Ord4} to generic values of $\alpha$ and $\beta$. Note that the discussion on the RG mixing in Section~\ref{mixinggrav} is insensitive to $\alpha$ and $\beta$.
%
\subsection{Results at $\mathcal{O}(\Mp^{-2})$}
%
The divergent parts of the 4-point amplitudes at $\mathcal{O}(\Mp^{-2})$ are sensitive to $\alpha$ and $\beta$. The corresponding expressions with generic $\alpha$ and $\beta$ are given by
\begin{align}
\amp_{\rm UV}(1_{\bar{\psi}_1},2_{{\psi}_2},3_{\phi_3},4_{\phi_4})&=-\frac{(190-\alpha+3\beta )T_{s}+3\, (20+\alpha- 3\beta)(Y_{t}+Y_{u})}{576\pi^2\Mp^2\epsilon}\la 13 \ra [23]
\,,\\
\amp_{\rm UV}(1_{\bar{\psi}_1},2_{\psi_2},3_{\psi_3},4_{\bar{\psi}_4})&=-\frac{11\,(5+\beta)\left(T_s+T_t\right)-12\, \beta\, Y_u}{192\pi^2\Mp^2\epsilon}\la 14 \ra [23]\,,\\
\left.\amp_{\rm UV}(1_{\phi_1},2_{\phi_2},3_{\phi_3},4_{\phi_4})\right|_{(A)}&=-\frac{5}{12\pi^2\Mp^2\epsilon}\frac{13-\alpha}{12}\left[(t-u)\,T_s+(u-s)\,T_t+(s-t)\,T_u\right]\,,\\
\left.\amp_{\rm UV}(1_{\bar\phi},2_{{\phi}},3_{{\phi}'},4_{\bar\phi'})\right|_{(B)}&=-\frac{s \, \alpha}{12\Mp^2\epsilon}( \gamma_0^{(\phi)}+\gamma_0^{(\phi')})\,.
\end{align}
Note that $\amp_{\rm UV}(1_{\bar{\psi}_1},2_{\psi_2},3_{\psi_3},4_{\bar{\psi}_4})$ contains a term proportional to $Y_u$ only for $\beta \neq 0$.
%
\subsection{Results at $\mathcal{O}(\Mp^{-4})$}
%
As we have discussed at length in Section~\ref{Ord4}, the $J\neq 2$ parts of the tree-level amplitudes only affect the running of the Wilson coefficients in the four-scalar and four-fermion amplitudes. For generic values of $\alpha$ and $\beta$ these take the form
\begin{align}
\gamma_{\phi\phi'}^{(2)} &= -\frac{1}{8\pi^2} \bigg( K + \frac{67}{10} + \frac{(1-\alpha)(19-\alpha)}{54}\bigg)\,,\qquad\qquad
\gamma_{\phi\phi'}^{(1)}=0\,,\\ 
\gamma_{\phi\phi'}^{(0)} &= -\frac{1}{48\pi^2} \bigg(\frac{N_\phi}{6} + \frac{7}{2} + \frac{\alpha - 1}{108}\big( 143 +18\, (1+\alpha)\, N_\phi - 17 \alpha \big)\bigg)\,, \label{eq:anomDim8ScalarNonMin}\\
\gamma_{\psi\psi'}^{(2)} &= -\frac{1}{8\pi^2}\bigg(K + \frac{181}{30}+\frac{\beta(5+\beta)}{12}\bigg)\,,\\
\gamma_{\psi\psi'}^{(1)} &= -\frac{1}{16\pi^2}\bigg(\frac{25}{24} +\frac{\beta}{24}\big( 15 +\beta N_\psi +4\beta \big)\bigg)\,.\label{eq:anomDim8FermionNonMin}
\end{align}
The remaining anomalous dimensions in Section~\ref{Ord4} are independent of $\alpha$ and $\beta$.
%
\section{Operator mixing and power counting}\label{opMix}
%
In Section~\ref{mixinggrav} we discussed the RG mixing among dimension-six operators in the GRSMEFT, that is contributions to anomalous dimensions of the form $\gamma_i^{(6)}\propto C_j^{(6)}$. However, due to the dimensionful nature of the gravitational coupling also mixing into higher-dimensional operators occurs, for example of the form $\gamma_i^{(8)}\propto C_j^{(6)}$. These mixing contributions, where the anomalous dimension is linear in the Wilson coefficients, can be again restricted with the help of the non-renormalization theorem in Eq.~\eqref{htildemixing}, i.e.~an operator $\mathcal{O}_i$ can only be renormalized by another operator $\mathcal{O}_j$ if the modified helicity and number of external legs of their minimal amplitudes satisfy $|\tilde{h}_i-\tilde{h}_j| \leq n_i-n_j$. This condition applies irrespectively of the operator dimensions. Thus in order to classify all allowed mixings it is necessary to consider operators of different dimension simultaneously in combination with a power counting rule.

As we already mentioned in Section~\ref{mixinggrav}, in order to have mixing among operators of the same dimension there can be no gravitons crossing the unitarity cut or internally in the factorized tree-level amplitudes as this would introduce additional powers of $1/\Mp$. This directly leads to the requirement that the number of external gravitons must stay constant or increase.
Instead, if there are internal gravitons the operator dimension jumps by $\Delta w = w_i - w_j$ which we found to be bounded by
\be \label{eq:pwrCount}
\Delta w \leq 2\, L + \Delta n_m\,,
\ee
where $L$ is the number of loops and $\Delta n_m = n_{m,\, i} - n_{m,\, j}$ is the change in the number of external matter particles. Equality is reached if all involved couplings are gravitational.
Indeed, in general the jump in operator dimension due to gravitational interactions is given by
\be \label{eq:pwrCountgen}
\Delta w = 2 L  + \Delta n_m - \sum_k v_k \left( n_{m,\, k} - 2 \right) \,,
\ee
where $v_k$ is the number of vertices with $n_{m,\, k} \,\geq 2$ matter particles (in our hypotheses, this means including all marginal/minimal-gravity interactions, but excluding the 3-point interactions with just gravitons).
If all $v_k$ are minimal 3-point gravitational interactions, with $n_{m,\, k} = 2$, then \eq{eq:pwrCount} becomes an equality.

We should note that in addition to RG mixing into higher-dimensional operators due to minimally coupled gravity there are also contributions from multiple insertions of effective operators, for example~$\gamma_i^{(w_i + 4)} \propto C_j^{(w_j + 4)}\, C_k^{(w_k + 4)}$ with $w_i = w_j + w_k$. If such a contribution exists and if there is a large separation between the cutoff of the EFT and the Planck scale, i.e.~$\Lambda \ll \Mp$, this contribution is enhanced with respect to the gravitational one by $(\Mp / \Lambda)^{\Delta w}$. 

\begin{figure}[htp]
\centering
\begin{tikzpicture}

\draw[thick,->] (-3,-0.5) -- (12.5,-0.5) node[anchor=west] {$n$};
\draw[thick,->] (-3,-0.5) -- (-3,14.5) node[anchor=south] {$\tilde{h}$};

\foreach \posx / \labelx in {-1/3,1.5/4,4/5,6.5/6,9/7,11.5/8}
   \draw (\posx cm,-0.4) -- (\posx cm,-0.6) node[anchor=north] {$\labelx$};

\foreach \posy / \labely in {1.5/0,4.5/1,7.5/2,10.5/3,13.5/4}
    \draw (-2.9,\posy cm) -- (-3.1,\posy cm) node[anchor=east] {$\labely$};


\node[align=center] (n3h3) at (-1,10.5) {{\color{blue}\small $F^3$};\\ {\color{blue}\small $C^3$}, {\color{blue}\small $C F^2$}}; 

\node[align=center] (n4h4) at (1.5,13.5) {{\small $F^4;$}\\ {\small $C^4, C^2 F^2, C F^3$}};

\node[align=center] (n4h2) at (1.5,7.5) {{\small\color{blue} $F\psi^2\phi$}, {\small\color{blue} $\psi^4$},\\ {\small\color{blue} $F^2 \phi^2$}; {\small \color{blue} $C^2 \phi^2$}\\[0.2cm] {\small $F^2 \phi^2 D^2, F^2\psi \bar{\psi} D,$} \\ {\small $F\psi^2\phi D^2, \psi^4 D^2;$}\\ {\small  $C^2 \phi^2 D^2, C F \psi\bar{\psi} D, $}\\ {\small $C F \phi^2 D^2, C \psi^2 \phi D^2$}};

\node[align=center] (n4h0) at (1.5,1.5) {{\small\color{blue} $\psi^2\bar{\psi}^2$},{\small\color{blue} $\psi\bar{\psi} \phi^2 D$}, {\small\color{blue} $\phi^4 D^2$}\\[0.2cm]{\small $F^2 \bar{F}^2, F\bar{F} \phi^2 D^2,$} \\ {\small $\phi^4 D^4,\bar{F}\psi^2\phi D^2,$}\\ {\small $F\bar{F} \psi \bar{\psi}D, \psi\bar{\psi}\phi^2 D^3,$}\\{\small $F\bar{\psi}^2\phi D^2, \psi^2\bar{\psi}^2 D^2;$}\\{\small $C^2 \bar{C}^2, C^2 \bar{F}^2, \bar{C}^2 F^2$}};

\node[align=center] (n5h3) at (4,10.5) {{\small $F^3\phi^2, F^2\psi^2\phi,F\psi^4;$}\\ {\small $C^3 \phi^2, C F^2\phi^2, C^2 \psi^2 \phi,$}\\ {\small $C F \psi^2 \phi , C \psi^4$}};

\node[align=center] (n5h1) at (4,4.5) {{\small\color{blue} $\psi^2\phi^3$}\\[0.2cm]{\small $F^2\bar{\psi}^2\phi, F\psi^2\bar{\psi}^2,$}\\{\small $F\psi\bar{\psi}\phi^2 D, F \phi^4 D^2,$}\\ {\small $\psi^3\bar{\psi}\phi D,\psi^2\phi^3 D^2;$}\\ {\small $C^2 \bar{\psi}^2\phi$}};

\node[align=center] (n6h2) at (6.5,7.5) {{\small $F^2\phi^4,F\psi^2\phi^3,\psi^4\phi^2;$} \\ {\small $C^2 \phi^4$}};

\node[align=center] (n6h0) at (6.5,1.5) {{\small\color{blue} $\phi^6$}\\[0.2cm]{\small $\phi^6 D^2, \psi\bar{\psi}\phi^4 D,\psi^2\bar{\psi}^2 \phi^2$}};

\node[align=center] (n7h1) at (9,4.5) {{\small $\psi^2\phi^5$}};

\node[align=center] (n8h0) at (11.5,1.5) {{\small $\phi^8$}};
\end{tikzpicture}
\caption{\emph{Classes of independent dimension-six (blue) and -eight (black) operators in the GRSMEFT labeled according to the modified helicity and number of external particles of the minimal amplitude they induce. The non-renormalization theorem in Eq. (\ref{htildemixing}) implies that a Wilson coefficient $C_j$ can only renormalize a $C_i$ that lies inside the cone $\tilde{h}_i \leq \tilde{h}_{j} \pm (n_i-n_j)$. This holds irrespectively of the considered operator dimensions and also allows mixing into higher-dimensional operators according to the power counting rule in Eq.~\eqref{eq:pwrCount}.}}
\label{fig:helOpPlotdim6and8}
\end{figure}
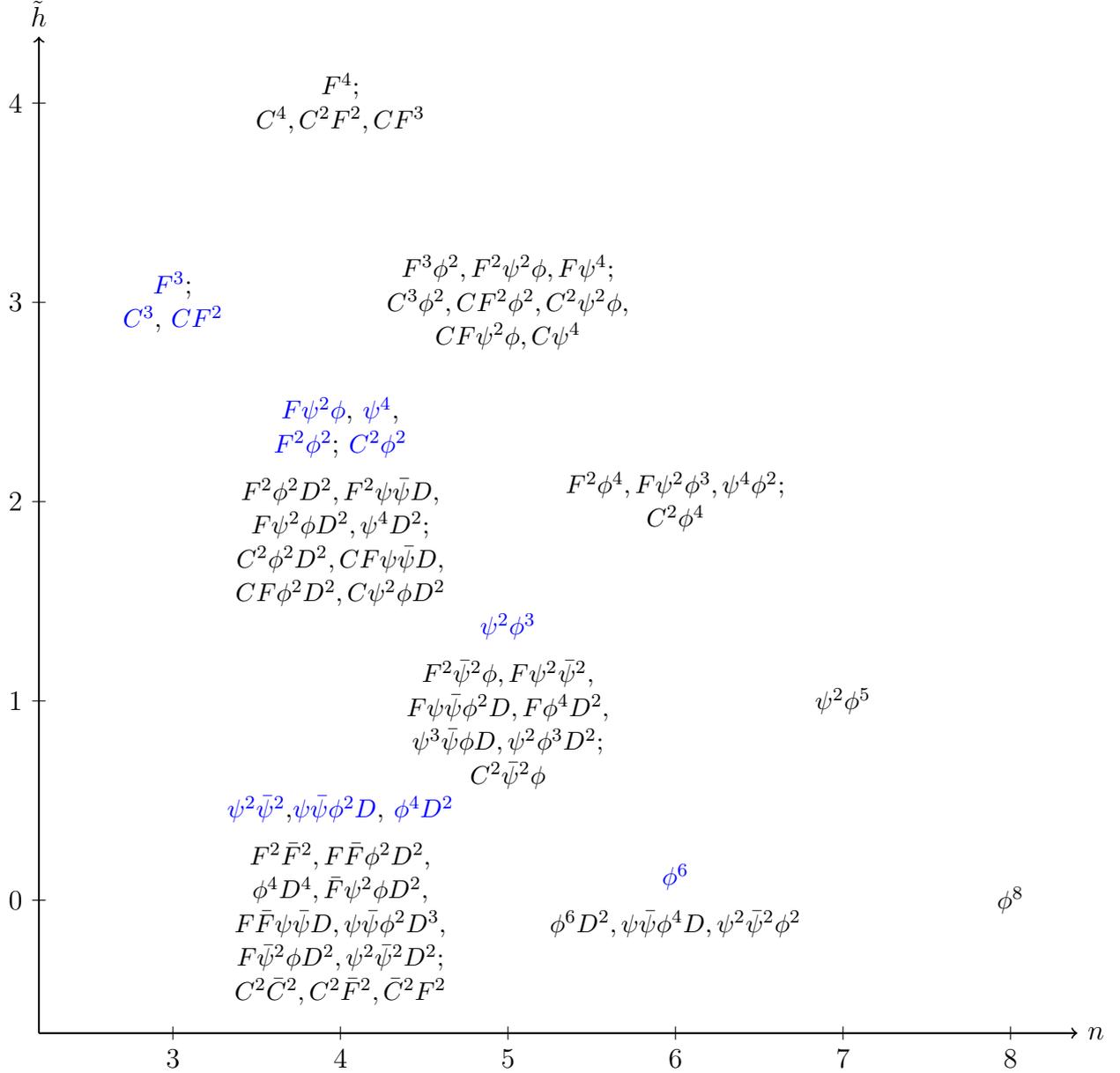

As an explicit example let us consider the dimension-six and -eight operator classes of the GRSMEFT which are shown in Fig.~\ref{fig:helOpPlotdim6and8} in blue and black, respectively. In Section~\ref{mixinggrav} we computed all allowed mixings among the gravitational dimension-six operators. 
For instance we found that the $C^3$ operator with coordinates $(n,\tilde{h}) = (3,3)$ cannot renormalize any other dimension-six operator since it has the maximal number of external graviton fields at dimension six. However, the picture changes when dimension-eight operators are included. Both the helicity selection rule and the power counting rule allow for it to renormalize, among others, $C^4$, $C^2 F^2$ $(4,4)$ and $C^2 \phi^2 D^2$ $(4,2)$, and we see that the number of external gravitons is now allowed to increase or decrease. Similar conclusions also hold for the other gravitational dimension-six operators.

Note that the power counting rule also allows the renormalization of dimension-eight pure matter operators by dimension-six pure matter operators. For instance, $\phi^4 D^2$~$(4,0)$ renormalizes $\phi^4 D^4$~$(4,0)$ and $\psi^2\phi^3$~$(5,1)$ renormalizes $\psi^2 \phi^3 D^2$~$(5,1)$.
Also note that, as mentioned above, these contributions are often subleading for $\Lambda \ll \Mp$, since in general there can be additional contributions with two insertions of dimension-six operators.
Indeed, the anomalous dimension of the dimension-eight Wilson coefficient is schematically of the form
\be
\gamma_i^{(8)} = a_1 \frac{\Lambda^2}{\Mp^2} C_j^{(6)} + a_2 C_k^{(6)}C_l^{(6)}\,,
\ee
where $a_1$ and $a_2$ are constants. In the limit $\Lambda \ll \Mp$ the first term is negligible. This is e.g.~the case for $\phi^4 D^2 \rightarrow \phi^4 D^4$, where the $\gamma_{\phi^4 D^4} \sim (C_{\phi^4 D^2})^2$ contribution dominates.

These arguments can be straightforwardly extended to higher operator dimensions.

\bibliographystyle{JHEP.bst}
\bibliography{AmplitudeDraft,AmplitudeDraftMax,AmplitudeDraftJavi}

\end{document}